\documentstyle[verbatim]{amsart} 
\numberwithin{equation}{section}
\newtheorem{thm}{Theorem}[section]
\newtheorem{lem}[thm]{Lemma}
\newtheorem{prop}[thm]{Proposition}
\newtheorem{cor}[thm]{Corollary}

\begin{document}

\def\({\left (}
\def\){\right )}
\def\[{\left[}
\def\]{\right]}

\def\supermn{{{\frak R}{}^{m|n}}}

\def\sdet{\operatorname{sdet\,}}
\def\str{\operatorname{str\,}}
\def\Mat{\operatorname{Mat\,}}
\def\tr{\operatorname{tr\,}}
\def\sgn{\operatorname{sgn\,}}
\def\euc{{\mathbb R}}
\def\fR{{\frak R}}
\def\fC{{\frak C}}
\def\cev{{\fC}_{\mathrm{ev}}}
\def\cod{{\fC}_{\mathrm{od}}}
\def\rev{{\fR}_{\mathrm{ev}}}
\def\rod{{\fR}_{\mathrm{od}}}
\def\elec{\varepsilon}
\def\frachi{{\frac{\hbar}i}}

\def\ccsl{{/\kern-0.5em{\mathcal  C}}}
\def\cbsl{{/\kern-0.65em{\mathcal  B}}} 
\def\cssl{{/\kern-0.6em{\mathcal  S}}} 
\def\cdsl{{/\kern-0.7em{\mathcal  D}}} 
\def\cesl{{/\kern-0.58em{\mathcal  E}}} 
\def\clsl{{/\kern-0.6em{\mathcal  L}}} 
\def\chsl{{/\kern-0.65em{\mathcal  H}}} 
\def\cpsl{{/\kern-0.6em{\mathcal  P}}} 

\def\where{\quad\text{where}\quad}
\def\when{\quad\text{when}\quad}
\def\with{\quad\text{with}\quad}
\def\for{\quad\text{for}\quad}
\def\forany{\quad\text{for any}\quad}
\def\foreach{\quad\text{for each}\quad}
\def\et{\quad\text{and}\quad}

\def\spin{\kbar}    
\def\kbar{{\mathchar'26\mkern-9mu k}}

\def\pdxi{\frac{\partial}{\partial x_i }}
\def\pdxj{\frac{\partial}{\partial x_j }}
\def\pdqi{\frac{\partial}{\partial q_i }}
\def\pdq{{\frac{\partial}{\partial {q}}}}
\def\pdqj{\frac{\partial}{\partial q_j }}
\def\pdqk{\frac{\partial}{\partial q_k }}
\def\pdxxij{\frac{\partial}{\partial \xi_j }}
\def\pdthk{\frac{\partial}{\partial \theta_k }}
\def\pdtheta{\frac{\partial}{\partial \theta}}
\def\pdppi{\frac{\partial}{\partial \pi}}
\def\pdxxi{\frac{\partial}{\partial \xi}}
\def\pdppik{\frac{\partial}{\partial \pi_k}}
\def\eps{\varepsilon}
\def\dx{d\over {dx}}
\def\dt{{d\over {dt}}}
\def\ds{d\over {ds}}
\def\pdx{{\frac{\partial}{\partial {x}}}}
\def\pdxk{{\frac{\partial}{\partial {x_k}}}}
\def\p2dxjk{\frac{\partial^2}{\partial {x_j}\partial {x_k}}}
\def\pdy{{\frac{\partial}{\partial {y}}}}
\def\pdt{{\frac{\partial}{\partial {t}}}}
\def\pds{{\frac{\partial}{\partial {s}}}}
\def\thea{{\theta_{\alpha}}}
\def\slim{s-\lim}
\def\fa{{\frak a}}
\def\fb{{\frak b}}
\def\fc{{\frak c}}

\def\bmid{|} %{\,|\!|\,}
\def\trp{{}^{t\!}}
\def\dl{(\!(}
\def\dr{)\!)}
\def\trl{(\!(\!(}
\def\trr{)\!)\!)}
\def\trs{|\kern-0.1em|\kern-0.1em|}
\def\sf{{\bold f}}

\def\unbx{\underline{x}}
\def\unbX{\underline{X}}
\def\unbt{\underline{t}}
\def\unbq{\underline{q}}
\def\unbeta{\underline{\eta}}
\def\unbxi{\underline{\xi}}
\def\unbXi{\underline{\Xi}}
\def\unbtheta{\underline{\theta}}
\def\unbth{\underline{\theta}}
\def\unbpi{\underline{\pi}}
\def\unbsigma{\underline{\sigma}}
\def\barx{\bar{x}}
\def\barX{\bar{X}}
\def\bart{\bar{t}}
\def\barxi{\bar{\xi}}
\def\barXi{\bar{\Xi}}
\def\bartheta{\bar{\theta}}
\def\barth{\bar{\theta}}
\def\barpi{\bar{\pi}}
\def\inidata{{\unbx},{\unbxi},{\unbtheta},{\unbpi}}
\def\findata{{\barx},{\barxi},{\bartheta},{\barpi}}
\def\mixdata{{\barx},{\unbxi},{\bartheta},{\unbpi}}
\def\mixxdata{{\barx},{\bartheta},{\unbxi},{\unbpi}}
\def\cbra(#1){\langle{#1}\rangle}
\def\ssm{{supersmooth}}

\def\Eta{\Upsilon}

\def\GL{\operatorname{GL}}

%%%

\title[A new treatise on a system of PDOs]
{An extension of the method of characteristic\\
to a system of Partial Differential Operators\\
-- an application\\
 to the Weyl equation with external field\\
by ``Super Hamiltonian path-integral method"\\}
\author[Atsushi INOUE]
{Atsushi INOUE}
\address{Department of Mathematics,
Tokyo Institute of Technology,
Oh-okayama, Meguro-ku, Tokyo, 152, Japan}

\dedicatory{Dedicated to the memory of Professor Masaya Yamaguti}
\thanks{This work is partially supported by Monbusho Grand-in-aid No.08304010}
\email{inoue@@math.titech.ac.jp}
\date {August 15, 1999, revised \today}
\maketitle
\begin{abstract}
{A system of PDOs(=Partial Differential Operators) 
has {\bf two non-commutativities}, 
(i) one from $[\partial_q,q]=1$
(Heisenberg relation),
(ii) the other from $[A,B]\neq 0$ for $A, B$ being matrices in general.
Non-commutativity from Heisenberg relation is 
nicely controlled by using Fourier transformations 
(i.e. the theory of $\Psi$DOs=pseudo-differential operators).
\par
Here, we give a {\bf new method of treating non-commutativity}
$[A,B]\neq 0$, and explain by taking the Weyl 
equation with external electro-magnetic potentials
as {\bf the simplest representative for a system of PDOs}.
More precisely, we construct a Fourier Integral Operator with
``matrix-like phase and amplitude'' which gives 
a parametrix for that Weyl equation. 
To do this, we first reduce the usual matrix valued Weyl equation
on the Euclidian space %$\euc^3\times \euc$ 
to the one on the superspace, 
called the super Weyl equation.
Using analysis on superspace, we may associate a function, called
the super Hamiltonian function, corresponding to that super Weyl equation.
Starting from this super Hamiltonian function, 
we define phase and amplitude functions
which are solutions of the Hamilton-Jacobi equation 
and the continuity equation on the superspace, respectively.
Then, we define a Fourier integral operator with these phase
and amplitude functions which gives a good parametrix
for the initial value problem of that super Weyl equation.
After taking the Lie-Trotter-Kato limit with respect to the time slicing, 
we get the desired evolutional operator of the super Weyl equation.
Bringing back this result to the matrix formulation,
we have the final result.
Therefore, we get a quantum mechanics with spin from a classical mechanics 
on the superspace which answers partly the problem of Feynman.
}\end{abstract}
\tableofcontents
\baselineskip=16pt

\section{Introduction and Result}

\subsection{The method of characteristics}
It is well-known that the following initial value problem 
on the region $\Omega$ in $\euc^{1+m}$
$$
\left\{
\begin{aligned}
&\pdt u(t,q)+\sum_{j=1}^m a_j(t,q){\pdqj}u(t,q)=b(t,q)u(t,q)+f(t,q),\\
&u({\unbt},q)={\underline{u}}(q),
\end{aligned}\right.
$$
is solved by the method of characteristics.
Denoting the solution of the characteristic equation
$$
\left\{
\begin{aligned}
&\frac d{dt} q_j(t)=a_j(t,q(t)),\\
&q_j({\unbt})={\unbq}_j\quad(j=1,{\cdots},m),
\end{aligned}\right.
$$
by
$$
q(t)=q(t,{\unbt};{\unbq})=(q_1(t), {\cdots}, q_m(t))\in\euc^m,
$$
we get
\begin{prop}
Let $a_j\in C^1(\Omega:\euc)$ and $b, f\in C(\Omega:\euc)$.
For any point $({\unbt},{\unbq})\in\Omega$,
we assume that ${\underline{u}}$ is  $C^1$ in a neighbourhood 
of ${\unbq}$.
\par
Then, in a neighbourhood of $({\unbt},{\unbq})$, 
there exists uniquely a
solution $u(t, q)$. More precisely, putting
$$
U(t,{\unbq})
=e^{\int_{{\unbt}}^t d\tau\, B(\tau,q)}
\left\{\int_{{\unbt}}^t ds\,
e^{-\int_{{\unbt}}^s d\tau\, B(\tau,{\unbq})}F(s,{\unbq})+
{\underline{u}}({\unbq})\right\},
$$
that solution is represented by
$$
u(t,\bar{q})=U(t,y(t,{\unbt};{\bar{q}}))
$$
where
$B(t,{\unbq})=b(t,q(t,{\unbt};{\unbq}))$,  
$F(t,{\unbq})=f(t,q(t,{\unbt};{\unbq}))$
and
${\unbq}=y(t,{\unbt};{\bar{q}})$ 
is a inverse function derived from
${\bar{q}}=q(t,{\unbt};{\unbq})$.
\end{prop}

{Problem}: Is it possible to {\bf extend the method of characteristics 
to the system of PDOs}?

\subsection{Weyl equation}
We take {\bf the Weyl equation
as the simplest representative of a system of PDOs} and we 
give an answer of the following problem. 
\par
{\bf Problem}: 
{\it Find if possible, an {\bf explicit representation} of 
$\psi(t,q):{\mathbb R}\times{\mathbb R}^3\to{\mathbb C}^2$ satisfying
\begin{equation}\tag{W}
\left\{
\begin{split}
&i\hbar \pdt \psi(t,q)={\mathbb H}(t)\psi(t,q),\\
& \psi(\unbt,q)={\underline{\psi}}(q),
\end{split}
\right.
\label{rt1.1}
\end{equation}
where, $\unbt$ is an arbitrarily fixed initial time and
\begin{equation}
{\mathbb H}(t)={\mathbb H}\Big(t,q,\frachi\pdq\Big)
=\sum_{k=1}^3 c\boldsymbol{\sigma}_k 
\Big(\frachi \frac{\partial}{\partial q_k}-\frac{\elec}{c}
A_k(t,q)\Big)+{\elec}A_0(t,q).
\label{rt1.1-2}
\end{equation}
}

In the above, the Pauli matrices $\{\boldsymbol{\sigma}_j\}$ 
is represented by
\begin{equation*}
\boldsymbol{\sigma}_1=\begin{pmatrix}
0&1\\
1&0
\end{pmatrix},\quad
\boldsymbol{\sigma}_2=\begin{pmatrix}
0&-i\\
i&0
\end{pmatrix},\quad
\boldsymbol{\sigma}_3=\begin{pmatrix}
1&0\\
0&-1
\end{pmatrix},\quad
\boldsymbol{\sigma}_0={\mathbb I}_2=\begin{pmatrix}
1&0\\
0&1
\end{pmatrix}.
%\label{rt1.2}
\end{equation*}

{\it Remark.} Though the meaning of {explicit representation} is not so clear,
but it's meaning will be clarified in the sequel.

{\bf Claim}:
We define a {\bf good parametrix} 
for the initial value problem for the Weyl equation with
a given external time-dependent electro-magnetic field 
$(A_0(t,q),A_1(t,q),A_2(t,q),$ $A_3(t,q))$ 
by {\bf Super Hamiltonian Path-Integral Method}.
Here, essential is to introduce
 ``the Hamiltonian mechanics corresponding to the Weyl equation",
and to define Fourier Integral Operators using quantities based on the
Hamilton-Jacobi equation.

\subsection{Important Remark}
In general, if one wants to study 
the precise properties of the solution of PDO,
one uses the properties of corresponding classical mechanics 
defined by the symbol of PDO to
mention the fine structure of the solution of PDOs.
But this theory of $\Psi$DOs
isn't applicable directly to a system of PDOs. For example, one doesn't
know {\bf how one
defines the Hamilton flow for a system of PDOs}. 
Therefore, 
one needs to apply the
technique of $\Psi$DOs after diagonalizing the given system.

In this paper, we treat the aforementioned
 {\bf two non-commutativities on equal footing}
by introducing odd variables, therefore, we treat matrices as they are.

Our object of this paper is {\bf not}
to prove the well-posedness of \eqref{rt1.1} as a symmetric hyperbolic system.
Rather, we treat a system of PDOs {\bf as it is},
in other word, we never concern with the properties of
characteristic roots of the given system, 
but we {\bf define the Hamilton flow} for that system 
after reformulating it on the superspace.
(As is well-known, we may apply the so-called energy methods
to a symmetric hyperbolic system without regarding
characteristic roots.)

\subsection{Superspace setting and Result}
By
{\bf introducing odd variables to decompose the matrix structure},
we first reduce the usual matrix valued Weyl equation \eqref{rt1.1}
on the Euclidian space $\euc\times\euc^3$ 
to the one on the superspace $\euc\times{\fR}^{3|2}$,
%with value in ${\fC}_{ev}$, 
called {\bf the super Weyl equation \eqref{rt1.3}} 
on the superspace $\euc\times{\fR}^{3|2}$:
\begin{equation}\tag{SW}
\left\{
{\begin{aligned}
&i\hbar \pdt u(t,x,\theta)
={\mathcal H}\Big(t,x,\frachi \pdx,\theta,{\pdtheta}\Big)u(t,x,\theta),\\
&u(\unbt,x,\theta)={\underline{u}}(x,\theta).
\end{aligned}}
\right.
\label{rt1.3}
\end{equation}

{\it Remark }: The most important thing here is that
every quantities appeared above \eqref{rt1.3} 
are {\bf like scalars though non-commutative}!
\vspace{2mm}

{\bf Claim}: There exists the classical mechanics corresponding to
the (super) Weyl equation and that a parametrix of it
is constructed as a Fourier integral operator
using phase and amplitude functions defined
by that classical mechanics. (We call this {\bf a good parametrix}
because
not only it gives a parametrix but also its dependence on
the quantities from classical mechanics is explicit.)

Therefore, the (super) Weyl equation is regarded as to be
obtained by {\bf quantizing that classical mechanics after Feynman's procedure}.
Because that (super) Weyl equation is 
``of first order" both in ``even and odd variables",
we need to modify Feynman's argument 
from Lagrangian to Hamiltonian formulated ``path-integral" 
(compare arguments between
Fujiwara \cite{Fu79} and Inoue \cite{In99-1} for the Schr\"odinger equation
and see also Inoue and Maeda \cite{IM85} 
for the heat equation with the scalar curvature term).

The precise meaning of the above claim is formulated as follows:
\begin{thm}
Let $\{A_j(t,q)\}_{j=0}^3\in C^\infty(\euc\times\euc^3:\euc)$ satisfy,
for any $k=0,1,2,{\cdots}$,
\begin{equation}  %\tag{G}
\trs A_j\trs_{k,\infty}
=\sup_{t,q,|\gamma|=k}|(1+|q|)^{|\gamma|-1}
\partial_q^\gamma A_j(t,q)|<\infty \for j=0,{\cdots},3.
\label{G}
\end{equation}
For $|t-\unbt|$ sufficiently small, 
we have {\bf a good parametrix} for \eqref{rt1.3} represented by
$$
{\mathcal U}(t,\unbt)\,u(x,\theta)
=(2\pi\hbar)^{-3/2}\hbar\int_{\fR^{3|2}}d\xi\,d\pi\,
{\mathcal D}^{1/2}({t},\unbt\,;{x,\theta,\xi,\pi})
e^{i\hbar^{-1}{\mathcal S}({t},\unbt\,; {x,\theta,\xi,\pi})}
{\mathcal F}{\underline{u}}(\xi,\pi),
$$
for $u(x,\theta)\in {\cdsl}_{\mathrm{SS,ev}}$ 
and is extended to $\clsl_{\mathrm{SS,ev}}^2({\frak R}^{3|2})$.
Here, ${\mathcal S}({t},\unbt\,; {x,\theta,\xi,\pi})$ and 
${\mathcal D}({t},\unbt\,; {x,\theta,\xi,\pi})$ satisfy 
the Hamilton-Jacobi equation and the continuity equation, respectively:
$$
({\mathrm{H}}\!-\!{\mathrm{J}})\;\left\{
{\begin{aligned}
&\pdt {\mathcal S} %(t,{\unbt};{x,\theta,\xi,\pi})
+{\mathcal H}\Big(t,{x},\frac{\partial {\mathcal S}}{\partial{x}},
{\theta},\frac{\partial {\mathcal S}}{\partial{\theta}}\Big)=0,\\
&{\mathcal S}(\unbt,\unbt\,;{x,\theta,\xi,\pi})=\langle {x}|{\xi}\rangle+
\langle{\theta}|{\pi}\rangle,
\end{aligned}}
\right.
\et
({\mathrm C})\;\left\{
{\begin{aligned}
&\pdt {\mathcal D}+
\frac{\partial}{\partial{x}}
\Big({\mathcal D}
\frac{\partial{\mathcal H}}{\partial {\xi}}\Big)
+\frac{\partial}{\partial {\theta}}
\Big({\mathcal D}
\frac{\partial{\mathcal H}}{\partial {\pi}}\Big)=0,\\
&{\mathcal D}(\unbt,\unbt\,;{x,\theta,\xi,\pi})=1.
\end{aligned}}
\right.
$$
\end{thm}
\par{\it Remark.} The assumption \eqref{G} garantees the suitable estimates
for phase and amplitude functions.

Using the identification maps
$$
\sharp:L^2(\euc^3:{\mathbb C}^2)\to \clsl_{\mathrm{SS,ev}}^2({\frak R}^{3|2}) \et 
\flat:\clsl_{\mathrm{SS,ev}}^2({\frak R}^{3|2})\to L^2(\euc^3:{\mathbb C}^2),
$$
we get
\begin{cor}
Let $\{A_j(t,q)\}_{j=0}^3\in C^\infty(\euc\times\euc^3:\euc)$ satisfy \eqref{G}.
For $|t-\unbt|$ sufficiently small, 
we have a good parametrix for \eqref{rt1.1} represented by
$$
{\mathbb U}(t,\unbt){\underline{\psi}}(q)=\flat
(2\pi\hbar)^{-3/2}\hbar\int_{\fR^{3|2}}d\xi\,d\pi\,
{\mathcal D}^{1/2}({t},\unbt\,; {x,\theta,\xi,\pi})
e^{i\hbar^{-1}{\mathcal S}({t},\unbt\,; {x,\theta,\xi,\pi})}
{\mathcal F}(\sharp{\underline{\psi}})(\xi,\pi)
\Big|_{x_{\mathrm B}=q}.
$$
\end{cor}

%%%

%%\input ams-Weyl1-2

\par
{\it Remarks. } (0) The result of this paper is announced in Inoue \cite{In99-3}.
Though the notion of superanalysis on
Fr\'echet-Grassmann algebra seems not so familiar even today, 
we have no space to present a part of elements of superanalysis in this paper.
If the reader is strange to the notion such as
even and odd variables,
elementary and real analysis on super-smooth functions, 
please consult,
\cite{In92,In98-1,In98-2,In99-3,IM91}. 
We hope the procedure employed here will help to 
study the propagation of 
singularities of a system of PDOs and others,
after developping  
theories of $\Psi$DOs and FIOs to those on superspaces.
\newline
(1) {\bf The problem how to regard the spin system 
following the Feynman's principle},
is posed in p.355 of the book of Feynman \& Hibbs \cite{FH65}.
J.L. Martin \cite{Mar59-1,Mar59-2} gave an idea 
of how to make the correspondence
from the Fermi oscillator to the classical objects on a non-commutative algebra. 
On the other hand, without knowing Martin's results, 
Berezin \& Marinov \cite{BM77} tried to construct the classical mechanics
which produces both boson and fermion equally by ``quantization".
\par
We answer a part of {\bf the problem of Feynman}
using superanalysis,
by taking the Weyl equation as the typical 
and the simplest example of the spin system.
\begin{quotation}
Feynman proposed that the solution of Schr\"odinger equation
$$
i\hbar\frac{\partial u}{\partial t}=-\frac{\hbar^2}{2M^2}
\sum_{i=1}^m\frac{\partial^2 u}{\partial q_i^2}+V(t,q)u
$$
is represented by
$$
u(t,q)=\int_{{\mathbb R}^m} dq'F(t,q,q')u(0,q')
$$
with
$$
\begin{aligned}
&F(t,q,q')=\int_{{\mathcal{P}}_{t,q,q'}}[d_F\gamma]\,{\exp}{\bigg[i\hbar^{-1}\int_0^t
ds\,L(s,\gamma(s),\dot{\gamma}(s))\bigg]},\\
&L(t,q,\dot{q})=\frac{\dot{q}^2}{2M^2}+V(t,q),\\
&{\mathcal{P}}_{t,q,q'}=\{\gamma\in AC([0,t]:\euc^m)\;|\;\gamma(0)=q',\,\gamma(t)=q\},\\
&\qquad\qquad\with d_F\gamma=\mbox{notorious Feynman measure on
${\mathcal{P}}_{t,q,q'}$}.
\end{aligned}
$$
Feynman-Hibbs \cite{FH65} posed a question whether this derivation of quantum mechanics
from classical quantities is applicable to spin systems, such as Dirac and Pauli
equations.

Since there exists no non-trivial Feynman measure $[d_F\gamma]$ on
${\mathcal{P}}_{t,q,q'}$, we try to give a mathematical meaning to the above expression.
Under certain condition on $V$, Fujiwara \cite{Fu79} defines
$$
F(t,0)u(q)=\int_{\euc^m}dq'\, D(t,0,q,q')^{1/2}e^{i\hbar^{-1}S(t,0,q,q')}u(q')
$$
with
$$
S(t,0,q,q')=\inf_{\gamma\in{\mathcal{P}}_{t,q,q'}}\int_0^t
ds\,L(s,\gamma(s),\dot{\gamma}(s)),\quad
D(t,0,q,q')=\det\bigg(\frac{\partial^2 S(t,0,q,q')}{\partial q_i\partial q'_j}\bigg),
$$
and he proves that $F(t,0)$ gives a good parametrix.
Moreover, Fujiwara \cite{Fu80} gives a kernel representation of the fundamental solution.

Since the above procedure is based on the Lagrangian formulation, we {\bf need to
reformulate} it in the Hamiltonian setting as Inoue \cite{In99-1},
called Hamiltonian path-integral method. 
After reformulating a certain first order system of PDOs in the superspace,
we claim to apply the above Hamiltonian path-integral procedure to that first order
system of PDOs. But this paper gives a partial answer to the Feynman's problem,
because we
have not yet constructed  an ``explicit integral representation" 
of the fundametal
solution itself. To do this, we need to prepare
more elaborated composition formulas of FIOs as in \cite{Fu80}.
\end{quotation}
\par\noindent
(2) We may extend our procedure to the case  
where the electro-magnetic potentials are valued in $2\times2$-matrices,
if the quantity $\tilde{A}_0$ defined below is real valued 
and the condition \eqref{G} is satisfied for all $\{A_j^{[*]}\}$.

In fact, by decomposing
$$
{\bold A}_j(t,q)=A_j^{[0]}{\Bbb I}_2+A_j^{[1]}{\pmb \sigma}_1
+A_j^{[2]}{\pmb \sigma}_2
+A_j^{[3]}{\pmb \sigma}_3\for j=0,\cdots,3,
\quad\mbox{with $A_j^{[*]}=A_j^{[*]}(t,q)\in \euc$},
$$
we have
$$
{\pmb \sigma}_j{\bold A}_j(t,q)+{\bold A}_0(t,q){\Bbb I}_2
={\pmb \sigma}_j\tilde{A}_j(t,q)+\tilde{A}_0(t,q){\Bbb I}_2,
$$
where
$$
\begin{gathered}
\tilde{A}_1=A_1^{[0]}+A_0^{[1]}+i(A_2^{[3]}-A_3^{[2]}),\;
\tilde{A}_2=A_2^{[0]}+A_0^{[2]}+i(A_3^{[1]}-A_1^{[3]}),\;
\tilde{A}_3=A_3^{[0]}+A_0^{[3]}+i(A_1^{[2]}-A_2^{[1]}),\\
\tilde{A}_0=A_0^{[0]}+A_1^{[1]}+A_2^{[2]}+A_3^{[3]}.
\end{gathered}
$$

\par\noindent
(3) Even if the electro-magnetic field 
is time-independent, that is, ${\Bbb H}(t)={\Bbb H}$, 
and the existence of a self-adjoint realization of $\Bbb H$ in
$L^2(\euc^3:{\Bbb C}^2)$ is assured,
our result above is new in the following sense.
It is well-known that 
$e^{-i\hbar^{-1}t{\Bbb H}}{\underline\psi}$ gives a solution of
\eqref{rt1.1} by Stone's theorem. Moreover, by the kernel theorem of Schwartz,
there exists a distribution ${\Bbb E}(t,q,q')$ such that
$e^{-i\hbar^{-1}t{\Bbb H}}{\underline\psi}(q)
=\langle {\Bbb E}(t,q,\cdot),{\underline\psi}(\cdot)\rangle$,
where $\langle\cdot,\cdot\rangle$ is 
the duality between ${\mathcal D}'$ and ${\mathcal D}$.
Therefore, our result here answers partly {\bf the problem how one may express 
the kernel of operator $e^{-i\hbar^{-1}t{\Bbb H}}$ 
using ``classical quantities"}.
\newline
(4) Especially, in case ${\elec}=0$, ${\Bbb U}(t,0)$ 
gives an explicit solution 
of \eqref{rt1.3} with %
\begin{equation*}%\tag*{(FSW)}
\begin{gathered}
\gamma_t=c\hbar^{-1}t|\unbxi|,\quad
\delta_t=|\unbxi|\cos\gamma_t-i\unbxi_3\sin\gamma_t,\\
{\mathcal S}_{{\elec}=0}(t,0\,;{\mixdata})=\langle\barx|\unbxi\rangle
+\frac{|\unbxi|\langle\bartheta|\unbpi\rangle
-\hbar(\unbxi_1+i\unbxi_2)\sin\gamma_t\,\bartheta_1\bartheta_2
-\hbar^{-1}(\unbxi_1-i\unbxi_2)\sin\gamma_t\,\unbpi_1\unbpi_2}{\delta_t},\\
{\mathcal D}_{{\elec}=0}(t,0\,;{\mixdata})=|\unbxi|^{-2}
\big[|\unbxi|\cos\gamma_t-i\unbxi_3\sin\gamma_t\big]^2.
\end{gathered}
\end{equation*}
It is worth remarking that ``classical quantities" above contain the parameter
$\hbar$. This means in a sense that ``a spinning particle has 
no classical counter-part" as Pauli claimed one day.
\newline
(5) We use Einstein's convention to summing up repeated indeces
unless there occurs confusion.

%%%

%%\input s-Weyl2

\section{Outline of Proof}

\subsection{Interpretation of the method of characteristics}
To explain the meaning of ``explicit representation", 
we reconsider the method of characteristics by taking the simplest example:
\begin{equation}
\left\{
{\begin{aligned}
&i\hbar\pdt u(t,q)=a\frachi{\pdq} u(t,q)+bqu(t,q),\\
&u(0,q)={\underline{u}}(q),
\end{aligned}}\right.
\for a,b\in\euc.
\label{sqm}
\end{equation}
From the right-hand side of above, we define a Hamiltonian as follows
(more precisely, the Weyl symbol should be derived):
$$
H(q,p)=e^{-i\hbar^{-1}qp}\Big(a\frachi{\pdq}+bq\Big)e^{i\hbar^{-1}qp}=ap+bq.
$$
The classical mechanics (or bicharacteristic) associated to this Hamiltonian
is given by
\begin{equation}
{\left\{
{\begin{aligned}
&\dot q(t)=H_p=a,\\
&\dot p(t)=-H_q=-b
\end{aligned}}\right.}
\;\mbox{with the initial data}\;
{\begin{pmatrix}
q(0)\\
p(0)
\end{pmatrix}
=\begin{pmatrix}
{\unbq}\\
{\underline{p}}
\end{pmatrix}}
\label{scm}
\end{equation}
which is readily solved as
$$
q(s)={\unbq}+as,\quad p(s)={\underline{p}}-bs.
$$
From above Proposition, putting ${\underline {t}}=0$, we get readily that
$$
U(t,{\unbq})=
{\underline{u}}({\unbq})
e^{-i\hbar^{-1}(b{\unbq}t+2^{-1}ab{t^2})}.
$$
As the inverse function of ${\overline{q}}=q(t,{\unbq})$ 
is given by ${\unbq}=y(t,\bar q)=\bar q-at$, we get
$$
u(t,{\overline{q}})=U(t,{\unbq})|_{{\unbq}=y(t,\bar q)}
={\underline{u}}({\overline{q}}-at)
e^{-i\hbar^{-1}(b{\overline{q}}t-2^{-1}ab{t^2})}.
$$
We remark that there is no flavor of classical mechanics of this expression
because we use only a part $q(\cdot)$ 
of bicharacteristics $(q(\cdot),p(\cdot))$.

Another point of view from {\bf Hamiltonian path-integral method}:
Put
$$
S_0(t,{\unbq},{\underline{p}})
=\int_0^t ds \,[\dot q(s)p(s)-H(q(s),p(s))]
=-b{\unbq}t-2^{-1}ab{t^2},
$$
$$
S(t,{\overline{q}},{\underline{p}})
=\bigg({\unbq}\,{\underline{p}}+S_0(t,{\unbq},{\underline{p}})\bigg)
\bigg|_{{\unbq}=y(t,{\overline{q}})}
={\overline{q}}{\underline{p}}-a{\underline{p}}t-b{\overline{q}}t+2^{-1}ab{t^2}.
$$
Then, the classical action
$S(t,{\bar{q}},{\underline{p}})$ 
satisfies the Hamilton-Jacobi equation.
$$
\left\{
\begin{aligned}
&\pdt S+H({\bar{q}},\partial_{\bar{q}}S)=0,\\
&S(0,{\bar{q}},{\underline{p}})={\bar{q}}\,{\underline{p}}.
\end{aligned}
\right.
$$
On the other hand, the van Vleck determinant
(though scalar in this case) is calculated as
$$
D(t,{\overline{q}},{\underline{p}})
={\frac{\partial^2S(t,{\overline{q}},{\underline{p}})}
{\partial{\overline{q}}\partial{\underline{p}}}}=1.
$$
This quantity satisfies the continuity equation:
$$
\left\{\begin{aligned}
&\pdt {D}+\partial_{\bar{q}}({D} H_p)=0
\quad\text{where
$ H_p=\frac{\partial H}{\partial {p}}
\big({\bar{q}},\frac{\partial S}{\partial \bar{q}}\big)$},\\
&{D}(0,{\bar{q}},{\underline{p}})=1.
\end{aligned}
\right.
$$

As an interpretation of Feynman's idea, 
we regard that {\bf the transition from classical to quantum mechanics}
is to study the following quantity or the one represented by this procedure
 (the term ``quantization" is not so well-defined mathematically):
\begin{equation}
u(t,{\overline{q}})=(2\pi\hbar)^{-1/2}\int_{\euc} d{\underline{p}}\,
D^{1/2}(t,{\overline{q}},{\underline{p}})
e^{i\hbar^{-1}S(t,{\overline{q}},{\underline{p}})}
\hat{\underline{u}}({\underline{p}}).
\label{*1}
\end{equation}
That is, in our case at hand, we should study the quantity defined by
$$
\begin{aligned}
u(t,{\overline{q}})
&=(2\pi\hbar)^{-1/2}\int_{\euc} d{\underline{p}}\,
e^{i\hbar^{-1}S(t,{\overline{q}},{\underline{p}})}
\hat{\underline{u}}({\underline{p}})\\
&=(2\pi\hbar)^{-1}\iint_{\euc^2} d{\underline{p}}d{\unbq}\,
e^{i\hbar^{-1}(S(t,{\overline{q}},{\underline{p}})-{\unbq}\,{\underline{p}})}
{\underline{u}}({\unbq})
={\underline{u}}({\overline{q}}-at)
e^{i\hbar^{-1}(-b{\overline{q}}t+2^{-1}ab{t^2})}.
\end{aligned}
$$
Therefore, we may say that this second construction \eqref{*1}
gives the explicit connection
between the solution \eqref{sqm} 
and the classical mechanics given by \eqref{scm}.
We feel the above expression ``good'' because there appear two classical
quantities $S(t,{\overline{q}},{\underline{p}})$ 
and $D(t,{\overline{q}},{\underline{p}})$ explicitly
and we regard this procedure as an honest follower of Feynman's spirit.

{\bf{Claim}}: 
Applying superanalysis,
we may extend the second argument above to a system of PDOs 
e.g.
quantum mechanical equations with spin such as Dirac, Weyl or Pauli equations,
(and if possible, any other system of PDOs),
after interpreting these equations
as those on superspaces.

\subsection{Our procedure}
(I) We identify a ``spinor"
$\psi(t,q)={}^t(\psi_1(t,q),\psi_2(t,q))
:{\Bbb R}\times{\Bbb R}^3\to {\Bbb C}^2$
with an even supersmooth function 
$u(t,x,\theta)=u_0(t,x)+u_1(t,x)\theta_1\theta_2:
{\Bbb R}\times{\frak R}^{3|2}\to \cev$. 
Here, ${\frak R}^{3|2}$ is the superspace and
$u_0(t,x)$, $u_1(t,x)$ are the Grassmann continuation of $\psi_1(t,q)$,
$\psi_2(t,q)$, respectively.
(The reason why we don't identify $\psi(t,q)$ 
with $u(t,x,\theta)=u_0(t,x)+u_1(t,x)\theta$
for one odd variable $\theta$, is clarified in \cite{In98-1}.)

(II) We represent matrices $\{\boldsymbol{\sigma}_j\}$ as (even) operators 
acting on $u(t,x,\theta)$ such that
\begin{equation}
\left\{
{\begin{aligned}
&\sigma_1\Big(\theta,
\frac{\partial}{\partial\theta}\Big)=
\theta_1\theta_2
-\frac{\partial^2}{\partial\theta_1\partial\theta_2},\\
&\sigma_2\Big(\theta,
\frac{\partial}{\partial\theta}\Big)=
i\Big(\theta_1\theta_2
+\frac{\partial^2}{\partial\theta_1\partial\theta_2}\Big),\\
&\sigma_3\Big(\theta,
\frac{\partial}{\partial\theta}\Big)=
1-\theta_1\frac{\partial}{\partial\theta_1}
-\theta_2\frac{\partial}{\partial\theta_2}.
\end{aligned}}\right.\label{rt2.1}
\end{equation}

(III) Therefore, we may introduce
the differential operator which corresponds to ${\Bbb H}(t,q,-i\hbar\partial_q)$:
\begin{equation}
{\mathcal H}\Big(t,x,\frachi \pdx,\theta,{\pdtheta}\Big)=\sum_{j=1}^3c\sigma_j\Big(\theta,
\frac{\partial}{\partial\theta}\Big)
\Big({\frachi}\frac{\partial}{\partial x_j}-\frac{\elec}{c} A_j(t,x)\Big)+{\elec}A_0(t,x).
\label{rt2.3}
\end{equation}
It yields the superspace version of the Weyl equation represented by
\begin{equation}
\left\{
{\begin{aligned}
&i\hbar \pdt u(t,x,\theta)
={\mathcal H}\Big(t,x,\frachi \pdx,\theta,{\pdtheta}\Big)u(t,x,\theta),\\
&u(\unbt,x,\theta)={\underline{u}}(x,\theta).
\end{aligned}}
\right.
\label{rt2.4}
\end{equation}

(IV) Using the Fourier transformation on superspace ${\fR}^{m|n}$,
we have the ``complete Weyl symbol'' of \eqref{rt2.3} as ordinary case. 
In our case, we put $\kbar=\hbar$, $n=2$ and $v(\theta)=v_0+v_1\theta_1\theta_2$,
$w(\pi)=w_0+w_1\pi_1\pi_2$:
$$
\begin{gathered}
(F_o v)(\pi) =\hbar
\int_{\fR^{0|2}} d\theta\,
e^{-i\hbar^{-1}\langle\theta|\pi\rangle}v(\theta)
=\hbar v_1+\hbar^{-1}v_0\pi_1\pi_2,\\ 
({\bar F}_o w)(\theta) =
\hbar
\int_{\fR^{0|2}} d\pi\,
e^{i\hbar^{-1}\langle\theta|\pi\rangle}w(\pi)
=\hbar w_1+\hbar^{-1}w_0\theta_1\theta_2.
\end{gathered}
$$ 

Therefore, the Weyl symbols of $\sigma_j(\theta,\partial_\theta)$ are given by
\begin{equation}
\left\{
{\begin{aligned}
&\sigma_1(\theta,\pi)=\theta_1\theta_2+\hbar^{-2}\pi_1\pi_2,\\
&\sigma_2(\theta,\pi)=i(\theta_1\theta_2-\hbar^{-2}\pi_1\pi_2),\\
&\sigma_3(\theta,\pi)=-i\hbar^{-1}(\theta_1\pi_1+\theta_2\pi_2).
\end{aligned}}\right.\label{rt2.2}
\end{equation}
Moreover, we put
\begin{equation}
{\mathcal H}(t,x,\xi,\theta,\pi)
=\sum_{j=1}^3 c\sigma_j(\theta,\pi)\big(\xi_j-\frac{\elec}{c}A_j(t,x)\big)
+{\elec}A_0(t,x).
\label{rt2.5}
\end{equation}

(V) As $\mathcal H$ is even, we may consider the classical mechanics corresponding to 
${\mathcal H}(t,x,\xi,\theta,\pi)$: % given by
\begin{equation}\tag*{$\rm (2.6){}_{\mathrm{ev}}$}
\allowdisplaybreaks
\left\{
{\begin{aligned}
&\dt x_j
=\frac{\partial{\mathcal H}(t,x,\xi,\theta,\pi)}{\partial \xi_j}=c\sigma_j(\theta,\pi),\\
&\dt \xi_j
=-\frac{\partial{\mathcal H}(t,x,\xi,\theta,\pi)}{\partial x_j}
=\sum_{k=1}^3{\elec}\sigma_k(\theta,\pi)\frac{\partial A_k(t,x)}{\partial x_j}
-{\elec}\frac{\partial A_0(t,x)}{\partial x_j},\\
\end{aligned}}
\right.\label{rt2.6-1}
\end{equation}
\begin{equation}\tag*{$\rm (2.6){}_{\mathrm{od}}$}
\allowdisplaybreaks
\left\{
{\begin{aligned}
&\dt \theta_1=-\frac{\partial{\mathcal H}(t,x,\xi,\theta,\pi)}{\partial \pi_1}
=-c\hbar^{-2}(\eta_1-i\eta_2)\pi_2-ic\hbar^{-1}\eta_3\theta_1,\\
&\dt \theta_2=-\frac{\partial{\mathcal H}(t,x,\xi,\theta,\pi)}{\partial \pi_2}
=c\hbar^{-2}(\eta_1-i\eta_2)\pi_1-ic\hbar^{-1}\eta_3\theta_2,\\
&\dt \pi_1=-\frac{\partial{\mathcal H}(t,x,\xi,\theta,\pi)}{\partial \theta_1}
=-c(\eta_1+i\eta_2)\theta_2+ic\hbar^{-1}\eta_3\pi_1,\\
&\dt \pi_2=-\frac{\partial{\mathcal H}(t,x,\xi,\theta,\pi)}{\partial \theta_2}
=c(\eta_1+i\eta_2)\theta_1+ic\hbar^{-1}\eta_3\pi_2.
\end{aligned}}
\right.\label{rt2.6-2}
\end{equation}
\addtocounter{equation}{1}
In the above, we put
\begin{equation}
\eta_j(t)=\xi_j(t)-\frac{\elec}{c}A_j(t,x(t)) \for j=1,2,3,
\label{rt2.7}
\end{equation}
and at time $t=\unbt$, the initial data are given by
\begin{equation}
(x(\unbt),\xi(\unbt),\theta(\unbt),\pi(\unbt))=({\inidata}).
\end{equation}

Then, we have the following existence theorem.
\begin{thm}\label{existenceCM}
Let $\{A_j(t,q)\}_{j=0}^3\in C^\infty(\euc\times\euc^3:\euc)$.
\newline
(0) For any $T>0$ and any initial data 
$({\inidata})
\in{\fR}^{6|4}={\mathcal T}^*{\fR}^{3|2}$,
there exists a unique solution 
$(x(t),\xi(t),\theta(t),\pi(t))$ of \ref{rt2.6-1} and \ref{rt2.6-2} on
$[-T,T]$.
\newline
(1) The solution $(x(t),\xi(t),\theta(t),\pi(t))$ 
of \ref{rt2.6-1} and \ref{rt2.6-2}
on $[-T,T]$ is ``s-smooth'' in $(t,{\inidata})$.
That is, smooth in $t$ for fixed $({\inidata})$ 
and supersmooth in $({\inidata})$ for fixed  $t$.
\newline
(2) Assume, moreover, that $\{A_j(t,q)\}_{j=0}^3$ satisfy \eqref{G}.
\par
(i) Then, we have, for $t, {\unbt}\in[-T,T]$, 
and $k=|\alpha+\beta|=0,1,2,\cdots,$
\begin{equation}
\left\{
\begin{aligned}
&{\bmid}\pi_{\mathrm B}\partial_{\unbx}^\alpha \partial_{\unbxi}^\beta 
(x(t,{\unbt}\,; {\inidata})-{\unbx}){\bmid} %({\unbx}^{[0]},{\unbxi}^{[0]},0,0)
=0,\\ 
&{\bmid}\pi_{\mathrm B}\partial_{\unbx}^\alpha \partial_{\unbxi}^\beta 
(\xi(t,{\unbt}\,; {\inidata})-{\unbxi}){\bmid}
\le {\elec}|t-{\unbt}|\delta_{0|\beta|}%\langle{\unbx}^{[0]}\rangle^{1-|\alpha|}
\trs A_0\trs_{|\alpha|+1,\infty}.
\end{aligned}
\right.
\label{rt3.24}
\end{equation}
\par
(ii) Let $|t-{\unbt}|\le1$.
If $|a+b|=1$ and $k=|\alpha+\beta|=0,1,2,\cdots,$ there exist constants
$C_{1}^{(k)}$ (with $C_{1}^{(0)}=0$) 
independent of $(t,\unbxi,\unbtheta)$ such that
\begin{equation}
\left\{
\begin{aligned}
&{\bmid}\pi_{\mathrm B}\partial_{\unbx}^\alpha \partial_{\unbxi}^\beta 
\partial_{\unbtheta}^a \partial_{\unbpi}^b 
(\theta(t,{\unbt}\,; {\inidata})-\unbtheta)){\bmid}
\le C^{(k)}_{1}|t-{\unbt}|^{(1/2)(1-(1-k)_+)},\\ 
&{\bmid}\pi_{\mathrm B}\partial_{\unbx}^\alpha \partial_{\unbxi}^\beta 
\partial_{\unbtheta}^a \partial_{\unbpi}^b 
(\pi(t,{\unbt}\,; {\inidata})-\unbpi)){\bmid}
\le C^{(k)}_{1}|t-{\unbt}|^{(1/2)(1-(1-k)_+)}. 
\end{aligned}
\right.
\label{rt3.25}
\end{equation}
\par
(iii) Let $|t-{\unbt}|\le1$.
If $|a+b|=2$ and $k=|\alpha+\beta|=0,1,2,\cdots,$ there exist constants
$C_{2}^{(k)}$ independent of $(t,\unbxi,\unbtheta)$ such that
\begin{equation}
\left\{
\begin{aligned}
&{\bmid}\pi_{\mathrm B}\partial_{\unbx}^\alpha \partial_{\unbxi}^\beta 
\partial_{\unbtheta}^a \partial_{\unbpi}^b 
(x(t,{\unbt}\,; {\inidata})-{\unbx}){\bmid}
\le C^{(k)}_{2}|t-{\unbt}|^{1+(1/2)(1-(1-k)_+)},\\ 
&{\bmid}\pi_{\mathrm B}\partial_{\unbx}^\alpha \partial_{\unbxi}^\beta 
\partial_{\unbtheta}^a \partial_{\unbpi}^b 
(\xi(t,{\unbt}\,; {\inidata})-{\unbxi}){\bmid}
\le C^{(k)}_{2}|t-{\unbt}|^{1+(1/2)(1-(1-k)_+)}.
\end{aligned}
\right.
\label{rt3.26}
\end{equation}
\par
(iv) Let $|t-{\unbt}|\le1$.
If $|a+b|=3$ and $k=|\alpha+\beta|=0,1,2,\cdots,$ there exist constants
$C_{3}^{(k)}$ independent of $(t,\unbxi,\unbtheta)$ such that
\begin{equation}
\left\{
\begin{aligned}
&{\bmid}\pi_{\mathrm B}\partial_{\unbx}^\alpha \partial_{\unbxi}^\beta 
\partial_{\unbtheta}^a \partial_{\unbpi}^b 
(\theta(t,{\unbt}\,; {\inidata})-\unbtheta)){\bmid}
\le C^{(k)}_{3}|t-{\unbt}|^{3/2+(1/2)(1-(1-k)_+)},\\ 
&{\bmid}\pi_{\mathrm B}\partial_{\unbx}^\alpha \partial_{\unbxi}^\beta 
\partial_{\unbtheta}^a \partial_{\unbpi}^b 
(\pi(t,{\unbt}\,; {\inidata})-\unbpi)){\bmid}
\le C^{(k)}_{3}|t-{\unbt}|^{3/2+(1/2)(1-(1-k)_+)}. 
\end{aligned}
\right.
\label{rt3.27}
\end{equation}
\par
(v) Let $|t-{\unbt}|\le1$.
If $|a+b|=4$ and $k=|\alpha+\beta|=0,1,2,\cdots,$ there exist constants
$C_{4}^{(k)}$ independent of $(t,\unbxi,\unbtheta)$ such that
\begin{equation}
\left\{
\begin{aligned}
&{\bmid}\pi_{\mathrm B}\partial_{\unbx}^\alpha \partial_{\unbxi}^\beta 
\partial_{\unbtheta}^a \partial_{\unbpi}^b 
(x(t,{\unbt}\,; {\inidata})-{\unbx}){\bmid}
\le C^{(k)}_{4}|t-{\unbt}|^{5/2+(1/2)(1-(1-k)_+)},\\ 
&{\bmid}\pi_{\mathrm B}\partial_{\unbx}^\alpha \partial_{\unbxi}^\beta 
\partial_{\unbtheta}^a \partial_{\unbpi}^b 
(\xi(t,{\unbt}\,; {\inidata})-{\unbxi}){\bmid}
\le C^{(k)}_{4}|t-{\unbt}|^{5/2+(1/2)(1-(1-k)_+)}.
\end{aligned}
\right.
\label{rt3.28-0}
\end{equation}
\end{thm}
\par{\it Remark. }
In the following, we denote the solution
$x(t)=(x_j(t))$ by $x(t,\unbt)=(x_j(t,\unbt))$ with
$x_j(t)=x_j(t,\unbt)=x_j(t,\unbt\,;{\inidata})$, etc. if necessary.

On the other hand, we have
\begin{thm}\label{inverseCM}
Let $\{A_j(t,q)\}_{j=0}^3\in C^\infty(\euc\times\euc^3:\euc)$ satisfy \eqref{G}.
\newline
For $x(t,{\unbt}\,;{\inidata}),\,\theta(t,{\unbt}\,;{\inidata}),\,
\xi(t,{\unbt}\,;{\inidata}),\,\pi(t,{\unbt}\,;{\inidata})$ 
solutions of \ref{rt2.6-1} and \ref{rt2.6-2} 
obtained in Theorem \ref{existenceCM},
there exists a constant $0<\delta\le 1$ such that the followings hold:
\par
(i) For any fixed $(t,{\unbt}\,;\unbxi,\unbpi)$, $|t-{\unbt}|<\delta$, 
the mapping 
\begin{equation}
{\fR}^{3|2}\ni(\unbx,\unbtheta)\mapsto(x=x(t,{\unbt}\,;{\inidata}),
\theta=\theta(t,{\unbt}\,;{\inidata}))\in{\fR}^{3|2}
\label{91}
\end{equation}
gives a {\ssm} diffeomorphism.
We denote the inverse mapping defined by
\begin{equation}
{\fR}^{3|2}\ni(\barx,\bartheta)\mapsto(y=y(t,{\unbt}\,;{\mixdata}),
\omega=\omega(t,{\unbt}\,;{\mixdata}))\in{\fR}^{3|2},
\end{equation}
which is {\ssm} 
in $(\mixdata)$ for fixed $(t,{\unbt})$.
\par
(ii) Let $|a+b|=0$. We have
\begin{equation}
\left\{
\begin{aligned}
&{\bmid}\pi_{\mathrm B}\partial_{\barx}^\alpha\partial_{\unbxi}^\beta
(y(t,{\unbt}\,;{\mixdata})-\barx){\bmid}
=0,\\
&|y^{[0]}(t,{\unbt}\,;{\barx}^{[0]},{\unbxi}^{[0]})-{\barx}^{[0]}| 
\leq C_2|t-{\unbt}|(1+|{\barx}^{[0]}|+|{\unbxi}^{[0]}|).
\end{aligned}\right.\label{3-57}
\end{equation}
\par
(iii) Let $|a+b|=1$. For $k=|\alpha+\beta|$, there exists a constant
${\tilde C}^{(k)}_1$ such that
\begin{equation}
{\bmid}\pi_{\mathrm B}\partial_{\barx}^\alpha\partial_{\unbxi}^\beta
\partial_{\bartheta}^a\partial_{\unbpi}^b
(\omega(t,{\unbt}\,;{\mixdata})-\bartheta){\bmid}
\le {\tilde C}^{(k)}_1|t-{\unbt}|^{(1/2)(1-(1-k)_+)}.
\end{equation}
\par
(iv) Let $|a+b|=2$. For $k=|\alpha+\beta|$, there exists a constant
${\tilde C}^{(k)}_2$ such that
\begin{equation}
{\bmid}\pi_{\mathrm B}\partial_{\barx}^\alpha\partial_{\unbxi}^\beta
\partial_{\bartheta}^a\partial_{\unbpi}^b
(y(t,{\unbt}\,;{\mixdata})-\barx){\bmid}
\le {\tilde C}^{(k)}_2|t-{\unbt}|^{1+(1/2)(1-(1-k)_+)}.
\end{equation}
\par
(v) Let $|a+b|=3$. For $k=|\alpha+\beta|$, there exists a constant
${\tilde C}^{(k)}_3$ such that
\begin{equation}
{\bmid}\pi_{\mathrm B}\partial_{\barx}^\alpha\partial_{\unbxi}^\beta
\partial_{\bartheta}^a\partial_{\unbpi}^b
(\omega(t,{\unbt}\,;{\mixdata})-\bartheta){\bmid}
\le {\tilde C}^{(k)}_3|t-{\unbt}|^{3/2+(1/2)(1-(1-k)_+)}.
\end{equation}
\par
(vi) Let $|a+b|=4$. For $k=|\alpha+\beta|$, there exists a constant
${\tilde C}^{(k)}_4$ such that
\begin{equation}
{\bmid}\pi_{\mathrm B}\partial_{\barx}^\alpha\partial_{\unbxi}^\beta
\partial_{\bartheta}^a\partial_{\unbpi}^b
(y(t,{\unbt}\,;{\mixdata})-\barx){\bmid}
\le {\tilde C}^{(k)}_4|t-{\unbt}|^{5/2+(1/2)(1-(1-k)_+)}.
\end{equation}
\end{thm}

(VI) Now, we put
\begin{equation}
{\mathcal S}_0(t,{\unbt}\,;{\inidata})
=\int_{\unbt}^t ds\,\{\langle \dot x(s)|\xi(s) \rangle
+\langle \dot\theta(s)|\pi(s)\rangle 
-{\mathcal H}(s,x(s),\xi(s),\theta(s),\pi(s))\},
\label{rt2.12}
\end{equation}
and
\begin{equation}
{\mathcal S}(t,{\unbt}\,;{\mixdata})=\bigg(\langle {\unbx}|{\unbxi}\rangle+
\langle {\unbtheta}|{\unbpi}\rangle
+{\mathcal S}_0(t,{\unbt}\,;{\inidata})\bigg)\bigg|
\begin{Sb}{\unbx}=y(t,{\unbt}\,;{\mixdata})\\
{\unbtheta}=\omega(t,{\unbt}\,;{\mixdata})\end{Sb}.
\label{rt2.13}
\end{equation}

\begin{thm}\label{HJ-estimates}
${\mathcal S}(t,{\unbt}\,;{\mixdata})$ satisfies the following Hamilton-Jacobi equation:
\begin{equation}
\left\{
\begin{aligned}
&\pdt {\mathcal S}+{\mathcal H}
\Big(t,\bar{x},\frac{\partial {\mathcal S}}{\partial\bar{x}},
\bar{\theta},\frac{\partial {\mathcal S}}{\partial \bar{\theta}}\Big)=0,\\
&{\mathcal S}({\unbt},{\unbt}\,;{\mixdata})=\langle \bar{x}|{\unbxi}\rangle+
\langle\bar{\theta}|{\unbpi}\rangle.
\end{aligned}
\right.
\label{rt2.14}
\end{equation}
Moreover, decomposing
\begin{equation}
{\mathcal S}(t,s\,;x,\xi,\theta,\pi)
= {\mathcal S}_{\mathrm B}(t,s\,;x,\xi)
+{\mathcal S}_{\mathrm S}(t,s\,;x,\xi,\theta,\pi)
\end{equation}
with 
\begin{equation}
\begin{aligned}
{\mathcal S}_{\mathrm B}(t,s\,;x,\xi)&={\mathcal S}(t,s\,;x,\xi,0,0)
={\mathcal S}_{\bar0\bar0}(t,s\,;x,\xi),\\
{\mathcal S}_{\mathrm S}(t,s\,;x,\theta,\xi,\pi)
&=
\sum_{|c|+|d|=even \ge 2}{\mathcal S}_{cd} (t,s\,;x,\xi) \theta^c\pi^d\\
&={\mathcal S}_{\bar1\bar0}\theta_1\theta_2
+\sum_{j,k=1}^2{\mathcal S}_{c_jd_k}\theta^{c_j}\pi^{d_k}
+{\mathcal S}_{\bar0\bar1}\pi_1\pi_2
+{\mathcal S}_{\bar1\bar1}\theta_1\theta_2\pi_1\pi_2,\\
&\where
\bar 0=(0,0),\;\bar 1=(1,1),\;c_1=(1,0)=d_1,\;c_2=(0,1)=d_2\in\{0,1\}^2,
\end{aligned}
\label{rt3.70}
\end{equation}
we get the following estimates:
For any $\alpha$, $\beta$, there exist constants
$C_{\alpha\beta}>0$ such that
\begin{equation}
{\begin{aligned}
&|\partial_x^\alpha\partial_\xi^\beta
({\mathcal S}_{\bar0\bar0}(t,s\,;x,\xi)-\langle x|\xi\rangle)|
\le C_{\alpha\beta}(1+|x|)^{(1-|\alpha|)_+}\delta_{0|\beta|}|t-s|\\
&|\partial_x^\alpha\partial_\xi^\beta{\mathcal S}_{\bar1\bar0}(t,s\,;x,\xi)|
\le C_{\alpha\beta}|t-s|,\\ 
&|\partial_x^\alpha\partial_\xi^\beta({\mathcal S}_{c_jd_j}(t,s\,;x,\xi)-1)|
\le C_{\alpha\beta}|t-s| \for j=1,2,\\
&|\partial_x^\alpha\partial_\xi^\beta{\mathcal S}_{\bar0\bar1}(t,s\,;x,\xi)|
\le C_{\alpha\beta}|t-s|,\\ 
&|\partial_x^\alpha\partial_\xi^\beta{\mathcal S}_{\bar1\bar1}(t,s\,;x,\xi)|
\le C_{\alpha\beta}|t-s|.
\end{aligned}}
\label{rt3.84}
\end{equation}
\end{thm}

Defining (sdet denotes the super-determinant)
\begin{equation}
{\mathcal D}(t,{\unbt}\,;{\mixdata})=\sdet  %(-1)^{3+2}\,
\begin{pmatrix}
\frac{\partial^2{\mathcal S}}{\partial {\barx}\,\partial {\unbxi}}&
\frac{\partial^2{\mathcal S}}{\partial {\barx}\,\partial {\unbpi}}\\
\frac{\partial^2{\mathcal S}}{\partial {\barth}\,\partial {\unbxi}}&
\frac{\partial^2{\mathcal S}}{\partial {\barth}\,\partial {\unbpi}}
\end{pmatrix}={\mathcal A}^2(t,{\unbt}\,;{\mixdata}),
\label{rt2.15}
\end{equation}
we get
\begin{thm}\label{continuity}
${\mathcal D}(t,{\unbt}\,;{\mixdata})$ or ${\mathcal A}(t,{\unbt}\,;{\mixdata})$  
satisfies the following continuity equation:
\begin{equation}
\left\{
\begin{aligned}
&\pdt {\mathcal D}+
\frac{\partial}{\partial{\barx}}
\Big({\mathcal D}
\frac{\partial{\mathcal H}}{\partial {\xi}}\Big)
+\frac{\partial}{\partial {\barth}}
\Big({\mathcal D}
\frac{\partial{\mathcal H}}{\partial {\pi}}\Big)=0,\\
&{\mathcal D}({\unbt},{\unbt}\,;{\mixdata})=1.
\end{aligned}
\right.
\label{rt2.16}
\end{equation}
Or, we have
\begin{equation}
\left\{
\begin{aligned}
&{\mathcal A}_t+\big\{{\mathcal A}_{x_j}{\mathcal H}_{\xi_j}+{\mathcal A}_{\theta_k}{\mathcal H}_{\pi_k}
+\frac12{\mathcal A}
[\partial_{x_j}{\mathcal H}_{\xi_j}+\partial_{\theta_k}{\mathcal H}_{\pi_k}]\big\}
=0,\\ 
&{\mathcal A}(s,s\,;x,\xi,\theta,\pi)=1.
\end{aligned}
\right.
\label{A00}
\end{equation}
Here, the argument of ${\mathcal D}$ or ${\mathcal A}$ is $(t,{\unbt}\,;{\mixdata})$,
those of $\partial_\xi{\mathcal H}$ and 
$\partial_\pi{\mathcal H}$ are $({\barx},\partial_{\bar{x}}
{\mathcal S}, \bar\theta, \partial_{\barth}{\mathcal S})$,
respectively.
\par
Decomposing
\begin{equation}
{\mathcal A}(t,s\,;x,\xi,\theta,\pi)
=\sum_{|c|+|d|=even \ge 0}
{\mathcal A}_{cd}(t,s\,;x,\xi)\theta^{c} \pi^{d}
={\mathcal A}_{\mathrm B}(t,s\,;x,\xi)+{\mathcal D}_{\mathrm S}(t,s\,;x,\xi,\theta,\pi),
\end{equation}
as before, we get the following:
If $|t-s|$ is sufficiently small, we have
\begin{equation}
{\begin{aligned}
&|\partial_x^\alpha\partial_\xi^\beta({\mathcal A}_{\bar0\bar0}(t,s\,;x,\xi)-1)|
\le C_{\alpha\beta}|t-s|,\\
&|\partial_x^\alpha\partial_\xi^\beta{\mathcal A}_{\bar1\bar0}(t,s\,;x,\xi)|
\le C_{\alpha\beta}|t-s|,\\
&|\partial_x^\alpha\partial_\xi^\beta{\mathcal A}_{c_jd_j}(t,s\,;x,\xi)|
\le C_{\alpha\beta}|t-s|\for j=1,2\\
&|\partial_x^\alpha\partial_\xi^\beta{\mathcal A}_{\bar0\bar1}(t,s\,;x,\xi)|
\le C_{\alpha\beta}|t-s|,\\
&|\partial_x^\alpha\partial_\xi^\beta{\mathcal A}_{\bar1\bar1}(t,s\,;x,\xi)|
\le C_{\alpha\beta}|t-s|.
\end{aligned}}
\label{rt3.99}
\end{equation}
\end{thm}

In the following argument of this section, we change the order of variables 
and rewrite them from $(t,{\unbt}\,;\mixdata)$ to
$(t,{\unbt}\,;\barx,\bartheta,\unbxi,\unbpi)$, and then to
$(t,s\,;x,\theta,\xi,\pi)$. 

(VII) We define an operator
\begin{equation}
\begin{aligned}
({\mathcal U}(t,{s})u)(x,\theta)
&=c_{3,2} %(2\pi\hbar)^{-3/2}\hbar
\iint_{\fR^{3|2}\times\fR^{3|2}}d{\xi}d{\pi}\,
{\mathcal A}(t,{s}\,;x,\theta,\xi,\pi)
e^{i\hbar^{-1}{{\mathcal S}}(t,{s}\,;x,\theta,\xi,\pi)}
{\mathcal F}u({\xi},{\pi}).
\end{aligned}
\label{rt2.17}
\end{equation}

On the other hand, we show readily 
\begin{equation}
{\mathcal H}\Big(t,x,{\frachi}{\pdx},\theta,{\pdtheta}\Big)
={\hat{\mathcal H}}(t)
\end{equation}
where ${\hat{\mathcal H}}(t)$ is a (Weyl type) pseudo-differential operator with
symbol ${\mathcal H}(t,x,\theta,\xi,\pi)$, that is, 
\begin{equation} 
\begin{aligned}
({\hat{\mathcal H}}(t)u)(x,\theta)=c_{3,2}^2
\iint_{\fR^{3|2}\times\fR^{3|2}} d\xi d\pi dyd\omega\,  
& e^{i\hbar^{-1}(\langle{x-y}|\xi\rangle +\langle\theta-\omega|\pi\rangle)}\\
&\qquad\times
{\mathcal H}\bigg(t,\frac{x+y}2,\frac{\theta+\omega}2,\xi,\pi\bigg)u(y,\omega).
\end{aligned}
\end{equation}

{\bf Claim}: The operator \eqref{rt2.17} is a good parametrix for \eqref{rt2.4}.
\par
[{\bf good parametrix}]: 
We call \eqref{rt2.17} as a good parametrix because it has the
following properties:
\newline
(a) This \eqref{rt2.17} gives a parametrix of the probelm \eqref{rt2.4},
which has {the explicit dependence on the classical quantities}. 
That is, the Bohr correspondence is examplified even with spin structure.
%Moreover, as a by-pruduct, we may check rather easily whether we may replace
%$\hbar$ in \eqref{rt2.17} with $-i$, in other word, 
%we may derive ``heat" from ``Schr\"odinger".
\newline
(b) As the infinitesimal generator of that parametrix, 
we have an operator ${\hat{\mathcal H}}(t)$
%${\mathcal H}(t,x,-i\hbar\partial_x,\theta,-i\hbar\partial_\theta)$
which corresponds to the {the Weyl quantization} for
the symbol ${\mathcal H}(t,x,\xi,\theta,\pi)$, that is, a certain symmetry
is preserved naturally.
\newline
(c) By Trotter-Kato's time-slicing method, products of \eqref{rt2.17} yield
a fundamental solution as is presented in the following theorem.

\begin{thm}
Let $\{A_j(t,q)\}_{j=0}^3\in C^\infty(\euc\times\euc^3:\euc)$ satisfy \eqref{G}.  
\newline
(1) There exists a positive number $\delta$ such that if $|t-s|<\delta$ then
${\mathcal U}(t,s)$ is a well defined  bounded operator in 
$\clsl_{{\mathrm {SS}}}^2({\fR}^{3|2})$.
Moreover, if $|t-r|+|r-s|<\delta$, then there exists a constant $C$ such that
\begin{equation}
\Vert {\mathcal U}(t,r){\mathcal U}(r,s)-{\mathcal U}(t,s)\Vert
_{{\Bbb B}({\not{\mathcal L}}_{\mathrm {SS}}^2({\fR}^{3|2}),\;
%\,\clsl_{{\mathrm {SS}}}^2({\fR}^{3|2}),\;
{\not{\mathcal L}}_{\mathrm {SS}}^2({\fR}^{3|2}))}
%\clsl_{{\mathrm {SS}}}^2({\fR}^{3|2}))}
 \le C(|t-r|^2+|r-s|^2).
\end{equation}
\newline
(2) Let $\Delta$ be an arbitrary subdivision of the interval $[{s},t]$
or $[t,{s}]$
for any $t, s\in\euc$, such that
$$
\Delta: {s}=t_0<t_1<\cdots<t_N=t
\quad\text{or}\quad
\Delta: {s}=t_0>t_1>\cdots>t_N=t
$$
with
$|\Delta|=\max_{1\le i\le N}|t_i-t_{i-1}|$.
We put 
$$
{\mathcal U}_\Delta(t,{s})={\mathcal U}(t_N,t_{N-1}){\mathcal U}(t_{N-1},t_{N-2})
\cdots{\mathcal U}(t_1,t_0).
$$ 
Then, ${\mathcal U}_\Delta(t,{s})$ converges when $|\Delta|\to 0$  to an
unitary operator ${\mathcal E}(t,{s})$ 
in the uniform operator topology in 
$\clsl_{{\mathrm {SS}}}^2({\frak R}^{3|2})$.
More precisely, there exist constants $\gamma_1, \gamma_2$ such that
\begin{equation}
\Vert {\mathcal E}(t,{s})-{\mathcal U}_\Delta(t,{s})\Vert
_{{\Bbb B}({\not{\mathcal L}}_{\mathrm {SS}}^2({\fR}^{3|2}),\;
%\,\clsl_{{\mathrm {SS}}}^2({\fR}^{3|2}),\;
{\not{\mathcal L}}_{\mathrm {SS}}^2({\fR}^{3|2}))}
%\clsl_{{\mathrm {SS}}}^2({\fR}^{3|2}))}
\le\gamma_1|\Delta|
e^{\gamma_2|\Delta|}.
\end{equation}
\newline
(3) 
(i) $\euc^2\ni (t,{s})\to {\mathcal E}(t,{s})\in {\Bbb
B}(\,\clsl_{{\mathrm {SS}}}^2({\fR}^{3|2}),\,\clsl_{{\mathrm {SS}}}^2({\fR}^{3|2}))$ 
is continuous and satisfies
$
{\mathcal E}(t,r){\mathcal E}(r,s)={\mathcal E}(t,s).
$
\newline
(ii) For $u\in \ccsl_{\mathrm {SS,0}}({\frak R}^{3|2})$, we have
\begin{equation}
\left\{
\begin{aligned}
&i\hbar\dt {\mathcal E}(t,{s})u={\hat{\mathcal H}}(t)\,{\mathcal E}(t,{s})u,\\
& {\mathcal E}({s},{s})u=u.
\end{aligned}
\right.
\end{equation}
\end{thm}
\par{\bf Remark}: The reason for preparing complicated estimates
in Theorems \ref{existenceCM}--\ref{continuity},
is to apply the known $L^2$-bounded theorem to our FIO.

On the other hand, remarking that
\begin{equation}
\flat{\hat{\mathcal H}}(t)\sharp \psi={\Bbb H}(t)\psi,
\end{equation}
and putting that
\begin{equation}
{\Bbb U}(t,{s})\psi=\flat\,{\mathcal U}(t,{s})\sharp\psi,\quad
{\Bbb U}_\Delta(t,{s})=\flat\,{\mathcal U}_\Delta(t,{s})\sharp,\quad
{\Bbb E}(t,{s})\psi=\flat\,{\mathcal E}(t,{s})\sharp\psi,
\end{equation}
we have
\begin{thm}
Let $\{A_j(t,q)\}_{j=0}^3\in C^\infty(\euc\times\euc^3:\euc)$ sastisfy \eqref{G}.  
\newline
(1) There exists a positive number $\delta$ such that if $|t-s|<\delta$ then
${\Bbb U}(t,s)$ is well defined  bounded operator in $L^2(\euc^3:{\Bbb C}^2)$.
Moreover, if $|t-r|+|r-s|<\delta$, then there exists a constant $C$ such that
\begin{equation}
\Vert {\Bbb U}(t,r){\Bbb U}(r,s)-{\Bbb U}(t,s)\Vert
_{{\Bbb B}(L^2(\euc^3:{\Bbb C}^2),L^2(\euc^3:{\Bbb C}^2))}
 \le C(|t-r|^2+|r-s|^2).
\end{equation}
\newline
(2) For any $(t,{s})\in \euc^2$,
${\Bbb U}_\Delta(t,{s})$ converges 
when $|\Delta|\to 0$ to an unitary
operator ${\Bbb E}(t,{s})$ in the uniform operator
topology in  $L^2(\euc^3:{\Bbb C}^2)$.
More precisely, there exist constants $\gamma_1, \gamma_2$ such that
\begin{equation}
\Vert {\Bbb E}(t,{s})-{\Bbb U}_\Delta(t,{s})\Vert
_{{\Bbb B}(L^2(\euc^3:{\Bbb C}^2),L^2(\euc^3:{\Bbb C}^2))}
\le\gamma_1|\Delta| e^{\gamma_2|\Delta|}.
\end{equation}
\newline
(3) 
(i) $\euc^2\ni (t,{s})\to {\Bbb E}(t,{s})\in {\Bbb
B}(L^2(\euc^3:{\Bbb C}^2),L^2(\euc^3:{\Bbb C}^2))$ is continuous
and satisfies
$
{\Bbb E}(t,r){\Bbb E}(r,s)={\Bbb E}(t,s).
$ 
\newline
$\quad$
(ii) For $\psi\in C_0^\infty(\euc^3:{\Bbb C}^2)$, we have
\begin{equation}
\left\{
\begin{aligned}
&i\hbar\dt {\Bbb E}(t,{s})\psi={\Bbb H}(t)\,{\Bbb E}(t,{s})\psi,\\
&{\Bbb E}({s},{s})\psi=\psi.
\end{aligned}
\right.
\end{equation}
\end{thm}

{\bf Problem}: Construct a kernel representation of 
${\Bbb E}(t,{s})$ (i.e. a fundamental solution). If we could do so properly, then
we would give an answer of the problem posed 
by Feynman in p.355 of \cite{FH65}.
To do so, we need to develop the theory of FIO on superspace more precisely.
For example, we must extend the $\#$-product formula for two FIP with different
phaases which is done for FIO on ordinary Euclidian space.

%%%

%%\input s-Weyl3-1-1

\section{Classical Mechanics: Proofs of Theorems 2.1--2.5.}
\subsection{Hamiltonian flows}
\subsubsection{Proof of Theorem \ref{existenceCM}(Existence)}
We rewrite \ref{rt2.6-2} as
\begin{equation}
\dt
\begin{pmatrix}
\theta_1\\
\theta_2\\
\pi_1\\
\pi_2
\end{pmatrix}
=ic\hbar^{-1}{\Bbb X}(t)
\begin{pmatrix}
\theta_1\\
\theta_2\\
\pi_1\\
\pi_2
\end{pmatrix}
\with
\begin{pmatrix}
\theta_1({\unbt})\\
\theta_2({\unbt})\\
\pi_1({\unbt})\\
\pi_2({\unbt})
\end{pmatrix}
=\begin{pmatrix}
\unbtheta_1\\
\unbtheta_2\\
\unbpi_1\\
\unbpi_2
\end{pmatrix},
\label{rt3.1}
\end{equation}
where
\begin{equation}
{\Bbb X}(t)=
\begin{pmatrix}
-\eta_3(t)&0&0&{i}\hbar^{-1}(\eta_1(t)-i\eta_2(t))\\
0&-\eta_3(t)&-{i}\hbar^{-1}(\eta_1(t)-i\eta_2(t))&0\\
0&{i}\hbar(\eta_1(t)+i\eta_2(t))&\eta_3(t)&0\\
-{i}\hbar(\eta_1(t)+i\eta_2(t))&0&0&\eta_3(t)
\end{pmatrix}.
\label{rt3.2}
\end{equation}

Moreover,  
defining $\sigma_j(t)=\sigma_j(\theta(t),\pi(t))$, 
we have, by simple calculations,
\begin{equation}
\dt
\begin{pmatrix}
\sigma_1\\
\sigma_2\\
\sigma_3
\end{pmatrix}
=2c\hbar^{-1}{\Bbb Y}(t)
\begin{pmatrix}
\sigma_1\\
\sigma_2\\
\sigma_3
\end{pmatrix}
\with
\begin{pmatrix}
\sigma_1({\unbt})\\
\sigma_2({\unbt})\\
\sigma_3({\unbt})
\end{pmatrix}
=\begin{pmatrix}
\unbsigma_1\\
\unbsigma_2\\
\unbsigma_3
\end{pmatrix}
=\begin{pmatrix}
\unbtheta_1\unbtheta_2+\hbar^{-2}\unbpi_1\unbpi_2\\
i(\unbtheta_1\unbtheta_2-\hbar^{-2}\unbpi_1\unbpi_2)\\
-i\hbar^{-1}(\unbtheta_1\unbpi_1+\unbtheta_2\unbpi_2)
\end{pmatrix}
\label{rt3.3}
\end{equation}
where
\begin{equation}
{\Bbb Y}(t)=\begin{pmatrix}
0& -\eta_3(t) &\eta_2(t)\\
\eta_3(t)&0&-\eta_1(t)\\
-\eta_2(t)&\eta_1(t)&0
\end{pmatrix}.
\label{rt3.4}
\end{equation}

Now, we start our existence proof.
We decompose variables, using degree, as follow:
\begin{equation}
x_j(t)=\sum_{\ell=0}^\infty x_j^{[2\ell]}(t), \quad 
\xi_j(t)=\sum_{\ell=0}^\infty \xi_j^{[2\ell]}(t),\quad
\theta_k(t)=\sum_{\ell=0}^\infty \theta_k^{[2\ell+1]}(t), \quad 
\pi_k(t)=\sum_{\ell=0}^\infty \pi_k^{[2\ell+1]}(t).
\label{rt3.5}
\end{equation}

Then, we get, for $m=0,1,2,\cdots$,
\begin{equation}
\left\{
\begin{aligned}
&\dt x_j^{[2m]}=c\sigma_j^{[2m]}\where \sigma_j^{[0]}=0,\\
&\dt \xi_j^{[2m]}={\elec}\sum_{\ell=1}^m\sum_{k=1}^3 \sigma_k^{[2\ell]}
\frac{\partial A_k^{[2m-2\ell]}}{\partial x_j}
-{\elec}\frac{\partial {A}_0^{[2m]}}{\partial x_j}
\end{aligned}\right.
\with
\begin{pmatrix}
x^{[2m]}({\unbt})\\
\xi^{[2m]}({\unbt})
\end{pmatrix}
=
\begin{pmatrix}
\unbx^{[2m]}\\
\unbxi^{[2m]}
\end{pmatrix},
\label{rt3.6}
\end{equation}
\begin{equation}
\dt
\begin{pmatrix}
\theta_1^{[2m+1]}\\
\theta_2^{[2m+1]}\\
\pi_1^{[2m+1]}\\
\pi_2^{[2m+1]}\\
\end{pmatrix}
=ic\hbar^{-1}\sum_{\ell=0}^m{\Bbb X}^{[2\ell]}(t)
\begin{pmatrix}
\theta_1^{[2m+1-2\ell]}\\
\theta_2^{[2m+1-2\ell]}\\
\pi_1^{[2m+1-2\ell]}\\
\pi_2^{[2m+1-2\ell]}\\
\end{pmatrix}
\with
\begin{pmatrix}
\theta_1^{[2m+1]}({\unbt})\\
\theta_2^{[2m+1]}({\unbt})\\
\pi_1^{[2m+1]}({\unbt})\\
\pi_2^{[2m+1]}({\unbt})\\
\end{pmatrix}
=\begin{pmatrix}
\unbtheta_1^{[2m+1]}\\
\unbtheta_2^{[2m+1]}\\
\unbpi_1^{[2m+1]}\\
\unbpi_2^{[2m+1]}\\
\end{pmatrix},
\label{rt3.7}
\end{equation}
and
\begin{equation}
\dt
\begin{pmatrix}
\sigma_1^{[2m]}\\
\sigma_2^{[2m]}\\
\sigma_3^{[2m]}
\end{pmatrix}
=\sum_{\ell=0}^{m-1}2c\hbar^{-1}{\Bbb Y}^{[2\ell]}(t)
\begin{pmatrix}
\sigma_1^{[2m-2\ell]}\\
\sigma_2^{[2m-2\ell]}\\
\sigma_3^{[2m-2\ell]}
\end{pmatrix}
\with
\begin{pmatrix}
\sigma_1^{[2m]}({\unbt})\\
\sigma_2^{[2m]}({\unbt})\\
\sigma_3^{[2m]}({\unbt})
\end{pmatrix}
=\begin{pmatrix}
\unbsigma_1^{[2m]}\\
\unbsigma_2^{[2m]}\\
\unbsigma_3^{[2m]}
\end{pmatrix}.
\label{rt3.8}
\end{equation}
Here, ${\Bbb X}^{[2\ell]}(t)$, ${\Bbb Y}^{[2\ell]}(t)$, 
$\sigma^{[2\ell]}(t)$ are the degree of $2\ell$ parts of ${\Bbb X}(t)$, 
${\Bbb Y}(t)$, $\sigma(t)$, respectively:
$$
\begin{gathered}
\eta_k^{[2\ell]}(t)=\xi_k^{[2\ell]}(t)-\frac{\elec}{c}A_k^{[2\ell]}(t,x),\\
A_k^{[2\ell]}(t,x)=\sum{\begin{Sb}|\alpha|\le2\ell\\
\ell_1+\ell_2+\ell_3=\ell\end{Sb}}
\frac1{\alpha!}\partial_q^\alpha A_k(t,x^{[0]})\cdot
(x_{1}^{\alpha_1})^{[2\ell_1]}
(x_{2}^{\alpha_2})^{[2\ell_2]}
(x_{3}^{\alpha_3})^{[2\ell_3]},\\
\frac{\partial{A}_0^{[2\ell]}(t,x)}{\partial x_j}=\sum{\begin{Sb}|\alpha|\le2\ell\\
\ell_1+\ell_2+\ell_3=\ell\end{Sb}}
\frac1{\alpha!}\partial_q^\alpha \partial_{q_j}{A}_0(t,x^{[0]})\cdot
(x_{1}^{\alpha_1})^{[2\ell_1]}
(x_{2}^{\alpha_2})^{[2\ell_2]}
(x_{3}^{\alpha_3})^{[2\ell_3]}
\end{gathered}
$$
and
$$
\left\{
\begin{aligned}
\sigma_1^{[2m]}&=\sum_{\ell=0}^{m-1}
\Big(\theta_1^{[2\ell+1]}\theta_2^{[2m-2\ell-1]}
+\hbar^{-2}\pi_1^{[2\ell+1]}\pi_2^{[2m-2\ell-1]}\Big),\\
\sigma_2^{[2m]}&=i\sum_{\ell=0}^{m-1}
\Big(\theta_1^{[2\ell+1]}\theta_2^{[2m-2\ell-1]}
-\hbar^{-2}\pi_1^{[2\ell+1]}\pi_2^{[2m-2\ell-1]}\Big),\\
\sigma_3^{[2m]}&=-i\hbar^{-1}\sum_{\ell=0}^{m-1}
\Big(\theta_1^{[2\ell+1]}\pi_1^{[2m-2\ell-1]}
+\theta_2^{[2\ell+1]}\pi_2^{[2m-2\ell-1]}\Big).
\end{aligned}
\right.
$$
\par
[0] From \eqref{rt3.6} with $m=0$, we get
\begin{equation*}
\dt x_j^{[0]}(t)=0\et 
\dt\xi_j^{[0]}(t)=-{\elec}\frac{\partial{A}_0^{[0]}(t,x^{[0]})}{\partial x_j}
=-{\elec}\partial_{q_j}{A}_0^{[0]}(t,x^{[0]})
\for j=1,2,3.
%\label{rt3.9}
\end{equation*}
Therefore, for any $t\in \euc$,
\begin{equation*}
x_j^{[0]}(t)=\unbx_j^{[0]}\et
\xi_j^{[0]}(t)=\unbxi_j^{[0]}
-{\elec}\int_{\unbt}^t ds\,\partial_{q_j}{A}_0^{[0]}(s,\unbx^{[0]})\for j=1,2,3 .
%\label{rt3.10}
\end{equation*}
\par
[1] We put these into \eqref{rt3.7} with $m=0$, to have
\begin{equation}
\dt
\begin{pmatrix}
\theta_1^{[1]}\\
\theta_2^{[1]}\\
\pi_1^{[1]}\\
\pi_2^{[1]}
\end{pmatrix}
=ic\hbar^{-1}{\Bbb X}^{[0]}(t)
\begin{pmatrix}
\theta_1^{[1]}\\
\theta_2^{[1]}\\
\pi_1^{[1]}\\
\pi_2^{[1]}
\end{pmatrix}
\with
\begin{pmatrix}
\theta_1^{[1]}({\unbt})\\
\theta_2^{[1]}({\unbt})\\
\pi_1^{[1]}({\unbt})\\
\pi_2^{[1]}({\unbt})
\end{pmatrix}
=\begin{pmatrix}
\unbtheta_1^{[1]}\\
\unbtheta_2^{[1]}\\
\unbpi_1^{[1]}\\
\unbpi_2^{[1]}
\end{pmatrix}.
\label{rt3.11}
\end{equation}
Here, ${\Bbb X}^{[0]}(t)$ is a $4\times 4$-matrix whose arguments
depend on  
$(t,{\unbt},\unbx^{[0]},\unbxi^{[0]},
\partial_q^\beta {A}_0,\partial_q^\alpha A\,;|\alpha|=0,|\beta|\le 1)$ 
with values in $\Bbb C$.
Or more precisely, ${\Bbb X}^{[0]}(t)$ has components given by
$$
\eta_j^{[0]}(t)=\xi_j^{[0]}(t)-\frac{\elec}{c}A_j(t,\unbx^{[0]})
=\unbxi_j^{[0]}-{\elec}\int_{\unbt}^t ds\,\partial_{q_j}{A}_0^{[0]}(s,\unbx^{[0]})
-\frac{\elec}{c}A_j^{[0]}(t,\unbx^{[0]}).
$$
As \eqref{rt3.11} is the linear ODE in $({\fR}^{0|1})^4$
with smooth coefficients in $t$,
there exists a unique global (in time) solution, 
which has the following dependence. Putting $A=(A_1,A_2,A_3)$, we have
\begin{equation}
\left\{
\begin{aligned}
&\theta_j^{[1]}(t)=\theta_j^{[1]}(t,\unbx^{[0]},\unbxi^{[0]},
\unbtheta^{[1]},\unbpi^{[1]},
\partial_q^\beta {A}_0,\partial_q^\alpha A\,;|\alpha|=0,|\beta|\le 1),\;
\text{linear in $\unbtheta^{[1]},\unbpi^{[1]}$},\\
&\pi_j^{[1]}(t)=\pi_j^{[1]}(t,\unbx^{[0]},\unbxi^{[0]},
\unbtheta^{[1]},\unbpi^{[1]},
\partial_q^\beta {A}_0,\partial_q^\alpha A\,;|\alpha|=0,|\beta|\le 1),\;
\text{linear in $\unbtheta^{[1]},\unbpi^{[1]}$}.
\end{aligned}
\right.\label{rt3.12}
\end{equation}
\par
[2] For \eqref{rt3.6} with $m=1$, we have
\begin{equation*}
\left\{
\begin{aligned}
&\dt
x_j^{[2]}=c\sigma_j^{[2]},\\
&\dt
\xi_j^{[2]}={\elec}\sum_{k=1}^3
\sigma_k^{[2]}\frac{\partial A_k^{[0]}}{\partial x_j}
-{\elec}\frac{\partial {A}_0^{[2]}}{\partial x_j}
\end{aligned}
\right.
\with
\begin{pmatrix}
x^{[2]}({\unbt})=\unbx^{[2]}\\
\xi^{[2]}({\unbt})=\unbxi^{[2]}
\end{pmatrix}.
%\label{rt3.13}
\end{equation*}
Then, using \eqref{rt3.8} with $m=1$ and \eqref{rt3.12},
we have, for $j=1,2,3$,
$$
\left\{
\begin{aligned}
\sigma_1^{[2]}&=\theta_1^{[1]}\theta_2^{[1]}+\hbar^{-2}\pi_1^{[1]}\pi_2^{[1]},\\
\sigma_2^{[2]}&=i(\theta_1^{[1]}\theta_2^{[1]}-\hbar^{-2}\pi_1^{[1]}\pi_2^{[1]}),\\
\sigma_3^{[2]}&=-i\hbar^{-1}(\theta_1^{[1]}\pi_1^{[1]}+\theta_2^{[1]}\pi_2^{[1]})
\end{aligned}
\right.
\et
\left\{
\begin{aligned}
{A}_0^{[2]}(x)&=\sum_{k=1}^3\partial_{q_k}{A}_0(x^{[0]})x_{k}^{[2]},\\
\frac{\partial {A}_0^{[2]}}{\partial x_j}&=
\sum_{k=1}^3\partial_{q_kq_j}{A}_0(x^{[0]})x_{k}^{[2]}.
\end{aligned}
\right.
$$
Therefore, we have, for $j=1,2,3$,
\begin{equation*}
\left\{
\begin{aligned}
&x_j^{[2]}(t)=x_j^{[2]}(t,\unbx^{[2\ell]},\unbxi^{[0]},
\unbtheta^{[1]},\unbpi^{[1]},
\partial_q^\beta {A}_0,\partial_q^\alpha A\,;0\le\ell\le 1,|\alpha|=0,|\beta|\le 1),\\
&\xi_j^{[2]}(t)=\xi_j^{[2]}(t,\unbx^{[2\ell]},\unbxi^{[2\ell]},
\unbtheta^{[1]},\unbpi^{[1]},
\partial_q^\beta {A}_0,\partial_q^\alpha A\,;0\le\ell\le 1,|\alpha|\le 1,|\beta|\le 2).
\end{aligned}
\right.
%\label{rt3.14}
\end{equation*}
\par
[3] For \eqref{rt3.7} with $m=1$, we get
\begin{equation*}
\dt
\begin{pmatrix}
\theta_1^{[3]}\\
\theta_2^{[3]}\\
\pi_1^{[3]}\\
\pi_2^{[3]}
\end{pmatrix}
=ic\hbar^{-1}{\Bbb X}^{[0]}(t)
\begin{pmatrix}
\theta_1^{[3]}\\
\theta_2^{[3]}\\
\pi_1^{[3]}\\
\pi_2^{[3]}
\end{pmatrix}
+ic\hbar^{-1}{\Bbb X}^{[2]}(t)
\begin{pmatrix}
\theta_1^{[1]}\\
\theta_2^{[1]}\\
\pi_1^{[1]}\\
\pi_2^{[1]}
\end{pmatrix}
\with
\begin{pmatrix}
\theta_1^{[3]}({\unbt})\\
\theta_2^{[3]}({\unbt})\\
\pi_1^{[3]}({\unbt})\\
\pi_2^{[3]}({\unbt})
\end{pmatrix}
=\begin{pmatrix}
\unbtheta_1^{[3]}\\
\unbtheta_2^{[3]}\\
\unbpi_1^{[3]}\\
\unbpi_2^{[3]}
\end{pmatrix}.
%\label{rt3.15}
\end{equation*}
Here, ${\Bbb X}^{[2]}(t)$ is a $4\times 4$-matrix whose arguments
depend on  
$(t,\unbx^{[2\ell]},\unbxi^{[2\ell]},
\unbtheta^{[1]},\unbpi^{[1]},
\partial_q^\beta {A}_0,\partial_q^\alpha A\,;
0\le\ell\le 1,|\alpha|\le 1,|\beta|\le 2)$ 
with values in $\cev$.

Therefore, we get, for $k=1,2$,
\begin{equation*}
\left\{
\begin{aligned}
&\theta_k^{[3]}(t)=\theta_k^{[3]}(t,\unbx^{[2\ell]},\unbxi^{[2\ell]},
\unbtheta^{[2\ell+1]},\unbpi^{[2\ell+1]},
\partial_q^\beta{A}_0,\partial_q^\alpha{A}\,;
0\le\ell\le 1,|\alpha|\le 1, |\beta|\le 2),\\
&\pi_k^{[3]}(t)=\pi_k^{[3]}(t,\unbx^{[2\ell]},\unbxi^{[2\ell]},
\unbtheta^{[2\ell+1]},\unbpi^{[2\ell+1]},
\partial_q^\beta{A}_0,\partial_q^\alpha{A}\,;
0\le\ell\le 1,|\alpha|\le 1, |\beta|\le 2).
\end{aligned}
\right.
%\label{rt3.16}
\end{equation*}
\par
[4] Proceeding inductively, we get, 
\begin{equation*}
\left\{
\begin{aligned}
&x^{[2m]}(t)=x^{[2m]}(t,\unbx^{[2\ell]},\unbxi^{[2\ell]},
\unbtheta^{[2\ell-1]},\unbpi^{[2\ell-1]},
\partial_q^\beta{A}_0, \partial_q^\alpha A\,;
0\le\ell\le m,|\alpha|\le m-1,|\beta|\le m),\\
&\xi^{[2m]}(t)=\xi^{[2m]}(t,\unbx^{[2\ell]},\unbxi^{[2\ell]},
\unbtheta^{[2\ell-1]},\unbpi^{[2\ell-1]},
\partial_q^\beta{A}_0, \partial_q^\alpha A\,;
0\le\ell\le m,|\alpha|\le m,|\beta|\le m+1),\\
&\theta^{[2m+1]}(t)=\theta^{[2m+1]}(t,\unbx^{[2\ell]},\unbxi^{[2\ell]},
\unbtheta^{[2\ell+1]},\unbpi^{[2\ell+1]},
\partial_q^\beta{A}_0, \partial_q^\alpha A\,;
0\le\ell\le m,|\alpha|\le m,|\beta|\le m+1),\\
&\pi^{[2m+1]}(t)=\pi^{[2m+1]}(t,\unbx^{[2\ell]},\unbxi^{[2\ell]},
\unbtheta^{[2\ell+1]},\unbpi^{[2\ell+1]},
\partial_q^\beta{A}_0, \partial_q^\alpha A\,;
0\le\ell\le m,|\alpha|\le m,|\beta|\le m+1).
\end{aligned}
\right. %\label{rt3.17}
\end{equation*}
This gives the existence proof (Proof of Theorem \ref{existenceCM}).
Remarking that at each degree, the solution of \eqref{rt3.6} and \eqref{rt3.7}
is defined uniquely, we have the uniqueness of the solution of \ref{rt2.6-1}
and \ref{rt2.6-2}. $\qed$

Moreover, we have easily
\begin{cor}
Let $(x(t),\xi(t),\theta(t),\pi(t))\in C^1(\euc:{\mathcal T}^*{\frak R}^{3|2})$
be a solution of \ref{rt2.6-1}
and \ref{rt2.6-2}.
Then, it satisfies
\begin{equation}
\dt {\mathcal H}(t,x(t),\xi(t),\theta(t),\pi(t))
=\frac{\partial{\mathcal H}}{\partial t}(t,x(t),\xi(t),\theta(t),\pi(t)).
\label{rt3.18}
\end{equation}
\end{cor}

Using \eqref{rt2.7} and putting
$$
B_{jk}(t,x)=\frac{\partial A_k(t,x)}{\partial x_j}
-\frac{\partial A_j(t,x)}{\partial x_k},
$$
we rewrite \ref{rt2.6-1} as
\begin{equation}\tag*{$\rm (3.6){}'_{\mathrm{ev}}$} 
\allowdisplaybreaks
\left\{
\begin{aligned}
&\dt x_j=c\sigma_j(\theta,\pi),\\
&\dt \eta_j
=\sum_{k=1}^3{\elec}\sigma_k(\theta,\pi)B_{jk}(t,x)
-{\elec}\frac{\partial {A}_0(t,x)}{\partial x_j}.\\
\end{aligned}
\right.\label{rt2.6-11}
\end{equation}

\begin{cor}
Let $\{A_j(t,q)\}_{j=0}^3\in C^\infty(\euc\times\euc^3:\euc)$ satisfy \eqref{G}.
There exists a unique solution 
$({\tilde x}(t),{\tilde \eta}(t),
{\tilde \theta}(t),{\tilde \pi}(t))\in C^1(\euc:{\mathcal T}^*{\frak R}^{3|2})$
of \ref{rt2.6-11} and \ref{rt2.6-2}
with initial data
$$
({\tilde x}({\unbt}),
{\tilde \eta}({\unbt}),
{\tilde \theta}({\unbt}),
{\tilde \pi}({\unbt}))=
(\unbx,\unbeta,\unbtheta,\unbpi) \where
\unbeta_j=\unbxi_j-\frac{\elec}{c}A_j({\unbt},\unbx).
$$
Moreover, they are related to $(x(t),\xi(t),\theta(t),\pi(t))$ as
\begin{equation*}
\left\{
\begin{aligned}
&x_j(t,{\unbt}\,;{\inidata})
={\tilde x}_j(t,{\unbt}\,;
\unbx,\unbxi-\frac{\elec}{c}A({\unbt},\unbx),\unbtheta,\unbpi),\\
&\xi_j(t,{\unbt}\,;{\inidata})
={\tilde\eta}_j(t,{\unbt}\,;
\unbx,\unbxi-\frac{\elec}{c}A({\unbt},\unbx),\unbtheta,\unbpi)
+\frac{\elec}{c}A_j(t,{\tilde x}(t,{\unbt}\,;
\unbx,\unbxi-\frac{\elec}{c}A({\unbt},\unbx),\unbtheta,\unbpi)),\\
&\theta_k(t,{\unbt}\,;{\inidata})
={\tilde\theta}_k(t,{\unbt}\,;
\unbx,\unbxi-\frac{\elec}{c}A({\unbt},\unbx),\unbtheta,\unbpi),\\
&\pi_k(t,{\unbt}\,;{\inidata})
={\tilde\pi}_k(t,{\unbt}\,;
\unbx,\unbxi-\frac{\elec}{c}A({\unbt},\unbx),\unbtheta,\unbpi).
\end{aligned}\right.
\end{equation*}
\end{cor}

%%%

%%\input s-Weyl3-1-31

\subsubsection{Smoothness: Proof of Theorem \ref{existenceCM} continued}

For notational simplicity, we represent 
$x(t,{\unbt}\,; {\inidata})$ as $x(t)$ or $x$, etc.
We investigate the smoothness of 
$(x(t),\xi(t),\theta(t),\pi(t))$ 
with respect to the initial data $({\inidata})$. 
In the following, we put $(1-k)_+=\max(0,1-k)$ for $k\ge0$.

{\bf s-smoothness}:
In oder to prove the smoothness w.r.t. the initial data, 
we differentiate \ref{rt2.6-1} and \ref{rt2.6-2} formally w.r.t. 
$(\inidata)$, which
gives us the following differential equation:
\begin{equation}
{d \over dt } {\mathcal J}^{\cbra(1)}(t) 
= {\mathcal H}^{\cbra(2)}(t){\mathcal J}^{\cbra(1)}(t) \with
{\mathcal J}^{\cbra(1)}(0)=I.
\label{rt3.28}
\end{equation}
Here
\begin{equation}
{\mathcal J}^{\cbra(1)}(t)= 
\begin{pmatrix}
\partial_{\unbx}x&\partial_{\unbxi}x
&\partial_{\unbtheta}x&\partial_{\unbpi}x\\
\partial_{\unbx}\xi&\partial_{\unbxi}\xi
&\partial_{\unbtheta}\xi&\partial_{\unbpi}\xi\\
 \partial_{\unbx}\theta&  \partial_{\unbxi}\theta
&\partial_{\unbtheta}\theta&\partial_{\unbpi}\theta\\
 \partial_{\unbx}\pi&  \partial_{\unbxi}\pi
&\partial_{\unbtheta}\pi&\partial_{\unbpi}\pi
\end{pmatrix},\;
\partial_{\unbx}x=
\begin{pmatrix}
\partial_{{\unbx}_1}x_1&\partial_{{\unbx}_2}x_1&\partial_{{\unbx}_3}x_1\\
\partial_{{\unbx}_1}x_2&\partial_{{\unbx}_2}x_2&\partial_{{\unbx}_3}x_2\\
\partial_{{\unbx}_1}x_3&\partial_{{\unbx}_2}x_3&\partial_{{\unbx}_3}x_3\\
\end{pmatrix}, 
\quad \text{etc.}
\label{rt3.29}
\end{equation}
with arguments $(t,{\unbt}\,;\inidata)$ and
\begin{equation}
{\mathcal H}^{\cbra(2)}(t)=
\begin{pmatrix}
  \partial_x\partial_{\xi}{\mathcal H} & 
  \partial_{\xi}\partial_{\xi}{\mathcal H} &
  -\partial_{\theta}\partial_{\xi}{\mathcal H} & 
  -\partial_{\pi}\partial_{\xi}{\mathcal H} \\
-\partial_x\partial_x{\mathcal H} & -\partial_{\xi}\partial_x{\mathcal H} &
\partial_{\theta}\partial_x{\mathcal H} &\partial_{\pi}\partial_x{\mathcal H} \\
  \partial_x\partial_{\pi}{\mathcal H} & 
  \partial_{\xi}\partial_{\pi}{\mathcal H} &
-\partial_{\theta}\partial_{\pi}{\mathcal H} & 
-\partial_{\pi}\partial_{\pi}{\mathcal H} \\
  \partial_x\partial_{\theta}{\mathcal H} & 
  \partial_{\xi}\partial_{\theta}{\mathcal H} &
-\partial_{\theta}\partial_{\theta}{\mathcal H} & 
-\partial_{\pi}\partial_{\theta}{\mathcal H}
\end{pmatrix}
\label{rt3.30}
\end{equation}
where
$$
\partial_x\partial_{\xi}{\mathcal H}=
\begin{pmatrix}
\partial_{x_1}\partial_{\xi_1}{\mathcal H}&\partial_{x_2}\partial_{\xi_1}{\mathcal H}
&\partial_{x_3}\partial_{\xi_1}{\mathcal H}\\
\partial_{x_1}\partial_{\xi_2}{\mathcal H}&\partial_{x_2}\partial_{\xi_2}{\mathcal H}
&\partial_{x_3}\partial_{\xi_2}{\mathcal H}\\
\partial_{x_1}\partial_{\xi_3}{\mathcal H}&\partial_{x_2}\partial_{\xi_3}{\mathcal H}
&\partial_{x_3}\partial_{\xi_3}{\mathcal H}\\
\end{pmatrix}, 
\quad \text{etc.}
$$
with arguments $(t,x(t),\xi(t),\theta(t),\pi(t))$.
Remarking that each component of ${\mathcal H}^{\cbra(2)}(t)$ is 
differentiable w.r.t. $t$ for fixed $(\inidata)$
and proceeding as in the proof of the first part of Theorem \ref{existenceCM}, 
we get the
unique global (in time) solution of \eqref{rt3.28}. 
On the other hand, taking the difference quotient of \ref{rt2.6-1} and \ref{rt2.6-2}
w.r.t. the small perturbation of the initial data, 
making that perturbation tends to $0$
and remarking that each component of 
${\mathcal H}^{\cbra(2)}(t)$ is continuous w.r.t. $(\inidata)$, 
we may prove that the solution of \ref{rt2.6-1} and \ref{rt2.6-2} is 
in fact differentiable w.r.t. $(\inidata)$ and satisfies \eqref{rt3.28}.
(This process is well-known for proving the continuity of the solution
of ODE w.r.t. the initial data.)

Furthermore, for each positive integer $k\ge 1$ and $\ell\ge 2$, putting 
\begin{equation}
{\mathcal J}^{\cbra(k)}(t)=
\left(\partial_{\unbx}^{\alpha }\partial_{\unbxi}^{\beta}\partial_{\unbtheta}^a
\partial_{\unbpi}^b 
\begin{pmatrix}x\\ \xi\\ \theta\\ \pi\end{pmatrix}
\right){\begin{Sb}|\alpha+\beta|+\\ \;|a+b|=k\end{Sb}}
\et
{\mathcal H}^{\cbra(\ell)}(t)
=\big(\partial_{x}^{\alpha }\partial_{\xi}^{\beta}\partial_{\theta}^a
\partial_{\pi}^b {\mathcal H}\big){\begin{Sb}|\alpha+\beta|+\\ \;|a+b|=\ell\end{Sb}}
%_{(|\alpha|+|\beta|+|a|+|b|=\ell)},
\end{equation}
respectively, we have the following differential equation for $k\ge 2$:
\begin{equation}
\begin{aligned}
&{d \over dt}{\mathcal J}^{\cbra(k)}(t)
={\mathcal H}^{\cbra(2)}(t){\mathcal J}^{\cbra(k)}(t)+R^{\cbra(k)}(t) 
\with {\mathcal J}^{\cbra(k)}(0)=0\\
& \where R^{(k)}(t)=\sum_{p=2}^k\sum_{k=k_1+\cdots k_p}
c_{p,k}{\mathcal H}^{\cbra(p+1)}(t){\mathcal J}^{\cbra(k_1)}(t)\otimes\cdots\otimes {\mathcal J}^{\cbra(k_p)}(t).
\end{aligned}
\end{equation}
Here, $c_{p,k}$ are suitable constants. 
It is inductively proved that the each component of 
$R^{(k)}(t)$ is continuous w.r.t. $(\inidata)$
and differentiable w.r.t. $t$.
As above, this equation has the unique solution and therefore
the solution of \ref{rt2.6-1} and \ref{rt2.6-2} is 
in fact $k$-times differentiable w.r.t. $(\inidata)$.

Therefore, we get the $s$-smoothness of 
the solution of \ref{rt2.6-1} and \ref{rt2.6-2} w.r.t. $(t,{\unbt}\,;\inidata)$;
\begin{equation}
\left\{
\begin{aligned}
x(t)&= \sum_{|a|+|b|=0,2,4} x_{ab}(t)\unbtheta^a\unbpi^b \where
x_{ab}(t)=\partial_{\unbpi_2}^{b_2}\partial_{\unbpi_1}^{b_1}
\partial_{\unbtheta_2}^{a_2}\partial_{\unbtheta_1}^{a_1} 
 x(t,{\unbt}\,;\unbx,\unbxi,0,0),\\
\xi(t)&= \sum_{|a|+|b|=0,2,4} \xi_{ab}(t)\unbtheta^a\unbpi^b \where
\xi_{ab}(t)=\partial_{\unbpi_2}^{b_2}\partial_{\unbpi_1}^{b_1}
\partial_{\unbtheta_2}^{a_2}\partial_{\unbtheta_1}^{a_1} 
 \xi(t,{\unbt}\,;\unbx,\unbxi,0,0),\\
\theta(t)&=\sum_{|a|+|b|=1,3} \theta_{ab}(t)\unbtheta^a\unbpi^b\where
\theta_{ab}(t)=\partial_{\unbpi_2}^{b_2}\partial_{\unbpi_1}^{b_1}
\partial_{\unbtheta_2}^{a_2}\partial_{\unbtheta_1}^{a_1} 
\theta(t,{\unbt}\,;\unbx,\unbxi,0,0),\\ 
\pi(t)&= \sum_{|a|+|b|=1,3} \pi_{ab}(t)\unbtheta^a\unbpi^b \where
\pi_{ab}(t)=\partial_{\unbpi_2}^{b_2}\partial_{\unbpi_1}^{b_1}
\partial_{\unbtheta_2}^{a_2}\partial_{\unbtheta_1}^{a_1} 
 \pi(t,{\unbt}\,;\unbx,\unbxi,0,0),
\end{aligned}
\right.\label{rt3.33}
\end{equation}
with $a=(a_1,a_2),\;b=(b_1,b_2)\in\{0,1\}^2$.

%%%

%%\input s-Weyl3-1-32

\noindent
{\bf Estimates}:
We remark, by the structure of ${\mathcal H}(t,x,\xi,\theta,\pi)$, the following:
{\allowdisplaybreaks
\begin{gather}
\partial_{x_i}\partial_{\xi_j}{\mathcal H}=0,\quad
\partial_{\xi_i}\partial_{\xi_j}{\mathcal H}=0,\label{rt3.34}\\
{\begin{aligned}
\partial_{x_i}\partial_{x_j}{\mathcal H}&={\elec}\partial_{x_i}\partial_{x_j}{A}_0
-{\elec}\sum_{k=1}^3\sigma_k\partial_{x_i}\partial_{x_j}A_k\\
&={\elec}\partial_{x_i}\partial_{x_j}{A}_0
-{\elec}\sum_{|a|+|b|=2}(\text{linear combination of $\partial_{x_i}\partial_{x_j}A_*$})
\theta^a\pi^b,
\end{aligned}\label{rt3.35}}\\
\partial_{\theta_k}\partial_{\xi_j}{\mathcal H}
={\elec}\frac{\partial\sigma_j}{\partial\theta_k}
={\elec}\sum_{|a|+|b|=1}\text{const}_*\theta^a\pi^b,\label{rt3.36}\\
\partial_{\pi_k}\partial_{\xi_j}{\mathcal H}
={\elec}\frac{\partial\sigma_j}{\partial\pi_k}
={\elec}\sum_{|a|+|b|=1}\text{const}_*\theta^a\pi^b,\label{rt3.37}\\
\partial_{\theta_k}\partial_{x_j}{\mathcal H}
=-{\elec}\sum_{\ell=1}^3\frac{\partial\sigma_\ell}{\partial\theta_k}
\frac{\partial A_\ell}{\partial x_j}
={\elec}\sum_{|a|+|b|=1}
(\text{linear combination of $\partial_{x_j}A_*$})\theta^a\pi^b,\label{rt3.38}\\
\partial_{\pi_k}\partial_{x_j}{\mathcal H}
=-{\elec}\sum_{\ell=1}^3\frac{\partial\sigma_\ell}{\partial\pi_k}
\frac{\partial A_\ell}{\partial x_j}
={\elec}\sum_{|a|+|b|=1}
(\text{linear combination of $\partial_{x_j}A_*$})\theta^a\pi^b,\label{rt3.39}\\
\partial_{\theta_k}\partial_{\pi_l}{\mathcal H}
=-i\hbar^{-1}(c\xi_3-{\elec}A_3)\delta_{kl}=-i\hbar^{-1}c\eta_3\delta_{kl}
=-\partial_{\pi_l}\partial_{\theta_k}{\mathcal H},\label{rt3.40}\\
\partial_{\pi_1}\partial_{\pi_2}{\mathcal H}
=-\hbar^{-2}(c(\xi_1-i\xi_2)-{\elec}(A_1-iA_2))=-\hbar^{-2}c(\eta_1-i\eta_2)
=-\partial_{\pi_2}\partial_{\pi_1}{\mathcal H},
\label{rt3.41}\\
\partial_{\theta_1}\partial_{\theta_2}{\mathcal H}
=-c(\xi_1+i\xi_2)+{\elec}(A_1+iA_2)=-c(\eta_1+i\eta_2)
=-\partial_{\theta_2}\partial_{\theta_1}{\mathcal H}\label{rt3.42}
\end{gather}}
for any $i,j=1,2,3$ and $k,l=1,2$.

On the other hand, by \eqref{rt3.33}, we must estimate, for any $\alpha,\,\beta$,
\begin{equation}
\pi_{\mathrm B}\partial_{\unbx}^\alpha\partial_{\unbxi}^\beta
\partial_{\unbtheta}^a\partial_{\unbpi}^b x_j,\;
\pi_{\mathrm B}\partial_{\unbx}^\alpha\partial_{\unbxi}^\beta
\partial_{\unbtheta}^a\partial_{\unbpi}^b \xi_j
\for |a+b|=0,2,4,\label{rt3.001}
\end{equation}
and
\begin{equation}
\pi_{\mathrm B}\partial_{\unbx}^\alpha\partial_{\unbxi}^\beta
\partial_{\unbtheta}^a\partial_{\unbpi}^b \theta_k,\;
\pi_{\mathrm B}\partial_{\unbx}^\alpha\partial_{\unbxi}^\beta
\partial_{\unbtheta}^a\partial_{\unbpi}^b \pi_k
\for |a+b|=1,3.\label{rt3.002}
\end{equation}
Since it is obvious that body parts of other terms are $0$.

{\bf The case $|a+b|=0$:}
From \eqref{rt3.24}, we have
$\pi_{\mathrm B}\dot x_j=\pi_{\mathrm B}c\sigma_j(\theta,\pi)=0$,
which implies $\pi_{\mathrm B}(x_j-\unbx_j)=0$.
By \eqref{rt3.24}, we have, for $|\alpha+\beta|\ge1$,
$$
\begin{aligned}
\partial_{\unbx}^\alpha\partial_{\unbxi}^\beta \dot x_j(t)
=\sum{\begin{Sb}|\alpha-\alpha'|+\\
|\beta-\beta'|\ge1 \end{Sb}}
{\binom{\alpha}{\alpha'}}{\binom{\beta}{\beta'}}\Big\{
&\partial_{\unbx}^{\alpha-\alpha'}\partial_{\unbxi}^{\beta-\beta'} x_k(t)\cdot
\partial_{\unbx}^{\alpha'}\partial_{\unbxi}^{\beta'}
{\mathcal H}_{x_k\xi_j}
+\partial_{\unbx}^{\alpha-\alpha'}\partial_{\unbxi}^{\beta-\beta'}\xi_k(t)\cdot
\partial_{\unbx}^{\alpha'}\partial_{\unbxi}^{\beta'}
{\mathcal H}_{\xi_k\xi_j}\\
&\quad+\partial_{\unbx}^{\alpha-\alpha'}
\partial_{\unbxi}^{\beta-\beta'}\theta_\ell(t)\cdot
\partial_{\unbx}^{\alpha'}\partial_{\unbxi}^{\beta'}
{\mathcal H}_{\theta_\ell\xi_j}
+\partial_{\unbx}^{\alpha-\alpha'}\partial_{\unbxi}^{\beta-\beta'} \pi_\ell(t)\cdot
\partial_{\unbx}^{\alpha'}\partial_{\unbxi}^{\beta'}
{\mathcal H}_{\pi_\ell\xi_j}\Big\}
\end{aligned}
$$
with argument of ${\mathcal H}_{{x_k}\xi_j}$, etc, being $(x(t),\xi(t),\theta(t),\pi(t))$.
On the other hand, as ${\mathcal H}_{x_k\xi_j}=0={\mathcal H}_{\xi_k\xi_j}$ and body parts of
$\partial_{\unbx}^{\alpha}\partial_{\unbxi}^{\beta}\theta_\ell(t)=0=
\partial_{\unbx}^{\alpha}\partial_{\unbxi}^{\beta}\pi_\ell(t)$, we have,
$$
{\dt}\,\partial_{\unbx}^\alpha\partial_{\unbxi}^\beta  x_j\,
(t,{\unbt}\,;\unbx^{[0]},\unbxi^{[0]},0,0)=0\quad\mbox{and therefore}\quad
\pi_{\mathrm B}\partial_{\unbx}^\alpha\partial_{\unbxi}^\beta (x_j-{\unbx}_j)=0.
$$
Analogously, as we get
$$
{\dt}\,\partial_{\unbx}^\alpha\partial_{\unbxi}^\beta \xi_j\,
(t,{\unbt}\,;\unbx^{[0]},\unbxi^{[0]},0,0)
=-{\elec}\partial_{\unbx}^\alpha\partial_{\unbxi}^\beta\partial_{x_j}{{A}_0}(\unbx^{[0]})
$$
we have
$$
|\pi_{\mathrm B}\partial_{\unbx}^\alpha\partial_{\unbxi}^\beta 
(\xi_j(t,{\unbt}\,;{\inidata})-{\unbxi}_j)|
\le {\elec}|t-{\unbt}|\delta_{0|\beta|}\trs A_0\trs_{|\alpha|+1,\infty}.
%m_0^{\cbra({|\alpha|+1})}.
$$
These give (2)-(i) of Theorem \ref{existenceCM}.

{\bf The case $|a+b|=1$:}
For notational simplicity, we denote by
\begin{equation}
\partial_{\unbpi}\dot\theta=-\partial_{\unbpi}{\mathcal H}_{\pi}
=-\partial_{\unbpi}{x}\cdot{\mathcal H}_{x\pi}
-\partial_{\unbpi}{\xi}\cdot{\mathcal H}_{\xi\pi}
-\partial_{\unbpi}{\theta}\cdot{\mathcal H}_{\theta\pi}
-\partial_{\unbpi}{\pi}\cdot{\mathcal H}_{\pi\pi},\quad\text{etc}
\label{01}
\end{equation}
which is the abbreviation of
$$
\partial_{\unbpi_k}\dot\theta_j=-\partial_{\unbpi_k}{\mathcal H}_{\pi_j}
=-\partial_{\unbpi_k}{x_m}\cdot{\mathcal H}_{x_m\pi_j}
-\partial_{\unbpi_k}{\xi_m}\cdot{\mathcal H}_{\xi_m\pi_j}
-\partial_{\unbpi_k}{\theta_n}\cdot{\mathcal H}_{\theta_n\pi_j}
-\partial_{\unbpi_k}{\pi_n}\cdot{\mathcal H}_{\pi_n\pi_j},\quad\text{etc}.
$$
From above, we have,
\begin{equation}
\dt {\mathcal J}^{\cbra({0|1})}_1(t)={\mathcal H}^{\cbra({0|2})}(t)
{\mathcal J}^{\cbra({0|1})}_1(t)
\with
{\mathcal J}^{\cbra({0|1})}_1(0)=I
\label{rt3.48}
\end{equation}
where
$$
{\mathcal J}^{\cbra({0|1})}_1(t)=
{\begin{pmatrix}
\partial_{\unbtheta} \theta& \partial_{\unbpi} \theta\\
\partial_{\unbtheta} \pi& \partial_{\unbpi} \pi
\end{pmatrix}}
(t,{\unbt}\,;\unbx^{[0]},\unbxi^{[0]},0,0),\quad
{\mathcal H}^{\cbra({0|2})}(t)=
{\begin{pmatrix}
-{\mathcal H}_{\theta\pi}&-{\mathcal H}_{\pi\pi}\\
-{\mathcal H}_{\theta\theta}&-{\mathcal H}_{\pi\theta}
\end{pmatrix}}
(x^{[0]}(t),\xi^{[0]}(t),0,0).
$$

More explicitly, a part of \eqref{rt3.48} 
with the argument $(t,{\unbt}\,;\unbx^{[0]},\unbxi^{[0]},0,0)$ abbreviated, is rewritten as
\begin{equation}
\dt \begin{pmatrix}
%\frac{\partial\theta_1}{\partial\unbtheta_k}
\partial_{\unbtheta_k}\theta_1\\
%\frac{\partial\pi_2}{\partial\unbtheta_k}
\partial_{\unbtheta_k}\pi_2
\end{pmatrix}
=\begin{pmatrix}
-ic\hbar^{-1}\eta_3&-c\hbar^{-2}(\eta_1-i\eta_2)\\
c(\eta_1+i\eta_2)&ic\hbar^{-1}\eta_3
\end{pmatrix}
\begin{pmatrix}
\partial_{\unbtheta_k}\theta_1\\
\partial_{\unbtheta_k}\pi_2
\end{pmatrix}.
\label{3-47}\end{equation}
Applying $\begin{pmatrix}\hbar&0\\0&1\end{pmatrix}$ to both sides of \eqref{3-47}
and taking the body parts,
we have
\begin{equation}
\dt Z^1_{2,\unbtheta_k}(t)={\Bbb X}^1_2(t) Z^1_{2,\unbtheta_k}(t)
\label{3-11}\end{equation}
where
$$
Z^1_{2,\unbtheta_k}(t)=\begin{pmatrix} %\pi_{\mathrm B}
\hbar\partial_{\unbtheta_k}\theta_1\\
\partial_{\unbtheta_k}\pi_2
\end{pmatrix}(t,{\unbt}\,;\unbx^{[0]},\unbxi^{[0]},0,0),\quad
{\Bbb X}^1_2(t)=c\hbar^{-1}\begin{pmatrix}
-i\eta_3&-\eta_1+i\eta_2\\
\eta_1+i\eta_2&i\eta_3
\end{pmatrix}(t,{\unbt}\,;\unbx^{[0]},\unbxi^{[0]},0,0).
$$

We prepare the following simple lemma:
\begin{lem}
Let $H$ be a Hilbert space over ${\Bbb C}$ with scalar product and norm denoted by
$(\cdot,\cdot)$ and $\Vert\cdot\Vert$, respectively.
Let $A(t)\in C([0,T]:{\Bbb B}(H))$
with $\Re (A(t)v,v)=0$ for any $v\in H$.
If $u(t)\in C^1([0,T]:H)$ satisfy
$$
\dot u(t)=A(t)u(t)+F(t),
$$
then, we have
$$
\Vert u(t)\Vert^2=\Vert u(0)\Vert^2+2\int_0^t ds \, \Re (F(s),u(s)).
$$
Moreover, we get
$$
\Vert u(t)\Vert^2\le e^t\Vert u(0)\Vert^2+e^t\int_0^t ds \, e^{-s}\Vert F(s)\Vert^2.
$$
\end{lem}

Now, applying this lemma to \eqref{3-11} with $H={\Bbb C}^2$, $A(t)={\Bbb X}^1_2(t)$,
$u(t)=Z^1_{2,\unbtheta_k}(t)$ and $F(t)=0$,
we have
\begin{equation}
\bigg|\hbar\frac{\partial\theta_1}{\partial\unbtheta_k}
(t,{\unbt}\,;\unbx^{[0]},\unbxi^{[0]},0,0)\bigg|^2
+\bigg|\frac{\partial\pi_2}{\partial\unbtheta_k}
(t,{\unbt}\,;\unbx^{[0]},\unbxi^{[0]},0,0)\bigg|^2
=\hbar^2 \delta_{1k} \for k=1,2.
\end{equation}
Analogously, as
\begin{equation}
\dt Z^1_{2,\unbpi_\ell}(t)={\Bbb X}^1_2(t)Z^1_{2,\unbpi_\ell}(t)
\with Z^1_{2,\unbpi_\ell}(t)=
\begin{pmatrix}
\hbar\partial_{\unbpi_\ell}\theta_1\\
\partial_{\unbpi_\ell}\pi_2
\end{pmatrix}(t,{\unbt}\,;\unbx^{[0]},\unbxi^{[0]},0,0),
\label{3-12}
\end{equation}
we have
\begin{equation}
\bigg|\hbar\frac{\partial\theta_1}{\partial\unbpi_\ell}(t,{\unbt}\,;\unbx^{[0]},\unbxi^{[0]},0,0)\bigg|^2+
\bigg|\frac{\partial\pi_2}{\partial\unbpi_\ell}(t,{\unbt}\,;\unbx^{[0]},\unbxi^{[0]},0,0)\bigg|^2
=\delta_{2\ell} \for \ell=1,2.
\end{equation}

By the same fashion, we have 
\begin{equation}
\begin{aligned}
&\bigg|\hbar\frac{\partial\theta_2}{\partial\unbtheta_k}(t,{\unbt}\,;\unbx^{[0]},\unbxi^{[0]},0,0)\bigg|^2+
\bigg|\frac{\partial\pi_1}{\partial\unbtheta_k}(t,{\unbt}\,;\unbx^{[0]},\unbxi^{[0]},0,0)\bigg|^2
=\hbar^2 \delta_{2k} \for k=1,2,\\
&\bigg|\hbar\frac{\partial\theta_2}{\partial\unbpi_\ell}(t,{\unbt}\,;\unbx^{[0]},\unbxi^{[0]},0,0)\bigg|^2+
\bigg|\frac{\partial\pi_1}{\partial\unbpi_\ell}(t,{\unbt}\,;\unbx^{[0]},\unbxi^{[0]},0,0)\bigg|^2
=\delta_{1\ell} \for \ell=1,2.
\end{aligned}
\end{equation}
This gives the proof of (ii) with $|a+b|=1$, $k=|\alpha+\beta|=0$ 
(here, we abbused the subscript $k$).

We put
$$
\begin{gathered}
{\mathcal J}_1^{\cbra({k|1})}(t)=\begin{pmatrix}
\partial_{\unbx}^\alpha\partial_{\unbxi}^\beta
\partial_{\unbtheta}^a\partial_{\unbpi}^b\theta\\[5pt]
\partial_{\unbx}^\alpha\partial_{\unbxi}^\beta
\partial_{\unbtheta}^a\partial_{\unbpi}^b\pi
\end{pmatrix}(t,{\unbt}\,;\unbx^{[0]},\unbxi^{[0]},0,0)\with
|a+b|=1 \et |\alpha+\beta|=k,\\
\et {\mathcal H}^{\cbra({\ell|2})}(t)=
{\begin{pmatrix}
-\partial_{\unbx}^\alpha\partial_{\unbxi}^\beta{\mathcal H}_{\theta\pi}
&-\partial_{\unbx}^\alpha\partial_{\unbxi}^\beta{\mathcal H}_{\pi\pi}\\
-\partial_{\unbx}^\alpha\partial_{\unbxi}^\beta{\mathcal H}_{\theta\theta}
&-\partial_{\unbx}^\alpha\partial_{\unbxi}^\beta{\mathcal H}_{\pi\theta}
\end{pmatrix}}
(t,x^{[0]}(t),\xi^{[0]}(t),0,0) \with |\alpha+\beta|=\ell.
\end{gathered}
$$
Then, we have
\begin{equation}
\dt{\mathcal J}_1^{\cbra({k|1})}(t)={\mathcal H}^{\cbra({0|2})}(t){\mathcal J}_1^{\cbra({k|1})}(t)
+{\mathcal F}_1^{(k)}(t) \with {\mathcal J}_1^{\cbra({k|1})}(0)=0,
\label{rt3.44}\end{equation}
where
\begin{equation}
{\mathcal F}_1^{(k)}(t)
=\sum_{\ell=1}^k{\mathcal H}^{\cbra({\ell|2})}(t){\mathcal J}_1^{\cbra({k-\ell|1})}(t)
={\mathcal H}^{\cbra({k|2})}(t){\mathcal J}_1^{\cbra({0|1})}(t)
+\cdots.\label{102}
\end{equation}

For example, when $k=1$, we have
$$
\begin{aligned}
\partial_{\unbx}
\partial_{\unbpi}\dot\theta
&=-\partial_{\unbx}\partial_{\unbpi}{\theta}\cdot{\mathcal H}_{\theta\pi}
-\partial_{\unbx}\partial_{\unbpi}{\pi}\cdot{\mathcal H}_{\pi\pi}
-**-\partial_{\unbpi}{\theta}(\partial_{\unbx}{\mathcal H}_{\theta\pi})
-\partial_{\unbpi}{\pi}(\partial_{\unbx}{\mathcal H}_{\pi\pi})+**,\\
&\partial_{\unbx}{\mathcal H}_{\theta\pi}=
\partial_{\unbx}x\cdot{\mathcal H}_{x\theta\pi}
+\partial_{\unbx}\xi\cdot{\mathcal H}_{\xi\theta\pi}+**,\;
\partial_{\unbx}{\mathcal H}_{\pi\pi}=
\partial_{\unbx}x\cdot{\mathcal H}_{x\pi\pi}
+\partial_{\unbx}\xi\cdot{\mathcal H}_{\xi\pi\pi}+**
\end{aligned}
$$
where terms $**$ represent whose body parts vanish.
Therefore, 
a part of \eqref{rt3.44} is rewritten as
\begin{equation}
\dt 
Z^1_{2,\unbx_j\unbtheta_k}(t)
={\Bbb X}^1_2(t)Z^1_{2,\unbx_j\unbtheta_k}(t)
+{\Bbb Y}^1_{2,\unbx_j}(t)Z^1_{2,\unbtheta_k}(t)
\end{equation}
where
$$
\begin{gathered}
Z^1_{2,\unbx_j\unbtheta_k}(t)=
\begin{pmatrix}
\partial_{\unbx_j}\partial_{\unbtheta_k}\theta_1\\
\partial_{\unbx_j}\partial_{\unbtheta_k}\pi_2
\end{pmatrix}
(t,{\unbt}\,;\unbx^{[0]},\unbxi^{[0]},0,0),\\
{\Bbb Y}^1_{2,\unbx_j}(t)=\hbar^{-1}
\begin{pmatrix}
\hbar&0\\
0&1\end{pmatrix}
\begin{pmatrix}
\partial_{\unbx_j}{\mathcal H}_{\theta_1\pi_1}&\partial_{\unbx_j}{\mathcal H}_{\pi_2\pi_1}\\
\partial_{\unbx_j}{\mathcal H}_{\theta_1\theta_2}&\partial_{\unbx_j}{\mathcal H}_{\pi_2\theta_2}
\end{pmatrix}(t,{\unbt}\,;\unbx^{[0]},\unbxi^{[0]},0,0)
\begin{pmatrix}
1&0\\
0&\hbar\end{pmatrix}.
\end{gathered}
$$
Using above estimate, we have
$$
\bmid {\Bbb Y}^1_{2,\unbx_j}(t)Z^1_{2,\unbtheta_k}(t)\bmid\le \tilde{C}_1^{(1)},
$$
where
$\tilde{C}_1^{(1)}$ depending on $\trs A\trs_{1,\infty}=\sup_{j=1,2,3}\trs A_j\trs_{1,\infty}, 
\trs A_0\trs_{2,\infty}$
(dependence on ${\elec},\,c,\,\hbar,\,T$ won't be clearly mentioned hereafter). 
By Lemma 3.4, we get
$$
\bmid Z^1_{2,\unbx_j\unbtheta_k}(t)\bmid \le C_1^{(1)}\delta_{1k}|t-{\unbt}|^{1/2}.
$$
Calculating analogously, we get
$$
\bmid{\mathcal F}_1^{(1)}(t)\bmid\le\tilde{C}_1^{(1)}\et
\bmid{\mathcal J}_1^{\cbra({1|1})}(t)\bmid
\le C_1^{(1)}|t-{\unbt}|^{1/2} \for |t-{\unbt}|\le 1.
$$

By induction w.r.t. $k$,
because of the first term of the rightest hand of \eqref{102} having the bounded body part,
there exists constant $\tilde{C}_1^{(k)}$ s.t.
\begin{equation}
\bmid {\mathcal F}_1^{(k)}(t)\bmid\le\tilde{C}_1^{(k)}\for |t-{\unbt}|\le1.
\end{equation}
Therefore, using Lemma 3.4, we have, 
$$
\bmid {\mathcal J}_1^{\cbra({k|1})}(t)\bmid\le {C}_1^{(k)}|t-{\unbt}|^{1/2}\for |t-{\unbt}|\le1,\; k\ge1.
$$
In the above, constants $\tilde{C}_1^{(k)}$ and ${C}_1^{(k)}$ 
are independent of $(t,\unbxi,\unbtheta)$ 
(this saying will be abbreviated if it is no need to stress this).

We proved (ii) with $|a+b|=1$.

{\bf The case $|a+b|=2$:}
As before, using ${\mathcal H}_{x\xi}=0={\mathcal H}_{\xi\xi}$, we get
\begin{equation}
\partial_{\unbtheta}\dot x=
\partial_{\unbtheta}\theta\cdot{\mathcal H}_{\theta\xi}
+\partial_{\unbtheta}\pi\cdot{\mathcal H}_{\pi\xi},\quad
\partial_{\unbpi}\dot x=
\partial_{\unbpi}\theta\cdot{\mathcal H}_{\theta\xi}
+\partial_{\unbpi}\pi\cdot{\mathcal H}_{\pi\xi}.
\label{03}
\end{equation}
Moreover,
\begin{equation}
\begin{aligned}
&\partial_{\unbtheta}^2\dot x
=\partial_{\unbtheta}^2\theta\cdot{\mathcal H}_{\theta\xi}
+\partial_{\unbtheta}\theta(\partial_{\unbtheta}{\mathcal H}_{\theta\xi})
+\partial_{\unbtheta}^2\pi\cdot{\mathcal H}_{\pi\xi}
+\partial_{\unbtheta}\pi(\partial_{\unbtheta}{\mathcal H}_{\pi\xi}),\\
&\partial_{\unbpi}^2\dot x
=\partial_{\unbpi}^2\theta\cdot{\mathcal H}_{\theta\xi}
+\partial_{\unbpi}\theta(\partial_{\unbpi}{\mathcal H}_{\theta\xi})
+\partial_{\unbpi}^2\pi\cdot{\mathcal H}_{\pi\xi}
+\partial_{\unbpi}\pi(\partial_{\unbpi}{\mathcal H}_{\pi\xi}),\quad
\partial_{\unbtheta}\partial_{\unbpi}\dot x=\cdots,\\
&\with
\partial_{\unbtheta}{\mathcal H}_{\theta\xi}=
\partial_{\unbtheta}\theta
\cdot{\mathcal H}_{\theta\theta\xi}
+\partial_{\unbtheta}\pi\cdot{\mathcal H}_{\pi\theta\xi},\;
\partial_{\unbtheta}{\mathcal H}_{\pi\xi}=
\partial_{\unbtheta}\theta
\cdot{\mathcal H}_{\theta\pi\xi}
+\partial_{\unbtheta}\pi\cdot{\mathcal H}_{\pi\pi\xi},\;\;\text{etc}.
\end{aligned}
\label{04}
\end{equation}
As we have
$$
\begin{gathered}
\partial_{\unbtheta}^2\theta\cdot{\mathcal H}_{\theta\xi}\bigg|_{\unbtheta=\unbpi=0}
=\partial_{\unbtheta}^2\pi\cdot{\mathcal H}_{\pi\xi}\bigg|_{\unbtheta=\unbpi=0}=0,\\
{\bmid}\pi_{\mathrm B}
\partial_{\unbtheta}\theta(\partial_{\unbtheta}{\mathcal H}_{\theta\xi}){\bmid}
\le \tilde{C}_2^{(0)},
\quad \mbox{etc.},
\end{gathered}
$$
using estimates already obtained, we get
$$
{\bmid}\pi_{\mathrm B}\partial_{\unbtheta}^2 x{\bmid},\;
{\bmid}\pi_{\mathrm B}\partial_{\unbpi}\partial_{\unbtheta} x{\bmid}
\le C_2^{(0)}|t-{\unbt}|.  %(t,{\unbt}\,;\unbx^{[0]},\unbxi^{[0]},0,0)
$$
Analogously, 
\begin{equation}
\partial_{\unbtheta}\dot\xi=
-\partial_{\unbtheta}x\cdot{\mathcal H}_{xx}
-\partial_{\unbtheta}\theta\cdot{\mathcal H}_{\theta{x}}
-\partial_{\unbtheta}\pi\cdot{\mathcal H}_{\pi{x}},\quad
\partial_{\unbpi}\dot\xi=
-\partial_{\unbpi}x\cdot{\mathcal H}_{xx}
-\partial_{\unbpi}\theta\cdot{\mathcal H}_{\theta{x}}
-\partial_{\unbpi}\pi\cdot{\mathcal H}_{\pi{x}},
\label{05}
\end{equation}
and
\begin{equation}
\begin{aligned}
&\partial_{\unbtheta}^2\dot \xi
=-\partial_{\unbtheta}^2 x\cdot{\mathcal H}_{xx}
-\partial_{\unbtheta}\theta
(\partial_{\unbtheta}{\mathcal H}_{\theta{x}})
-\partial_{\unbtheta}\pi
(\partial_{\unbtheta}{\mathcal H}_{\theta\pi{x}})
-\partial_{\unbtheta}^2x\cdot{\mathcal H}_{xx}
-\partial_{\unbtheta}^2\theta\cdot{\mathcal H}_{\theta{x}}
-\partial_{\unbtheta}^2\pi\cdot{\mathcal H}_{\pi{x}},\\
&\partial_{\unbtheta}\partial_{\unbpi}\dot \xi=\cdots,\quad
\partial_{\unbpi}^2\dot \xi=\cdots,\;\;\text{etc}.
\end{aligned}
\label{06}
\end{equation}
which implies
$$
{\bmid}\pi_{\mathrm B}\partial_{\unbtheta}^2 \xi{\bmid},\;
{\bmid}\pi_{\mathrm B}\partial_{\unbtheta}\partial_{\unbpi}\xi{\bmid}
\le C_2^{(0)}|t-{\unbt}|.
$$

For $k\ge1$, putting
$$
{\mathcal J}_0^{\cbra({k|2})}(t)=\begin{pmatrix}
\partial_{\unbx}^\alpha\partial_{\unbxi}^\beta
\partial_{\unbtheta}^a\partial_{\unbpi}^b x\\[5pt]
\partial_{\unbx}^\alpha\partial_{\unbxi}^\beta
\partial_{\unbtheta}^a\partial_{\unbpi}^b\xi
\end{pmatrix}(t,{\unbt}\,;\unbx^{[0]},\unbxi^{[0]},0,0)\with
|a+b|=2 \et |\alpha+\beta|=k,
$$
we have
$$
\dot {\mathcal J}_0^{\cbra({k|2})}(t)={\mathcal F}^{(k)}_2(t)\with
\bmid{\mathcal F}^{(k)}_2(t)\bmid\le\tilde{C}_2^{(k)}|t-{\unbt}|^{1/2}\when |t-{\unbt}|\le1,
$$
which yields
$$
\bmid{\mathcal J}_0^{\cbra({k|2})}(t)\bmid\le{C}_2^{(k)}|t-{\unbt}|^{3/2}\when |t-{\unbt}|\le1.
$$

{\bf The case $|a+b|=3$:}
$$
{\mathcal J}_1^{\cbra({k|3})}(t)=\begin{pmatrix}
\partial_{\unbx}^\alpha\partial_{\unbxi}^\beta
\partial_{\unbtheta}^a\partial_{\unbpi}^b\theta\\
\partial_{\unbx}^\alpha\partial_{\unbxi}^\beta
\partial_{\unbtheta}^a\partial_{\unbpi}^b\pi
\end{pmatrix}(t,{\unbt}\,;\unbx^{[0]},\unbxi^{[0]},0,0)\with
|a+b|=3 \et |\alpha+\beta|=k
$$
satisfies
\begin{equation}
\dot{\mathcal J}_1^{\cbra({k|3})}(t)
={\mathcal H}^{\cbra({0|2})}(t){\mathcal J}_1^{\cbra({k|3})}(t)
+{\mathcal F}_3^{(k)}(t),
\end{equation}
with
$$
{\mathcal F}_3^{(k)}(t)
=\sum_{\ell=1}^k{\mathcal H}^{\cbra({\ell|2})}(t){\mathcal J}_1^{\cbra({k-\ell|3})}(t)
={\mathcal H}^{\cbra({k|2})}(t){\mathcal J}_1^{\cbra({0|3})}(t)
+\cdots.
$$

For example, when $k=0$, we have
$$
\partial_{\unbpi}^2\dot\theta
=-\partial_{\unbpi}^2{x}\cdot{\mathcal H}_{x\pi}
+\partial_{\unbpi}{x}(\partial_{\unbpi}{\mathcal H}_{x\pi})
-\partial_{\unbpi}^2{\xi}\cdot{\mathcal H}_{\xi\pi}
+\partial_{\unbpi}{\xi}(\partial_{\unbpi}{\mathcal H}_{\xi\pi})
-\partial_{\unbpi}^2{\theta}\cdot{\mathcal H}_{\theta\pi}
-\partial_{\unbpi}{\theta}(\partial_{\unbpi}{\mathcal H}_{\theta\pi})
-\partial_{\unbpi}^2{\pi}\cdot{\mathcal H}_{\pi\pi}
-\partial_{\unbpi}{\pi}(\partial_{\unbpi}{\mathcal H}_{\pi\pi}),
$$
and
$$
\begin{aligned}
\partial_{\unbtheta}\partial_{\unbpi}^2\dot\theta
&=
-\partial_{\unbpi}^2{x}(\partial_{\unbtheta}{\mathcal H}_{x\pi})
+\partial_{\unbtheta}\partial_{\unbpi}{x}(\partial_{\unbpi}{\mathcal H}_{x\pi})
-\partial_{\unbpi}^2{\xi}(\partial_{\unbtheta}{\mathcal H}_{\xi\pi})
+\partial_{\unbtheta}\partial_{\unbpi}{\xi}(\partial_{\unbpi}{\mathcal H}_{\xi\pi})\\
&\qquad
-\partial_{\unbtheta}\partial_{\unbpi}^2{\theta}\cdot{\mathcal H}_{\theta\pi}
-\partial_{\unbpi}{\theta}(\partial_{\unbtheta}\partial_{\unbpi}{\mathcal H}_{\theta\pi})
-\partial_{\unbtheta}\partial_{\unbpi}^2{\pi}\cdot{\mathcal H}_{\pi\pi}
-\partial_{\unbpi}{\pi}(\partial_{\unbtheta}\partial_{\unbpi}{\mathcal H}_{\pi\pi})
+\{**\},\quad\text{etc},
\end{aligned}
$$
where $\{**\}$ has no body part, because
$$
\begin{aligned}
\{**\}&=-\partial_{\unbtheta}\partial_{\unbpi}^2{x}\cdot{\mathcal H}_{x\pi}
-\partial_{\unbpi}{x}(\partial_{\unbtheta}\partial_{\unbpi}{\mathcal H}_{x\pi})
-\partial_{\unbtheta}\partial_{\unbpi}^2{\xi}\cdot{\mathcal H}_{\xi\pi}
-\partial_{\unbpi}{\xi}(\partial_{\unbtheta}\partial_{\unbpi}{\mathcal H}_{\xi\pi})\\
&\qquad
-\partial_{\unbpi}^2{\theta}(\partial_{\unbtheta}{\mathcal H}_{\theta\pi})
-\partial_{\unbtheta}\partial_{\unbpi}{\theta}(\partial_{\unbpi}{\mathcal H}_{\theta\pi})
-\partial_{\unbpi}^2{\pi}(\partial_{\unbtheta}{\mathcal H}_{\pi\pi})
-\partial_{\unbtheta}\partial_{\unbpi}{\pi}(\partial_{\unbpi}{\mathcal H}_{\pi\pi}).
\end{aligned}
$$
Then, the body part of a part of ${\mathcal F}^{(0)}_3(t)$ is estimated by
$$
{\bmid}\pi_{\mathrm B}\partial_{\unbpi}^2{x} \partial_{\unbtheta}{\mathcal H}_{x\pi}{\bmid}
\le C|t-{\unbt}|,
\;\mbox{and therefore,} \;
\bmid{\mathcal F}_3^{(0)}(t)\bmid\le \tilde{C}|t-{\unbt}|.
$$
Using Lemma 3.4 and the inequality above, we have
$$
\bmid{\mathcal J}_1^{\cbra({0|3})}(t)\bmid
\le C_3^{(0)}|t-{\unbt}|^{3/2}.
$$
Moreover, when $k\ge1$, we have
$$
\bmid{\mathcal F}_3^{(k)}(t)\bmid\le \tilde{C}|t-{\unbt}|^{3/2}\et
\bmid{\mathcal J}_1^{\cbra({k|3})}(t)\bmid
\le C_3^{(k)}|t-{\unbt}|^2.
$$

{\bf The case $|a+b|=4$:}
Let $k=0$. From \eqref{03}, we have
\begin{equation*}
\begin{aligned}
&\partial_{\unbpi}^2\dot x
=\partial_{\unbpi}^2\theta\cdot{\mathcal H}_{\theta\xi}
+\partial_{\unbpi}\theta(\partial_{\unbpi}{\mathcal H}_{\theta\xi})
+\partial_{\unbpi}^2\pi\cdot{\mathcal H}_{\pi\xi}
+\partial_{\unbpi}\pi(\partial_{\unbpi}{\mathcal H}_{\pi\xi}),\\
&\with\partial_{\unbpi}{\mathcal H}_{\theta\xi}
=\partial_{\unbpi}\theta\cdot{\mathcal H}_{\theta\theta\xi}
+\partial_{\unbpi}\pi\cdot{\mathcal H}_{\pi\theta\xi},\;
\partial_{\unbpi}{\mathcal H}_{\pi\xi}
=\partial_{\unbpi}\theta\cdot{\mathcal H}_{\theta\pi\xi}
+\partial_{\unbpi}\pi\cdot{\mathcal H}_{\pi\pi\xi}.
\end{aligned}
%\label{06}
\end{equation*}
Remarking that $\partial_{\unbtheta}{\mathcal H}_{\theta\theta\xi}
=\partial_{\unbtheta}{\mathcal H}_{\theta\pi\xi}
=\partial_{\unbtheta}{\mathcal H}_{\pi\pi\xi}=0$,
we have
\begin{equation*}
\begin{aligned}
&\partial_{\unbtheta}\partial_{\unbpi}^2\dot x
=\partial_{\unbtheta}\partial_{\unbpi}^2\theta\cdot{\mathcal H}_{\theta\xi}
-\partial_{\unbpi}^2\theta(\partial_{\unbtheta}{\mathcal H}_{\theta\xi})
+\partial_{\unbtheta}\partial_{\unbpi}\theta(\partial_{\unbpi}{\mathcal H}_{\theta\xi})
+\partial_{\unbpi}\theta(\partial_{\unbtheta}\partial_{\unbpi}{\mathcal H}_{\theta\xi})\\
&\qquad\quad
+\partial_{\unbtheta}\partial_{\unbpi}^2\pi\cdot{\mathcal H}_{\pi\xi}
-\partial_{\unbpi}^2\pi(\partial_{\unbtheta}{\mathcal H}_{\pi\xi})
+\partial_{\unbtheta}\partial_{\unbpi}\pi(\partial_{\unbpi}{\mathcal H}_{\pi\xi})
+\partial_{\unbpi}\pi(\partial_{\unbtheta}\partial_{\unbpi}{\mathcal H}_{\pi\xi}),\\
&\with
\partial_{\unbtheta}\partial_{\unbpi}{\mathcal H}_{\theta\xi}
=\partial_{\unbtheta}\partial_{\unbpi}\theta\cdot{\mathcal H}_{\theta\theta\xi}
+\partial_{\unbtheta}\partial_{\unbpi}\pi\cdot{\mathcal H}_{\pi\theta\xi},\;
\partial_{\unbtheta}\partial_{\unbpi}{\mathcal H}_{\pi\xi}
=\partial_{\unbtheta}\partial_{\unbpi}\theta\cdot{\mathcal H}_{\theta\pi\xi}
+\partial_{\unbtheta}\partial_{\unbpi}\pi\cdot{\mathcal H}_{\pi\pi\xi}.
\end{aligned}.
%\label{07}
\end{equation*}
Finally, we have
\begin{equation*}
\begin{aligned}
&\partial_{\unbtheta}^2\partial_{\unbpi}^2\dot x
=\partial_{\unbtheta}^2\partial_{\unbpi}\theta
(\partial_{\unbpi}{\mathcal H}_{\theta\xi})
+\partial_{\unbpi}\theta
(\partial_{\unbtheta}^2\partial_{\unbpi}{\mathcal H}_{\theta\xi})
+\partial_{\unbtheta}^2\partial_{\unbpi}\pi
(\partial_{\unbpi}{\mathcal H}_{\pi\xi})
+\partial_{\unbpi}\pi
(\partial_{\unbtheta}^2\partial_{\unbpi}{\mathcal H}_{\pi\xi})\\
&\qquad\qquad\quad
+\partial_{\unbtheta}^2\partial_{\unbpi}^2\theta\cdot{\mathcal H}_{\theta\xi}
+\partial_{\unbpi}^2\theta(\partial_{\unbtheta}^2{\mathcal H}_{\theta\xi})
+\partial_{\unbtheta}^2\partial_{\unbpi}^2\pi\cdot{\mathcal H}_{\pi\xi}
+\partial_{\unbpi}^2\pi(\partial_{\unbtheta}^2{\mathcal H}_{\pi\xi})\\
&\with
\partial_{\unbtheta}^2\partial_{\unbpi}{\mathcal H}_{\theta\xi}
=\partial_{\unbtheta}^2\partial_{\unbpi}\theta\cdot{\mathcal H}_{\theta\theta\xi}
+\partial_{\unbtheta}^2\partial_{\unbpi}\pi\cdot{\mathcal H}_{\pi\theta\xi}+**,\;
\partial_{\unbtheta}^2\partial_{\unbpi}{\mathcal H}_{\pi\xi}
=\partial_{\unbtheta}^2\partial_{\unbpi}\theta\cdot{\mathcal H}_{\theta\pi\xi}
+\partial_{\unbtheta}^2\partial_{\unbpi}\pi\cdot{\mathcal H}_{\pi\pi\xi}+**.
\end{aligned}
%\label{08}
\end{equation*}
Therefore, we get
$$
\bmid\mbox{body part of the right hand side above}\bmid
\le\tilde{C}_4^{(0)}|t-{\unbt}|^{3/2} \et
{\bmid}\pi_{\mathrm B}\partial_{\unbtheta}^2\partial_{\unbpi}^2 x{\bmid}
\le C_4^{(0)}|t-{\unbt}|^{5/2}.
$$

By the same talk, we get
$$
{\bmid}\pi_{\mathrm B}\partial_{\unbtheta}^2\partial_{\unbpi}^2 \xi{\bmid}
\le C_4^{(0)}|t-{\unbt}|^{5/2}.
$$

Proceeding as before, we get
$$
{\bmid}\pi_{\mathrm B}\partial_{\unbx}^\alpha\partial_{\unbxi}^\beta
\partial_{\unbtheta}^2\partial_{\unbpi}^2 x{\bmid},\quad
{\bmid}\pi_{\mathrm B}\partial_{\unbx}^\alpha\partial_{\unbxi}^\beta
\partial_{\unbtheta}^2\partial_{\unbpi}^2 \xi{\bmid}
\le C_4^{(k)}|t-{\unbt}|^3 \for k=|\alpha+\beta|\ge1.
$$
Therefore, Theorem \ref{existenceCM} has been proved.  $\qed$

%%%

%%\input s-Weyl3-1-41

\subsubsection{Inverse map: Proof of Theorem \ref{inverseCM}}

Supersmoothness of \eqref{91} is already proved.

{\bf Existence of the inverse map.} 
For notational simplicity, we put
$$
\begin{gathered}
X_i(\unbx,\unbtheta)=x_i(t,{\unbt}\,;{\unbx,\unbxi,\unbtheta,\unbpi}),\quad 
\Theta_\ell(\unbx,\unbtheta)
=\theta_\ell(t,{\unbt}\,;{\unbx,\unbxi,\unbtheta,\unbpi}),\\
Y_j(\barx,\bartheta)=y_j(t,{\unbt}\,;{\mixdata}),\quad
\Omega_{m}(\barx,\bartheta)
=\omega_{m}(t,{\unbt}\,;{\mixdata}),
\end{gathered}
$$
and we consider $(t,{\unbt}\,;\unbxi,\unbpi)$ as parameters 
which will not be represented explicitly.

For any fixed $(t,{\unbt}\,;\unbxi,\unbpi)$ and any given $(\barx,\bartheta)$,
we want to find $(\unbx,\unbtheta)$ 
such that
$$
\barx_i=X_i(\unbx,\unbtheta),\quad
\bartheta_\ell=\Theta_\ell(\unbx,\unbtheta).
$$
Denoting this $(\unbx,\unbtheta)$ as $(Y(\unbx,\unbtheta), 
\Omega(\unbx,\unbtheta))$, then we should have
\begin{equation}
\unbx_i=Y_i(X(\unbx,\unbtheta),\Theta(\unbx,\unbtheta)),\quad
\unbtheta_{\ell}=\Omega_{\ell}(X(\unbx,\unbtheta),\Theta(\unbx,\unbtheta)).
\label{E1}
\end{equation}

By the supersmoothness proved in (1) of Theorem \ref{existenceCM}, we have
$$
\begin{aligned}
X_i(\unbx,\unbtheta)&=X_{i,0}(\unbx)+X_{i,1}(\unbx)\unbtheta_1
+X_{i,2}(\unbx)\unbtheta_2+X_{i,3}(\unbx)\unbtheta_1\unbtheta_2
=\sum_{|a|\le2}\partial_{\unbtheta}^aX_i(\unbx,0)\unbtheta^a,\\
\Theta_k(\unbx,\unbtheta)&=\Theta_{k,0}(\unbx)+\Theta_{k,1}(\unbx)\unbtheta_1
+\Theta_{k,2}(\unbx)\unbtheta_2+\Theta_{k,3}(\unbx)\unbtheta_1\unbtheta_2
=\sum_{|a|\le2}\partial_{\unbtheta}^a\Theta_i(\unbx,0)\unbtheta^a.
\end{aligned}
$$
We assume (and prove by construction) that we may put also
$$
\begin{aligned}
Y_j(\barx,\bartheta)
&=Y_{j,0}(\barx)+Y_{j,1}(\barx)\bartheta_1
+Y_{j,2}(\barx)\bartheta_2+Y_{j,3}(\barx)\bartheta_1\bartheta_2
=\sum_{|b|\le2}\partial_{\bartheta}^b Y_j(\barx,0)\bartheta^b,\\
\Omega_{\ell}(\barx,\bartheta)
&=\Omega_{\ell,0}(\barx)+\Omega_{\ell,1}(\barx)\bartheta_1
+\Omega_{\ell,2}(\barx)\bartheta_2+\Omega_{\ell,3}(\barx)\bartheta_1\bartheta_2
=\sum_{|b|\le2}\partial_{\bartheta}^b \Omega_j(\barx,0)\bartheta^b.
\end{aligned}
$$

Denoting
$$
\begin{gathered}
{\tilde X}=({\tilde X}_1,{\tilde X}_2,{\tilde X}_3),\quad
{\tilde X}_{\mathrm B}
=({\tilde X}_{1,\mathrm B},{\tilde X}_{2,\mathrm B},{\tilde X}_{3,\mathrm B})\\
\with {\tilde X}_i={\tilde X}_{i,\mathrm B}+{\tilde X}_{i,\mathrm S}
=X_{i,0,\mathrm B}(\unbx_{\mathrm B})+X_{i,0,\mathrm S}(\unbx)
=X_i(\unbx,0)\in \rev,\\
{\tilde\Theta}=({\tilde\Theta}_1,{\tilde\Theta}_2) \with
{\tilde\Theta}_\ell=
\Theta_{\ell,0}(\unbx)=\Theta_\ell(\unbx,0)\in\rod,
\end{gathered}
$$
we have
$$
\begin{aligned}
Y_j({\tilde X},{\tilde\Theta})
&=Y_{j,0}({\tilde X})
+Y_{j,1}({\tilde X}){\tilde\Theta}_1
+Y_{j,2}({\tilde X}){\tilde\Theta}_2
+Y_{j,3}({\tilde X}){\tilde\Theta}_1{\tilde\Theta}_2,\\
&\quad\with
Y_{j,0}({\tilde X}),\;Y_{j,3}({\tilde X})\in\rev,\quad
Y_{j,1}({\tilde X}),\;Y_{j,2}({\tilde X})\in\rod,\\
\Omega_{\ell}({\tilde X},{\tilde\Theta})
&=\Omega_{\ell,0}({\tilde X})
+\Omega_{\ell,1}({\tilde X}){\tilde\Theta}_1
+\Omega_{\ell,2}({\tilde X}){\tilde\Theta}_2
+\Omega_{\ell,3}({\tilde X})
{\tilde\Theta}_1{\tilde\Theta}_2,\\
&\quad\with
\Omega_{\ell,0}({\tilde X}),\;
\Omega_{\ell,3}({\tilde X})\in\rod,\quad
\Omega_{\ell,1}({\tilde X}),\;
\Omega_{\ell,2}({\tilde X})\in\rev.
\end{aligned}
$$

{\bf Claim I}: From the first equation of \eqref{E1}, 
we construct $Y_{i,*}({\tilde X}_{\mathrm B})$ for each degree.
\newline
(I-0) Restricting $(\unbx,\unbtheta)$ to $(\unbx,0)$ in \eqref{E1},
we have
\begin{equation*}
\unbx_{i}=Y_{i,0}({\tilde X})+
Y_{i,1}({\tilde X}){\tilde\Theta}_1
+Y_{i,2}({\tilde X}){\tilde\Theta}_2
+Y_{i,3}({\tilde X})
{\tilde\Theta}_1{\tilde\Theta}_2.
\end{equation*}
Or more precisely, putting $Y_{*}^{[-1]}({\tilde X})=0$, we have
\begin{equation}
\begin{aligned}
Y_{i,0}^{[2p]}({\tilde X})=&\unbx_{i}^{[2p]}
-\sum_{q=0}^{p-1}Y_{i,1}^{[2p-2q-1]}({\tilde X}){\tilde\Theta}_1^{[2q+1]}
-\sum_{q=0}^{p-1}Y_{i,2}^{[2p-2q-1]}({\tilde X}){\tilde\Theta}_2^{[2q+1]}\\
&\qquad\quad
-\sum_{q=0}^{p}Y_{i,3}^{[2p-2q]}({\tilde X})
\sum_{r=0}^{q-1}{\tilde\Theta}_1^{[2r+1]}{\tilde\Theta}_2^{[2q-2r-1]}
\for p=0,1,2,\cdots.
\label{E2}
\end{aligned}
\end{equation}
(I-1) Differentiating \eqref{E1} w.r.t. ${\unbtheta_1}$ or ${\unbtheta_2}$
and restricting as above,
we have
for each $i=1,2,3$,
\begin{equation*}
\begin{pmatrix}
\Theta_{1,1}(\unbx)&\Theta_{2,1}(\unbx)\\
\Theta_{1,2}(\unbx)&\Theta_{2,2}(\unbx)
\end{pmatrix}
\begin{pmatrix}
Y_{i,1}({\tilde X})\\
Y_{i,2}({\tilde X})
\end{pmatrix}
=
\begin{pmatrix}
-X_{j,1}(\unbx)\partial_{\barx_j}Y_{i}({\tilde X},{\tilde\Theta})\\
-X_{j,2}(\unbx)\partial_{\barx_j}Y_{i}({\tilde X},{\tilde\Theta})
\end{pmatrix}.
\label{E2-1}
\end{equation*}
Or, we have
\begin{equation}
\sum_{q=0}^{p-1}
\begin{pmatrix}
\Theta_{1,1}^{[2q]}(\unbx)&\Theta_{2,1}^{[2q]}(\unbx)\\
\Theta_{1,2}^{[2q]}(\unbx)&\Theta_{2,2}^{[2q]}(\unbx)
\end{pmatrix}
\begin{pmatrix}
Y_{i,1}^{[2p-2q-1]}({\tilde X})\\
Y_{i,2}^{[2p-2q-1]}({\tilde X})
\end{pmatrix}
=\sum_{q=0}^{p-1}
\begin{pmatrix}
-X_{j,1}^{[2p-2q-1]}(\unbx)
\partial_{\barx_j}Y_{i}^{[2q]}({\tilde X},{\tilde\Theta})\\
-X_{j,2}^{[2p-2q-1]}(\unbx)
\partial_{\barx_j}Y_{i}^{[2q]}({\tilde X},{\tilde\Theta})
\end{pmatrix}.
\label{E3}
\end{equation}
(I-2) Differentiating \eqref{E1} w.r.t. ${\unbtheta_1}$ and ${\unbtheta_2}$, 
we have
\begin{equation*}
\begin{aligned}
-[\Theta_{1,1}(\unbx)\Theta_{2,2}(\unbx)
-\Theta_{2,1}(\unbx)\Theta_{1,2}(\unbx)]Y_{i,3}({\tilde X})
=&X_{j,3}(\unbx)\partial_{\barx_j}Y_{i}({\tilde X},{\tilde\Theta})
-X_{j,1}(\unbx)X_{k,2}(\unbx)
\partial^2_{\barx_k\barx_j}Y_{i}({\tilde X},{\tilde\Theta})\\
&\quad+[-X_{j,1}(\unbx)\Theta_{1,2}(\unbx)
+\Theta_{1,1}(\unbx)X_{j,2}(\unbx)]
\partial_{\barx_j}Y_{i,1}({\tilde X})\\
&\quad+[-X_{j,1}(\unbx)\Theta_{2,2}(\unbx)
+\Theta_{2,1}(\unbx)X_{j,2}(\unbx)]
\partial_{\barx_j}Y_{i,2}({\tilde X}).
\end{aligned}
\end{equation*}
Therefore, we have
\begin{equation}
\begin{aligned}
-\sum_{q=0}^p &\sum_{r=0}^q[\Theta_{1,1}^{[2q-2r]}(\unbx)\Theta_{2,2}^{[2r]}(\unbx)
-\Theta_{2,1}^{[2q-2r]}(\unbx)\Theta_{1,2}^{[2r]}(\unbx)]Y_{i,3}^{[2p-2q]}({\tilde X})\\
=&\sum_{q=0}^pX_{j,3}^{[2q]}(\unbx)\partial_{\barx_j}Y_{i}^{[2p-2q]}({\tilde X},{\tilde\Theta})
-\sum_{q=0}^{p}\sum_{r=0}^{q-1}X_{j,1}^{[2q-2r-1]}(\unbx)X_{k,2}^{[2r+1]}(\unbx)
\partial^2_{\barx_k\barx_j}Y_{i}^{[2p-2q]}({\tilde X},{\tilde\Theta})\\
&\quad+
\sum_{q=0}^{p-1}\sum_{r=0}^q
[-X_{j,1}^{[2q-2r+1]}(\unbx)\Theta_{1,2}^{[2r]}(\unbx)
+\Theta_{1,1}^{[2r]}(\unbx)X_{j,2}^{[2q-2r+1]}(\unbx)]
\partial_{\barx_j}Y_{i,1}^{[2p-2q-1]}({\tilde X})\\
&\quad+
\sum_{q=0}^{p-1}\sum_{r=0}^q
[-X_{j,1}^{[2q-2r+1]}(\unbx)\Theta_{2,2}^{[2r]}(\unbx)
+\Theta_{2,1}^{[2r]}(\unbx)X_{j,2}^{[2q-2r+1]}(\unbx)]
\partial_{\barx_j}Y_{i,2}^{[2p-2q-1]}({\tilde X}).
\end{aligned}
\label{E4}
\end{equation}

(1) For any fixed $\barx_{\mathrm B}$, we consider the map
$$
\unbx_{\mathrm B}\to F(\unbx_{\mathrm B})
=\barx_{\mathrm B}+\unbx_{\mathrm B}-X_{\mathrm B}(\unbx_{\mathrm B}),
$$
which satisfies
$$
F(\unbx_{\mathrm B})- F(\unbx'_{\mathrm B})=
(\unbx_{\mathrm B}-\unbx'_{\mathrm B})\int_0^1d\tau
\bigg(I-\frac{\partial X_{\mathrm B}}{\partial \unbx}
(\tau\unbx_{\mathrm B}+(1-\tau)\unbx'_{\mathrm B})\bigg).
$$
As $\partial_{\unbx_j}X_{i,0}(\unbx)-\delta_{ij}=0$,
$F(\unbx_{\mathrm B})$ is the contraction map, 
therefore, there exists a unique $\unbx_{\mathrm B}$ such that
$$
\barx_{\mathrm B}=X_{\mathrm B}(\unbx_{\mathrm B}).
$$
We denote this as $\unbx_{\mathrm B}=Y_{\mathrm B}(\barx_{\mathrm B})$,
that is, $Y^{[0]}({\tilde X}_{\mathrm B})$, and therefore, 
$Y^{[0]}({\tilde X})$ is defined, which satisfies \eqref{E2} with $p=0$.

(2) From \eqref{E3} with $p=1$, we have, for each $i=1,2,3,$
\begin{equation}
\begin{pmatrix}
\Theta_{1,1}^{[0]}(\unbx)&\Theta_{2,1}^{[0]}(\unbx)\\
\Theta_{1,2}^{[0]}(\unbx)&\Theta_{2,2}^{[0]}(\unbx)
\end{pmatrix}
\begin{pmatrix}
Y_{i,1}^{[1]}({\tilde X})\\
Y_{i,2}^{[1]}({\tilde X})
\end{pmatrix}
=
\begin{pmatrix}
-X_{j,1}^{[1]}(\unbx)\partial_{\barx_j}Y_{i}^{[0]}({\tilde X},{\tilde\Theta})\\
-X_{j,2}^{[1]}(\unbx)\partial_{\barx_j}Y_{i}^{[0]}({\tilde X},{\tilde\Theta})
\end{pmatrix}.
\label{E3-0}
\end{equation}
As $\partial_{\barx_j}Y_{i}^{[0]}({\tilde X},{\tilde\Theta})=
\partial_{\barx_j}Y_{i,0}^{[0]}({\tilde X}_{\mathrm B})$, 
the right-hand side above is given by the step (1) above. 
On the other hand, 
when $|t-\unbt|<\delta$, $t,\unbt\in[-T,T]$, for any $(\unbxi,\unbpi)$, we have
$$
\det
\begin{pmatrix}
\Theta_{1,1}^{[0]}(\unbx_{\mathrm B})&\Theta_{2,1}^{[0]}(\unbx_{\mathrm B})\\
\Theta_{1,2}^{[0]}(\unbx_{\mathrm B})&\Theta_{2,2}^{[0]}(\unbx_{\mathrm B})
\end{pmatrix}
\neq0\quad\mbox{because}\quad
\Theta_{k,j}(\unbx)=\partial_{\unbtheta_j}\theta_k(t,{\unbt}\,;\unbx,\unbxi,0,0).
$$
Solving \eqref{E3-0}, we get the degree 1 part of 
$Y_{i,*}^{[1]}({\tilde X}_{\mathrm B})$, that is, 
$Y_{i,*}^{[1]}({\tilde X})$ for $i=1,2,3,\; *=0,1,2,3.$

(3) Putting $p=0$ in \eqref{E4}, we have
$$
-[\Theta_{1,1}^{[0]}(\unbx)\Theta_{2,2}^{[0]}(\unbx)
-\Theta_{2,1}^{[0]}(\unbx)\Theta_{1,2}^{[0]}(\unbx)]Y_{i,3}^{[0]}({\tilde X})
=X_{j,3}^{[0]}(\unbx)\partial_{\barx_j}Y_{i,0}^{[0]}({\tilde X}).
$$
Therefore, we get $Y_{i,3}^{[0]}({\tilde X})$.

Returning back to (1) with $p=1$ in \eqref{E2} 
and then (2) with $p=2$ in \eqref{E3} and lastly (3) with $p=2$ in \eqref{E4}.
This process determines $Y_{i,0}^{[2]}({\tilde X}),\;Y_{i,1}^{[3]}({\tilde X}),\;
Y_{i,2}^{[3]}({\tilde X})$ and $Y_{i,3}^{[2]}({\tilde X})$.
Proceeding recursively, we determine $Y_{i,*}({\tilde X})$.

%%%

%%\input s-Weyl3-1-42
{\bf Claim II}: Analogously as above, we determine $\Omega_{\ell,*}({\tilde X})$ 
as follows.
\newline
(II-0) Restricting $(\unbx,\unbtheta)$ to $(\unbx,0)$ in 
the second equation of \eqref{E1}, we have
\begin{equation*}
0=\Omega_{\ell,0}({\tilde X})+\Omega_{\ell,1}({\tilde X}){\tilde\Theta}_1
+\Omega_{\ell,2}({\tilde X}){\tilde\Theta}_2
+\Omega_{\ell,3}({\tilde X}){\tilde\Theta}_1{\tilde\Theta}_2.
\end{equation*}
In other word, we have
\begin{equation}
\begin{aligned}
\Omega_{\ell,0}^{[2p+1]}({\tilde X})=&
-\sum_{q=0}^{p}\Omega_{\ell,1}^{[2q]}({\tilde X}){\tilde\Theta}_1^{[2p-2q+1]}
-\sum_{q=0}^{p}\Omega_{\ell,2}^{[2q]}({\tilde X}){\tilde\Theta}_2^{[2p-2q+1]}\\
&\qquad
-\sum_{q=0}^{p}\Omega_{\ell,3}^{[2p-2q+1]}({\tilde X})
\sum_{r=0}^{q}{\tilde\Theta}_1^{[2q-2r-1]}{\tilde\Theta}_2^{[2r+1]}.
\end{aligned}
\label{E5}
\end{equation}
\newline
(II-1) Differentiating \eqref{E1} w.r.t. ${\unbtheta_1}$,
we have,
\begin{equation*}
\begin{pmatrix}
\Theta_{1,1}(\unbx)&\Theta_{2,1}(\unbx)\\
\Theta_{1,2}(\unbx)&\Theta_{2,2}(\unbx)
\end{pmatrix}
\begin{pmatrix}
\Omega_{1,1}({\tilde X})\\
\Omega_{1,2}({\tilde X})
\end{pmatrix}
=
\begin{pmatrix}
1-X_{j,1}(\unbx)\partial_{\barx_j}\Omega_1({\tilde X})\\
-X_{j,2}(\unbx)\partial_{\barx_j}\Omega_1({\tilde X})
\end{pmatrix},
\end{equation*}
that is,
\begin{equation}
\sum_{q=0}^{p}\begin{pmatrix}
\Theta_{1,1}^{[2q]}(\unbx)&\Theta_{2,1}^{[2q]}(\unbx)\\
\Theta_{1,2}^{[2q]}(\unbx)&\Theta_{2,2}^{[2q]}(\unbx)
\end{pmatrix}
\begin{pmatrix}
\Omega_{1,1}^{[2p-2q]}({\tilde X})\\
\Omega_{1,2}^{[2p-2q]}({\tilde X})
\end{pmatrix}
=
\begin{pmatrix}
1^{[2p]}-\sum_{q=0}^{p}X_{j,1}^{[2q-1]}(\unbx)
\partial_{\barx_j}\Omega_1^{[2p-2q+1]}({\tilde X})\\
-\sum_{q=0}^{p}X_{j,2}^{[2q-1]}(\unbx)
\partial_{\barx_j}\Omega_1^{[2p-2q+1]}({\tilde X})
\end{pmatrix}.
\label{E6}
\end{equation}
Analogously, differentiating \eqref{E1} w.r.t. ${\unbtheta_2}$,
\begin{equation}
\sum_{q=0}^{p}\begin{pmatrix}
\Theta_{1,1}^{[2q]}(\unbx)&\Theta_{2,1}^{[2q]}(\unbx)\\
\Theta_{1,2}^{[2q]}(\unbx)&\Theta_{2,2}^{[2q]}(\unbx)
\end{pmatrix}
\begin{pmatrix}
\Omega_{2,1}^{[2p-2q]}({\tilde X})\\
\Omega_{2,2}^{[2p-2q]}({\tilde X})
\end{pmatrix}
=
\begin{pmatrix}
-\sum_{q=0}^{p}X_{j,1}^{[2q-1]}(\unbx)
\partial_{\barx_j}\Omega_2^{[2p-2q+1]}({\tilde X})\\
1^{[2p]}-\sum_{q=0}^{p}X_{j,2}^{[2q-1]}(\unbx)
\partial_{\barx_j}\Omega_2^{[2p-2q+1]}({\tilde X})
\end{pmatrix}.
\label{E7}
\end{equation}
(II-2) Differentiating \eqref{E1} w.r.t. ${\unbtheta_1}$ and ${\unbtheta_2}$,
we have, for $\ell=1,2$,
\begin{equation*}
\begin{aligned}
0=&X_{i,3}(\unbx)\partial_{\barx_i}\Omega_\ell({\tilde X},{\tilde\Theta})
-X_{i,1}(\unbx)X_{j,2}(\unbx)
\partial^2_{\barx_j\barx_i}\Omega_\ell({\tilde X},{\tilde\Theta})\\
&+\Theta_{1,3}(\unbx)\Omega_{\ell,1}({\tilde X})
+\Theta_{2,3}(\unbx)\Omega_{\ell,2}({\tilde X})
+[-X_{i,1}(\unbx)\Theta_{1,2}(\unbx)
+\Theta_{1,1}(\unbx)X_{i,2}(\unbx)]
\Omega_{\ell,1}({\tilde X})\\
&\quad
+[-X_{i,1}(\unbx)\Theta_{2,2}(\unbx)
+\Theta_{2,1}(\unbx)X_{i,2}(\unbx)]
\Omega_{\ell,2}({\tilde X})
+[\Theta_{1,1}(\unbx)\Theta_{2,2}(\unbx)
-\Theta_{2,1}(\unbx)\Theta_{1,2}(\unbx)]
\Omega_{\ell,3}({\tilde X}).
\end{aligned}
\end{equation*}
Rewriting above, we have, for each $p=0,1,2,\cdots$,
\begin{equation}
\begin{aligned}
-\sum_{q=0}^{p}&
\sum_{r=0}^{q}[\Theta_{1,1}^{[2q-2r]}(\unbx)\Theta_{2,2}^{[2r]}(\unbx)
-\Theta_{2,1}^{[2q-2r]}(\unbx)\Theta_{1,2}^{[2r]}(\unbx)]
\Omega_{\ell,3}^{[2p-2q+1]}({\tilde X})\\
&=\sum_{q=0}^{p}X_{i,3}^{[2q]}(\unbx)\partial_{\barx_i}
\Omega_\ell^{[2p-2q+1]}({\tilde X},{\tilde\Theta})\\
&\quad-\sum_{q=0}^{p}
\sum_{r=0}^{q}X_{i,1}^{[2r+1]}(\unbx)X_{j,2}^{[2q-2r-1]}(\unbx)
\partial^2_{\barx_j\barx_i}\Omega_\ell^{[2p-2q+1]}({\tilde X},{\tilde\Theta})\\
&\qquad+\sum_{q=0}^{p}\Theta_{1,3}^{[2q]}(\unbx)\Omega_{\ell,1}^{[2p-2q+1]}({\tilde X})
+\sum_{q=0}^{p}\Theta_{2,3}^{[2q]}(\unbx)\Omega_{\ell,2}^{[2p-2q+1]}({\tilde X})\\
&\quad\qquad+\sum_{q=0}^{p}
\sum_{r=0}^{q}[-X_{i,1}^{[2r+1]}(\unbx)\Theta_{1,2}^{[2q-2r]}(\unbx)
+\Theta_{1,1}^{[2q-2r]}(\unbx)X_{i,2}^{[2r+1]}(\unbx)]
\Omega_{\ell,1}^{[2p-2q+1]}({\tilde X})\\
&\qquad\qquad+\sum_{q=0}^{p}
\sum_{r=0}^{q}[-X_{i,1}^{[2r+1]}(\unbx)\Theta_{2,2}^{[2q-2r]}(\unbx)
+\Theta_{2,1}^{[2q-2r]}(\unbx)X_{i,2}^{[2r+1]}(\unbx)]
\Omega_{\ell,2}^{[2p-2q+1]}({\tilde X}).
\end{aligned}
\label{E8}
\end{equation}

By the same procedure which we employed to determine $Y_{i,*}({\tilde X})$, 
we may define
$\Omega_{\ell,*}({\tilde X})$.

Therefore, there exist a constant $\delta>0$ and functions 
$y(t,{\unbt}\,;{\mixdata})$, 
$\omega(t,{\unbt}\,;{\mixdata})$ 
such that for $|t-{\unbt}|<\delta$
\begin{equation}
\begin{aligned}
&\unbx=y(t,{\unbt}\,;x(t,{\unbt}\,;{\inidata}),\unbxi,
\theta(t,{\unbt}\,;{\inidata}),\unbpi),\\
&\unbtheta=\omega(t,{\unbt}\,;x(t,{\unbt}\,;{\inidata}),\unbxi,
\theta(t,{\unbt}\,;{\inidata}),\unbpi),
\end{aligned}
\end{equation}
and therefore
\begin{equation}
\begin{aligned}
&\barx=x(t,{\unbt}\,;y(t,{\unbt}\,;{\mixdata}),\unbxi,
\omega(t,{\unbt}\,;{\mixdata}),\unbpi),\\
&\bartheta=\theta(t,{\unbt}\,;y(t,{\unbt}\,;{\mixdata}),\unbxi,
\omega(t,{\unbt}\,;{\mixdata}),\unbpi).
\end{aligned}
\label{CMI-121}
\end{equation}

%%%

%%\input s-Weyl3-1-43

{\bf Estimates of the inverse mapping}:
When $|a+b|=0$, differentiating once the first equation of \eqref{CMI-121} w.r.t. $\barx$
or $\unbxi$, we get
$$
\begin{aligned}
&\delta_{jk}=\frac{\partial \barx_j}{\partial \barx_k}
=\frac{\partial y_\ell}{\partial \barx_k}
\frac{\partial x_j}{\partial \unbx_\ell}
+\frac{\partial \omega_m}{\partial \barx_k}
\frac{\partial x_j}{\partial \unbtheta_m},\\
&0=\frac{\partial \barx_j}{\partial \unbxi_k}
=\frac{\partial y_\ell}{\partial \unbxi_k}
\frac{\partial x_j}{\partial \unbx_\ell}
+\frac{\partial x_j}{\partial \unbxi_k}
+\frac{\partial \omega_m}{\partial \unbxi_k}
\frac{\partial x_j}{\partial \unbtheta_m}.
\end{aligned}
$$
Taking the body part and remarking \eqref{rt3.24}, we get
$$
\bmid\pi_{\mathrm B}\partial_{\barx}
(y(t,{\unbt}\,;{\mixdata})-\barx)\bmid
=0   %\le C|t-{\unbt}|
\quad\mbox{if}\quad |t-{\unbt}|\le\delta.
%\where \delta=(2C_0^{(1)})^{-1}.
$$
Here, we used the fact 
\begin{equation}
\pi_{\mathrm B}\frac{\partial x_j}{\partial \unbx_\ell}=
\pi_{\mathrm B}\bigg(\frac{\partial (x_j-\unbx_j)}{\partial \unbx_\ell}+
\frac{\partial \unbx_j}{\partial \unbx_\ell}\bigg)
=\delta_{j\ell} %\ge 1-C_0^{(1)}|t-{\unbt}|\ge \frac12 
\quad\mbox{if}\quad |t-{\unbt}|\le\delta,
\end{equation}
which follows from \eqref{rt3.24}.
Analogously, we have   %, with some constant ${\tilde C}_0^{(1)}>0$,
$$
\bmid\pi_{\mathrm B}\partial_{\unbxi}
(y(t,{\unbt}\,;{\mixdata})-\barx)\bmid
=0 %\le {\tilde C}_0^{(1)}|t-{\unbt}|
\quad\mbox{if}\quad |t-{\unbt}|\le\delta.
$$
By the same procedure, we have %, with some constant ${\tilde C}_0^{(k)}>0$,
$$
\bmid\pi_{\mathrm B}\partial^\alpha_{\barx}\partial^\beta_{\unbxi}
(y(t,{\unbt}\,;{\mixdata})-\barx)\bmid
=0 %\le {\tilde C}_0^{(k)}|t-{\unbt}|
\quad\mbox{if}\quad 
k=|\alpha+\beta|\ge2 \et |t-{\unbt}|\le\delta.
$$

The case when $|a+b|=1$, we have
\begin{equation}
\begin{aligned}
&\delta_{\ell m}=\frac{\partial \bartheta_\ell}{\partial \bartheta_m}
=\frac{\partial y_j}{\partial \bartheta_m}
\frac{\partial \theta_\ell}{\partial \unbx_j}
+\frac{\partial \omega_k}{\partial \bartheta_m}
\frac{\partial \theta_\ell}{\partial \unbtheta_k},\\
&0=\frac{\partial \bartheta_\ell}{\partial \unbpi_m}
=\frac{\partial y_j}{\partial \unbpi_m}
\frac{\partial \theta_\ell}{\partial \unbx_j}
+\frac{\partial \omega_k}{\partial \unbpi_m}
\frac{\partial \theta_\ell}{\partial \unbtheta_k}
+\frac{\partial \theta_\ell}{\partial \unbpi_m}.
\end{aligned}
\label{III}
\end{equation}
Taking the body part and using \eqref{rt3.25}, we get
$$
I=(\delta_{\ell m})=XY,\quad
X=\bigg(\pi_{\mathrm B}\frac{\partial \omega_k}{\partial \bartheta_m}\bigg),\;
Y=\bigg(\pi_{\mathrm B}\frac{\partial \theta_\ell}{\partial \unbtheta_k}\bigg).
$$
Since
\begin{equation}
\bigg|\pi_{\mathrm B}
\frac{\partial(\theta_\ell-\unbtheta_\ell)}{\partial \unbtheta_k}\bigg|
\le C_1^{(1)}|t-\unbt|^{1/2},\quad
\pi_{\mathrm B}
\frac{\partial\unbtheta_\ell}{\partial \unbtheta_k}
=\delta_{k\ell},
\end{equation}
if we take $|t-\unbt|\le \delta\le (4C_1^{(1)})^{-1}$, then
we have, as operators
$$
{\bmid}X{\bmid}={\bmid}Y^{-1}{\bmid}\le 2 
\quad\mbox{if}\quad |t-\unbt|\le \delta.
$$
Therefore, using $(X-I)Y=I-Y$, we have, as operators
$$
{\bmid}X-I{\bmid}\le {\bmid}I-Y{\bmid}\,{\bmid}Y^{-1}{\bmid}
\le 2C|t-\unbt|^{1/2},
$$
that is,
$$
{\bmid}\pi_{\mathrm B}\partial_{\bartheta}
(\omega_k(t,{\unbt}\,;{\mixdata})-\bartheta_k){\bmid}
\le {\tilde C}_1^{(0)}|t-{\unbt}|^{1/2}.
$$
From the second equality of \eqref{III} combined with \eqref{rt3.25},
$$
{\bmid}\pi_{\mathrm B}\partial_{\unbpi}
(\omega_k(t,{\unbt}\,;{\mixdata})-\bartheta_k){\bmid}
\le {\tilde C}_1^{(0)}|t-{\unbt}|^{1/2}.
$$
Analogously, we get, when $|a+b|=1$,
$$
\bmid\pi_{\mathrm B}\partial_{\barx}^\alpha\partial_{\unbxi}^\beta
\partial^a_{\bartheta}\partial^b_{\unbpi}
(\omega_k(t,{\unbt}\,;{\mixdata})-\bartheta_k)\bmid
\le {\tilde C}_1^{(k)}|t-{\unbt}|^{(1/2)(1-(1-k)_+)}
\quad\mbox{if}\quad k=|\alpha+\beta|,\quad |t-{\unbt}|\le\delta.
$$

Proceeding analogously as we did in proving Theorem \ref{existenceCM}, 
we have the desired results for $|a+b|\ge2$, which are
abbreviated here.
The second inequality in \eqref{3-57} is given in the next subsection.
$\qed$

Analogously, we have
\begin{prop}
(i) For any fixed $(t,{\unbt}\,;\unbx,\unbtheta)$, $|t-{\unbt}|<\delta$, 
the mapping
\begin{equation}
{\fR}^{3|2}\ni(\unbxi,\unbpi)\mapsto(\xi=\xi(t,{\unbt}\,;{\inidata}),
\pi=\pi(t,{\unbt}\,;{\inidata}))\in{\fR}^{3|2}
\end{equation}
is {\ssm}. 
The inverse mapping defined by
\begin{equation}
{\fR}^{3|2}\ni(\barxi,\barpi)\mapsto
(\eta=\eta(t,{\unbt}\,;{\unbx,\barxi,\unbtheta,\barpi}),
\rho=\rho(t,{\unbt}\,;{\unbx,\barxi,\unbtheta,\barpi}))\in{\fR}^{3|2} ,
\end{equation}
is {\ssm} in
$({\unbx,\barxi,\unbtheta,\barpi})$ 
for fixed $(t,{\unbt})$. 
\par
(ii) Let $|a+b|=0$. We have
\begin{equation}
\left\{
\begin{aligned}
&{\bmid}\pi_{\mathrm B}
\partial_{\unbx}^\alpha\partial_{\barxi}^\beta
(\eta(t,{\unbt}\,;{\unbx,\barxi,\unbtheta,\barpi})-\barxi){\bmid}
=0,\\
&
|\eta^{[0]}(t,{\unbt}\,;\unbx^{[0]},\barxi^{[0]})-\barxi^{[0]}| 
\leq C_2|t-{\unbt}|(1+|\unbx^{[0]}|+|\barxi^{[0]}|).
\end{aligned}\right. %\label{3-57}
\end{equation}
\par
(iii) Let $|a+b|=1$. For $k=|\alpha+\beta|$, there exists a constant
${\tilde C}^{(k)}_1$ such that
\begin{equation}
{\bmid}\pi_{\mathrm B}
\partial_{\unbx}^\alpha\partial_{\barxi}^\beta
\partial_{\unbtheta}^a\partial_{\barpi}^b
(\rho(t,{\unbt}\,;{\unbx,\barxi,\unbtheta,\barpi})-\barpi){\bmid}
\leq {\tilde C}^{(k)}_1|t-{\unbt}|^{(1/2)(1-(1-k)_+)}.
\end{equation}
\par
(iv) Let $|a+b|=2$. For $k=|\alpha+\beta|$, there exists a constant
${\tilde C}^{(k)}_2$ such that
\begin{equation}
{\bmid}\pi_{\mathrm B}
\partial_{\unbx}^\alpha\partial_{\barxi}^\beta
\partial_{\unbtheta}^a\partial_{\barpi}^b
(\eta(t,{\unbt}\,;{\unbx,\barxi,\unbtheta,\barpi})-\barxi){\bmid}
\leq {\tilde C}^{(k)}_2|t-{\unbt}|^{1+(1/2)(1-(1-k)_+)}.
\end{equation}
\par
(v) Let $|a+b|=3$. For $k=|\alpha+\beta|$, there exists a constant
${\tilde C}^{(k)}_3$ such that
\begin{equation}
{\bmid}\pi_{\mathrm B}
\partial_{\unbx}^\alpha\partial_{\barxi}^\beta
\partial_{\unbtheta}^a\partial_{\barpi}^b
(\rho(t,{\unbt}\,;{\unbx,\barxi,\unbtheta,\barpi})-\barpi){\bmid}
\leq {\tilde C}^{(k)}_3|t-{\unbt}|^{(3/2)+(1/2)(1-(1-k)_+)}.
\end{equation}
\par
(vi) Let $|a+b|=4$. For $k=|\alpha+\beta|$, there exists a constant
${\tilde C}^{(k)}_4$ such that
\begin{equation}
{\bmid}\pi_{\mathrm B}
\partial_{\unbx}^\alpha\partial_{\barxi}^\beta
\partial_{\unbtheta}^a\partial_{\barpi}^b
(\eta(t,{\unbt}\,;{\unbx,\barxi,\unbtheta,\barpi})-\barxi){\bmid}
\leq {\tilde C}^{(k)}_4|t-{\unbt}|^{(5/2)+(1/2)(1-(1-k)_+)}.
\end{equation}
\end{prop}

%%%

%%\input s-Weyl3-1-5

\subsubsection{Time reversing}
As the Hamilton equation \ref{rt2.6-1} and \ref{rt2.6-2} may be solved
backward in time, we denote, for $\unbt\le t\le\bart$, that
$x({t},{\bart}\,;\barx,\bartheta,\barxi,\barpi)$,
$\xi({t},{\bart}\,;\barx,\bartheta,\barxi,\barpi)$,
$\theta({t},{\bart}\,;\barx,\bartheta,\barxi,\barpi)$,
$\pi({t},{\bart}\,;\barx,\bartheta,\barxi,\barpi)$
are solutions at time $t$ of \ref{rt2.6-1} and \ref{rt2.6-2} 
with the initial time $t=\bart$ 
and the initial data $(\barx,\bartheta,\barxi,\barpi)$.

Proceeding as in previous sections, we have the following:
\begin{equation}
\left\{
\begin{aligned}
&\unbx=x({\unbt},{\bart}\,;\barx,\bartheta,\barxi,\barpi),\;
\barx=x({\bart},{\bart}\,;\barx,\bartheta,\barxi,\barpi),\\
&\unbtheta=\theta({\unbt},{\bart}\,;\barx,\bartheta,\barxi,\barpi),\;
\bartheta=\theta({\bart},{\bart}\,;\barx,\bartheta,\barxi,\barpi),\\
&\unbxi=\xi({\unbt},{\bart}\,;\barx,\bartheta,\barxi,\barpi),\;
\barxi=\xi({\bart},{\bart}\,;\barx,\bartheta,\barxi,\barpi),\\
&\unbpi=\pi({\unbt},{\bart}\,;\barx,\bartheta,\barxi,\barpi),\;
\barpi=\pi({\bart},{\bart}\,;\barx,\bartheta,\barxi,\barpi)
\end{aligned}\right.
\end{equation}
For the inverse mappings, we have, if $|{\bart}-{\unbt}|<\delta$,
\begin{equation}
\left\{
\begin{aligned}
&\barx=y({\unbt},{\bart}\,;\unbx,\barxi,\unbtheta,\barpi),\;
\unbx=y({\bart},{\bart}\,;\unbx,\barxi,\unbtheta,\barpi),\\
&\bartheta=\omega({\unbt},{\bart}\,;\unbx,\barxi,\unbtheta,\barpi),\;
\unbtheta=\omega({\bart},{\bart}\,;\unbx,\barxi,\unbtheta,\barpi),\\
&\barxi=\eta({\unbt},{\bart}\,;\unbx,\barxi,\unbtheta,\barpi),\;
\unbxi=\eta({\bart},{\bart}\,;\unbx,\barxi,\unbtheta,\barpi),\\
&\barpi=\rho({\unbt},{\bart}\,;\unbx,\barxi,\unbtheta,\barpi),\;
\unbpi=\rho({\bart},{\bart}\,;\unbx,\barxi,\unbtheta,\barpi).
\end{aligned}\right.
\end{equation}

Therefore, we get for $|{\bart}-{\unbt}|<\delta$,
\begin{equation}
\left\{
\begin{aligned}
&y({\unbt},{\bart}\,;\unbx,\barxi,\unbtheta,\barpi)
=x({\bart},{\unbt}\,;\unbx,\eta({\bart},{\unbt}\,;\unbx,\barxi,\unbtheta,\barpi),
\unbtheta,\rho({\bart},{\unbt}\,;\unbx,\barxi,\unbtheta,\barpi)),\\
&\omega({\unbt},{\bart}\,;\unbx,\barxi,\unbtheta,\barpi)
=\theta({\bart},{\unbt}\,;\unbx,\eta({\bart},{\unbt}\,;\unbx,\barxi,\unbtheta,\barpi),
\unbtheta,\rho({\bart},{\unbt}\,;\unbx,\barxi,\unbtheta,\barpi)),\\
&\eta({\unbt},{\bart}\,;\barx,\unbxi,\bartheta,\unbpi)
=\xi({\bart},{\unbt}\,;y({\bart},{\unbt}\,;\barx,\unbxi,\bartheta,\unbpi),\unbxi,
\omega({\bart},{\unbt}\,;\barx,\unbxi,\bartheta,\unbpi),\unbpi),\\
&\rho({\unbt},{\bart}\,;\barx,\unbxi,\bartheta,\unbpi)
=\pi({\bart},{\unbt}\,;y({\bart},{\unbt}\,;\barx,\unbxi,\bartheta,\unbpi),\unbxi,
\omega({\bart},{\unbt}\,;\barx,\unbxi,\bartheta,\unbpi),\unbpi).
\end{aligned}\right.
\label{3-65}
\end{equation}

{\it Proof of Theorem \ref{inverseCM} continued.}
The second inequality in \eqref{3-57} is proved using
$$
\begin{aligned}
y({\bart},{\unbt}\,;\barx,\unbxi,0,0)-\barx
&=\int_0^1 dr\, \unbxi\cdot\partial_{\unbxi}
y({\bart},{\unbt}\,;\barx,r\unbxi,0,0)\\
&\qquad+\int_0^1 dr\, \barx\cdot\partial_{\barx}
(y({\bart},{\unbt}\,;r\barx,0,0,0)-\barx)+y({\bart},{\unbt}\,;0,0,0,0).
\end{aligned}
$$
In fact, from above and the first inequality in \eqref{3-57}, we have
$$
|y^{[0]}({\bart},{\unbt}\,;\barx^{[0]},\unbxi^{[0]})-\barx^{[0]}|
\le C|{\bart}-{\unbt}|(|\barx^{[0]}|+|\unbxi^{[0]}|)
+|y^{[0]}({\bart},{\unbt}\,;0,0)|.
$$
Replacing $({\bart},{\unbt})$ in the first equality of \eqref{3-65} 
by $({\unbt},{\bart})$ and 
using \eqref{rt3.26}, we have
$$
|y^{[0]}({\bart},{\unbt}\,;0,0)|
=|x({\unbt},{\bart}\,;0,\eta^{[0]}({\unbt},{\bart}\,;0,0),0,0)|
\le C|{\bart}-{\unbt}|,
$$
since $\eta^{[0]}({\unbt},{\bart}\,;0,0)$ is continuous in ${\unbt}$, ${\bart}$.
Combining these, we get the desired inequality. $\qed$

%%%

%%\input s-Weyl3-2

\subsection{Action integral.}
We prepare the following lemma in a slightly general situation:
\begin{lem}
Let $(x(\unbx,\unbtheta),\theta(\unbx,\unbtheta),
u(\unbx,\unbtheta),\xi(\unbx,\unbtheta),\pi(\unbx,\unbtheta))$ 
be supersmooth functions of
$(\unbx,\unbtheta)\in {\fR}^{m|n}$ satisfying
$$
\left\{
\begin{aligned}
&\frac{\partial u}{\partial{\unbx}_j}
=\sum_{a=1}^m\frac{\partial x_a}{\partial{\unbx}_j}\xi_a(\unbx,\unbtheta)
+\sum_{b=1}^n\frac{\partial \theta_b}{\partial{\unbx}_j}\pi_b(\unbx,\unbtheta)
\for j=1,2,\cdots,m,\\
&\frac{\partial u}{\partial{\unbtheta}_\ell}
=\sum_{a=1}^m\frac{\partial x_a}{\partial{\unbtheta}_\ell}\xi_a(\unbx,\unbtheta)
+\sum_{b=1}^n\frac{\partial \theta_b}{\partial{\unbtheta}_\ell}\pi_b(\unbx,\unbtheta)
\for \ell=1,2,\cdots,n.
\end{aligned}\right.
$$
Assuming that
$$
\sdet
\begin{pmatrix}
\frac{\partial x_a}{\partial{\unbx}_j}
&\frac{\partial x_a}{\partial{\unbtheta}_\ell}\\[3pt]
\frac{\partial \theta_b}{\partial{\unbx}_j}
&\frac{\partial \theta_b}{\partial{\unbtheta}_\ell}
\end{pmatrix}(\unbx_{\mathrm B},0)\neq 0
\et \barx=x(\unbx,\unbtheta),\quad\bartheta=\theta(\unbx,\unbtheta),
$$
we have:
\newline
(i) There exist inverse functions $y(\barx,\bartheta)$, $\omega(\barx,\bartheta)$ such that
$$
x(y(\barx,\bartheta),\omega(\barx,\bartheta))=\barx,\quad
\theta(y(\barx,\bartheta),\omega(\barx,\bartheta))=\bartheta.
$$
\newline
(ii) Moreover, putting $w(\barx,\bartheta)=u(y(\barx,\bartheta),\omega(\barx,\bartheta))$, 
we have, for $j=1,2,\cdots, m$ and $\ell=1,2,\cdots,n$,
\begin{equation}
\frac{\partial w}{\partial {\barx}_j}=\xi_j(y(\barx,\bartheta),\omega(\barx,\bartheta)),\quad
\frac{\partial w}{\partial {\bartheta}_\ell}=\pi_\ell(y(\barx,\bartheta),\omega(\barx,\bartheta)).
\label{G1}
\end{equation}
\end{lem}
\par{\it Proof. }
By the inverse function theorem, we get (i).
From this, we have
$$
\begin{aligned}
&\delta_{a k}=\frac{\partial{\barx}_a}{\partial{\barx}_k}
=\frac{\partial y_j}{\partial{\barx}_k}
\frac{\partial x_a}{\partial{\unbx}_j}
+\frac{\partial \omega_\ell}{\partial{\barx}_k}
\frac{\partial x_a}{\partial{\unbtheta}_\ell},
\quad
0=\frac{\partial{\barx}_a}{\partial{\bartheta}_m}
=\frac{\partial y_j}{\partial{\bartheta}_m}
\frac{\partial x_a}{\partial{\unbx}_j}
+\frac{\partial \omega_\ell}{\partial{\bartheta}_m}
\frac{\partial x_a}{\partial{\unbtheta}_\ell},\\
&0=\frac{\partial{\bartheta}_b}{\partial{\barx}_k}
=\frac{\partial y_j}{\partial{\barx}_k}
\frac{\partial {\bartheta}_b}{\partial{\unbx}_j}
+\frac{\partial \omega_\ell}{\partial{\barx}_k}
\frac{\partial {\bartheta}_b}{\partial{\unbtheta}_\ell},
\quad
\delta_{b m}=\frac{\partial{\bartheta}_b}{\partial{\bartheta}_m}
=\frac{\partial y_j}{\partial{\bartheta}_m}
\frac{\partial{\bartheta}_b}{\partial{\unbx}_j}
+\frac{\partial \omega_\ell}{\partial{\bartheta}_m}
\frac{\partial{\bartheta}_b}{\partial{\unbtheta}_\ell}.
\end{aligned}
$$
Using these, we get readily that
$$
\begin{aligned}
\frac{\partial w}{\partial {\barx}_k}
&=\frac{\partial y_j}{\partial {\barx}_k}
\frac{\partial u}{\partial {\unbx}_j}
+\frac{\partial \omega_\ell}{\partial {\barx}_k}
\frac{\partial u}{\partial {\unbtheta}_\ell}\\
&=\frac{\partial y_j}{\partial {\barx}_k}
\Big(\frac{\partial x_a}{\partial{\unbx}_j}\xi_a(\unbx,\unbtheta)
+\frac{\partial \theta_b}{\partial{\unbx}_j}\pi_b(\unbx,\unbtheta)\Big)
+\frac{\partial \omega_\ell}{\partial {\barx}_k}
\Big(\frac{\partial x_a}{\partial{\unbtheta}_\ell}\xi_a(\unbx,\unbtheta)
+\frac{\partial \theta_b}{\partial{\unbtheta}_\ell}\pi_b(\unbx,\unbtheta)\Big)
\Bigg|{\begin{Sb}\unbx=y(\barx,\bartheta),\\
\unbtheta=\omega(\barx,\bartheta)\end{Sb}}\\
&=\Big(\frac{\partial y_j}{\partial {\barx}_k}\frac{\partial x_a}{\partial{\unbx}_j}
+\frac{\partial \omega_\ell}{\partial {\barx}_k}\frac{\partial x_a}{\partial{\unbtheta}_\ell}
\Big)\xi_a(\unbx,\unbtheta)
+\Big(\frac{\partial y_j}{\partial{\barx}_k}\frac{\partial \theta_b}{\partial{\unbx}_j}
+\frac{\partial \omega_\ell}{\partial {\barx}_k}\frac{\partial \theta_b}{\partial{\unbtheta}_\ell}
\Big)\pi_b(\unbx,\unbtheta)\Bigg|{\begin{Sb}\unbx=y(\barx,\bartheta),\\
\unbtheta=\omega(\barx,\bartheta)\end{Sb}}\\
&=\xi_k(y(\barx,\bartheta),\omega(\barx,\bartheta)).
\end{aligned}
$$
Analogously, we get the second equality in (ii).  $\qed$

\par{\it Proof of Theorem \ref{HJ-estimates}.}
For fixed $(\unbxi,\unbpi)$, we put
$$
\tilde{\mathcal S}({\bart},{\unbt}\,;\inidata)
=\langle \unbx|\unbxi\rangle
+\langle\unbtheta|\unbpi\rangle
+{\mathcal S}_0({\bart},{\unbt}\,;\inidata).
$$
Then, we have, using integration by parts w.r.t. $s$ in \eqref{rt2.13},
$$
\begin{aligned}
\frac{\partial\tilde{\mathcal S}}{\partial\unbx_j}
&=\unbxi_j+\int_{\unbt}^{\bart} ds\Big[
\frac{\partial \dot x_k}{\partial\unbx_j}\xi_k
+\dot x_k\frac{\partial \xi_k}{\partial\unbx_j}
+\frac{\partial \dot \theta_{m}}{\partial\unbx_j}\pi_{m}
+\dot \theta_m\frac{\partial \pi_{m}}{\partial\unbx_j}\\
&\qquad\qquad\qquad
-\Big(\frac{\partial x_k}{\partial\unbx_j}\frac{\partial {\mathcal H}}{\partial x_k}
+\frac{\partial \xi_k}{\partial\unbx_j}\frac{\partial {\mathcal H}}{\partial \xi_k}
+\frac{\partial\theta_{m}}{\partial\unbx_j}\frac{\partial {\mathcal H}}{\partial \theta_{m}}
+\frac{\partial\pi_{m}}{\partial\unbx_j}\frac{\partial {\mathcal H}}{\partial\pi_{m}}\Big)
\Big]\\
&=\unbxi_j+\frac{\partial x_k}{\partial\unbx_j}\xi_k\Big|_{\unbt}^{t}
+\frac{\partial \theta_{m}}{\partial\unbx_j}\pi_{m}\Big|_{\unbt}^{t}
=\frac{\partial x_k}{\partial\unbx_j}\xi_k
+\frac{\partial\theta_{m}}{\partial\unbx_j}\pi_{m}.
\end{aligned}
$$
Analogously, we get
$$
\frac{\partial\tilde{\mathcal S}}{\partial\unbtheta_\ell}
=\frac{\partial x_k}{\partial\unbtheta_\ell}\xi_k
+\frac{\partial\theta_{m}}{\partial\unbtheta_\ell}\pi_{m}.
$$
As we have already proved that if $|{\bart}-{\unbt}|\le\delta$, we have
$$
\pi_{\mathrm B}\sdet
\begin{pmatrix}
\frac{\partial x({\bart})}{\partial \unbx}
&\frac{\partial x({\bart})}{\partial \unbtheta}\\[4pt]
\frac{\partial \theta({\bart})}{\partial \unbx}&\frac{\partial \theta({\bart})}{\partial \unbtheta}
\end{pmatrix}\neq 0, %(\unbx_{\mathrm B},\unbxi_{\mathrm B},0,0)
$$
we may apply the above lemma.
Therefore, putting 
$$
{\mathcal S}({\bart},{\unbt}\,;\mixdata)
=\tilde{\mathcal S}({\bart},{\unbt}\,;y({\bart},{\unbt}\,;\mixdata),
\unbxi,\omega({\bart},{\unbt}\,;\mixdata),\unbpi),
$$
we have, by \eqref{G1}, 
\begin{equation}
\begin{aligned}
\frac{\partial{\mathcal S}}{\partial\barx_j}
&=\xi_j({\bart},{\unbt}\,;y({\bart},{\unbt}\,;\mixdata),\unbxi,\omega({\bart},{\unbt}\,;\mixdata),\unbpi),\\
\frac{\partial{\mathcal S}}{\partial\bartheta_\ell}
&=\pi_\ell({\bart},{\unbt}\,;y({\bart},{\unbt}\,;\mixdata),\unbxi,\omega({\bart},{\unbt}\,;\mixdata),\unbpi),
\end{aligned}
\label{G2}\end{equation}
and
$$
\frac{\partial{\mathcal S}}{\partial{\bart}}
=\frac{\partial\tilde{\mathcal S}}{\partial{\bart}}
+\frac{\partial y_j}{\partial{\bart}}
\frac{\partial\tilde{\mathcal S}}{\partial \unbx_j}
+\frac{\partial \omega_\ell}{\partial{\bart}}
\frac{\partial\tilde{\mathcal S}}{\partial \unbtheta_\ell}.
$$
On the other hand, 
$$
\frac{\partial}{\partial{\bart}}\,
\tilde{\mathcal S}({\bart},{\unbt}\,;\inidata)
=\langle\dot x({\bart})|\xi({\bart})\rangle
+\langle\dot\theta({\bart})|\pi({\bart})\rangle
-{\mathcal H}({\bart},x({\bart}),\theta({\bart}),\xi({\bart}),\pi({\bart})).
$$
Combining these with simple calculations, 
we get the desired Hamilton-Jacobi equation. 

Moreover, using \eqref{G2} and \eqref{3-65}, we have
\begin{equation}
\begin{matrix}
&{\partial_{\barx_j}}{\mathcal S}({\bart},{\unbt}\,;\mixdata)
=\eta_j({\unbt},{\bart}\,;\mixdata),\;
&{\partial_{\bartheta_\ell}}{\mathcal S}({\bart},{\unbt}\,;\mixdata)
=\rho_\ell({\unbt},{\bart}\,;\mixdata),\\
&{\partial_{\unbxi_k}}{\mathcal S}({\bart},{\unbt}\,;\mixdata)=y_k({\bart},{\unbt}\,;\mixdata),\;
&{\partial_{\unbpi_m}}{\mathcal S}({\bart},{\unbt}\,;\mixdata)=\omega_m({\bart},{\unbt}\,;\mixdata).
\end{matrix}   \qed
\label{3-66}\end{equation}

From here on, we change the notation:
$$
({\bart},{\unbt}\,;\mixdata)\to
(t,s\,;x,\xi,\theta,\pi).
$$

Then, putting 
$$
\tilde{\mathcal H}_*={\mathcal H}_*(t,x,{\mathcal S}_x(t,s\,;x,\xi,\theta,\pi),\theta,
{\mathcal S}_\theta(t,s\,;x,\xi,\theta,\pi))\big|_{\theta=\pi=0},
$$ 
from the Hamilton-Jacobi equation 
and ${\mathcal S}(s,s\,;x,\xi,\theta,\pi)=\langle x|\xi\rangle
+\langle \theta|\pi\rangle$, 
we get 
\begin{lem} Using the decomposition \eqref{rt3.70}, we get
\begin{gather}
{\mathcal S}_{\bar0\bar0,t}+\tilde{\mathcal H}=0\with 
{\mathcal S}_{\bar0\bar0}(s,s\,;x,\xi)=\langle x|\xi\rangle,\label{3-71}\\
{\mathcal S}_{\bar1\bar0,t}+\tilde{\mathcal H}_{\pi_2\pi_1}{\mathcal S}_{\bar1\bar0}^2
+(\tilde{\mathcal H}_{\pi_1\theta_1}+\tilde{\mathcal H}_{\pi_2\theta_2}){\mathcal S}_{\bar1\bar0}
+\tilde{\mathcal H}_{\theta_2\theta_1}=0\with
{\mathcal S}_{\bar1\bar0}(s,s\,;x,\xi)=0,\label{3-72}\\
{\mathcal S}_{c_1d_1,t}+
(\tilde{\mathcal H}_{\pi_1\theta_1}
+{\mathcal S}_{\bar1\bar0}\tilde{\mathcal H}_{\pi_2\pi_1}){\mathcal S}_{c_1d_1}=0
\with {\mathcal S}_{c_1d_1}(s,s\,;x,\xi)=1,\\
{\mathcal S}_{c_2d_2,t}+
(\tilde{\mathcal H}_{\pi_2\theta_2}
+{\mathcal S}_{\bar1\bar0}\tilde{\mathcal H}_{\pi_2\pi_1}){\mathcal S}_{c_2d_2}=0
\with {\mathcal S}_{c_2d_2}(s,s\,;x,\xi)=1,\\
{\mathcal S}_{c_1d_2,t}+
(\tilde{\mathcal H}_{\pi_1\theta_1}
+{\mathcal S}_{\bar1\bar0}\tilde{\mathcal H}_{\pi_2\pi_1}){\mathcal S}_{c_1d_2}=0
\with {\mathcal S}_{c_1d_2}(s,s\,;x,\xi)=0,\\
{\mathcal S}_{c_2d_1,t}+
(\tilde{\mathcal H}_{\pi_2\theta_2}
+{\mathcal S}_{\bar1\bar0}\tilde{\mathcal H}_{\pi_2\pi_1}){\mathcal S}_{c_2d_1}=0
\with {\mathcal S}_{c_2d_1}(s,s\,;x,\xi)=0.
\end{gather}
\end{lem}
\par{\it Proof. }
\eqref{3-71} is obtained by restricting (H-J) to $\theta=\pi=0$.
Integrating (H-J) w.r.t. $d\theta_1 d\theta_2$, we get \eqref{3-72}. 
In fact, as we have
$$
\begin{gathered}
\partial_{\theta_1}{\mathcal H} %=\frac{\partial}{\partial\theta_1}{\mathcal H}
=\frac{\partial{\mathcal S}_{x_j}}{\partial\theta_1}{\mathcal H}_{\xi_j}
+{\mathcal H}_{\theta_1}
+\frac{\partial{\mathcal S}_{\theta_2}}{\partial\theta_1}{\mathcal H}_{\pi_2},\\
\partial_{\theta_2}\partial_{\theta_1}{\mathcal H}
%=\frac{\partial^2}{\partial\theta_2\partial\theta_1}{\mathcal H}
=\frac{\partial^2{\mathcal S}_{x_j}}{\partial\theta_2\partial\theta_1}{\mathcal H}_{\xi_j}
-\frac{\partial{\mathcal S}_{x_j}}{\partial\theta_1}
\partial_{\theta_2}{\mathcal H}_{\xi_j}
+\partial_{\theta_2}{\mathcal H}_{\theta_1}
+\frac{\partial{\mathcal S}_{\theta_2}}{\partial\theta_1}
\partial_{\theta_2}{\mathcal H}_{\pi_2},
\end{gathered}
$$
where
$$
\begin{gathered}
\partial_{\theta_2}{\mathcal H}_{\xi_j}
={\mathcal H}_{\theta_2\xi_j}+\frac{\partial{\mathcal S}_{\theta_1}}{\partial\theta_2}
{\mathcal H}_{\pi_1\xi_j},\\
\partial_{\theta_2}{\mathcal H}_{\theta_1}
=\frac{\partial{\mathcal S}_{x_j}}{\partial\theta_2}{\mathcal H}_{\xi_j\theta_1}
+{\mathcal H}_{\theta_2\theta_1}
+\frac{\partial{\mathcal S}_{\theta_1}}{\partial\theta_2}{\mathcal H}_{\pi_1\theta_1},\\
\partial_{\theta_2}{\mathcal H}_{\pi_2}
=\frac{\partial{\mathcal S}_{x_j}}{\partial\theta_2}{\mathcal H}_{\xi_j\pi_2}
+{\mathcal H}_{\theta_2\pi_2}
+\frac{\partial{\mathcal S}_{\theta_1}}{\partial\theta_2}{\mathcal H}_{\pi_1\pi_2},
\end{gathered}
$$
remarking $\tilde{\mathcal H}_{\xi_j}=0$, ${\mathcal H}_{\xi_j\xi_k}=0$,
and restricting $\partial_{\theta_2}\partial_{\theta_1}{\mathcal H}$  
to $\theta=\pi=0$, we get the desired equality.
Other equalities are obtained in the same manner.  $\qed$

Since $\tilde{\mathcal H}={\elec}{A}_0(t,x)$, we get readily
\begin{equation}
{\mathcal S}_{\bar 0 \bar 0}(t,s\,;x,\xi)
=\langle x|\xi\rangle-{\elec}\int_s^t dr{A}_0(r,x).\label{S1}
\end{equation}

Putting 
\begin{equation}
w_0(t,s\,;x,\xi)
=\tilde{\mathcal H}_{\pi_1\theta_1}+{\mathcal S}_{\bar1\bar0}\tilde{\mathcal H}_{\pi_2\pi_1}
=\tilde{\mathcal H}_{\pi_2\theta_2}+{\mathcal S}_{\bar1\bar0}\tilde{\mathcal H}_{\pi_2\pi_1},
\label{w0}
\end{equation}
with
$$
\begin{aligned}
&\tilde{\mathcal H}_{\pi_1\theta_1}=-i\hbar^{-1}(c\xi_3-{\elec}A_3
-c{\elec}\int_s^t dr\partial_{x_3}{A}_0)
=\tilde{\mathcal H}_{\pi_2\theta_2}\\
&\tilde{\mathcal H}_{\pi_2\pi_1}=\hbar^{-2}
[c\xi_1-{\elec}A_1-c{\elec}\int_s^t dr\partial_{x_1}{A}_0-i(c\xi_2-{\elec}A_2
-c{\elec}\int_s^t dr\partial_{x_2}{A}_0)],
\end{aligned}
$$
we get
\begin{lem} For $|t-s|\le\delta$,
\begin{gather}
{\mathcal S}_{c_1d_1}(t,s\,;x,\xi)=e^{-\int_s^t dr\,w_0(r,s\,;x,\xi)}
={\mathcal S}_{c_2d_2}(t,s\,;x,\xi),\label{3-78}\\
{\mathcal S}_{c_1d_2}(t,s\,;x,\xi)={\mathcal S}_{c_2d_1}(t,s\,;x,\xi)=0.
\label{3-79}
\end{gather}
\end{lem}

Remarking \eqref{3-79}, we have
\begin{lem}
\begin{gather}
{\mathcal S}_{\bar0\bar1,t}+{\mathcal S}_{c_1d_1}{\mathcal S}_{c_2d_2}\tilde{\mathcal H}_{\pi_2\pi_1}=0
\with {\mathcal S}_{\bar0\bar1}(s,s\,;x,\xi)=0,\label{3-80}\\
{\mathcal S}_{\bar1\bar1,t}+2w_0{\mathcal S}_{\bar1\bar1}+w_1=0
\with {\mathcal S}_{\bar1\bar1}(s,s\,;x,\xi)=0,\label{3-81}
\end{gather}
where we put
$$
\begin{aligned}
w_1(t,s\,;x,\xi)
&=({\mathcal S}_{\bar1\bar0}{\mathcal S}_{\bar0\bar1,x_j}
-{\mathcal S}_{c_1d_1}{\mathcal S}_{c_2d_2,x_j})\tilde{\mathcal H}_{\xi_j\pi_1\theta_1}
+({\mathcal S}_{\bar1\bar0}{\mathcal S}_{\bar0\bar1,x_j}
-{\mathcal S}_{c_1d_1,x_j}{\mathcal S}_{c_2d_2})\tilde{\mathcal H}_{\xi_j\pi_2\theta_2}\\
&\qquad
+[({\mathcal S}_{\bar1\bar0}^2+{\mathcal S}_{c_1d_1}{\mathcal S}_{c_2d_2}){\mathcal S}_{\bar0\bar1,x_j}
-{\mathcal S}_{\bar1\bar0}({\mathcal S}_{c_1d_1}{\mathcal S}_{c_2d_2})_{x_j}]
\tilde{\mathcal H}_{\xi_j\pi_2\pi_1}
+{\mathcal S}_{\bar0\bar1,x_j}\tilde{\mathcal H}_{\xi_j\theta_2\theta_1}\\
&=c\hbar^{-2}\big[{\mathcal S}_{\bar0\bar1,x_1}-i{\mathcal S}_{\bar0\bar1,x_2}
+2{\mathcal S}_{\bar1\bar0}\int_s^td\tau(w_{0,x_1}-iw_{0,x_2})\big]
e^{-2\int_s^tdr\, w_0(r,s\,;x,\xi)}\\
&\qquad
+c\hbar^{-2}{\mathcal S}_{\bar1\bar0}^2
({\mathcal S}_{\bar0\bar1,x_1}-i{\mathcal S}_{\bar0\bar1,x_2})
-2ic\hbar^{-1}\big({\mathcal S}_{\bar1\bar0}{\mathcal S}_{\bar0\bar1,x_3}
+\int_s^tdr\, w_{0,x_3}e^{-2\int_s^rd\tau w_0(\tau,s\,;x,\xi)}\big)\\
&\qquad\qquad
+c({\mathcal S}_{\bar0\bar1,x_1}+i{\mathcal S}_{\bar0\bar1,x_2}).
\end{aligned}
$$
\end{lem}
\par{\it Proof. }
To get \eqref{3-80}, we used \eqref{3-79}.
Remarking 
$$
\tilde{\mathcal H}_{\pi_2\pi_1\xi_1}=c\hbar^{-2},\;\tilde{\mathcal H}_{\pi_2\pi_1\xi_2}=-ic\hbar^{-2},\;
\tilde{\mathcal H}_{\theta_2\theta_1\xi_1}=c,\;\tilde{\mathcal H}_{\theta_2\theta_1\xi_2}=ic,\;
\tilde{\mathcal H}_{\pi_1\theta_1\xi_3}=\tilde{\mathcal H}_{\pi_2\theta_2\xi_3}=-ic\hbar^{-1},
$$ 
and \eqref{3-78}, we have \eqref{3-81}.  $\qed$

Therefore, we get the representation
\begin{equation}
\begin{aligned}
&{\mathcal S}_{\bar0\bar1}(t,s\,;x,\xi)=-\int_s^t dr \,
\tilde{\mathcal H}_{\pi_2\pi_1}(r,s\,;x,\xi)
e^{-2\int_s^r d\tau w_0(\tau,s\,;x,\xi)},\\
&{\mathcal S}_{\bar1\bar1}(t,s\,;x,\xi)
=-\int_s^t dr \, w_1(r,s\,;x,\xi)e^{-\int_s^rd\tau w_0(\tau,s\,;x,\xi)}.
\end{aligned}\end{equation}

Combining estimates in Theorems \ref{existenceCM}, \ref{inverseCM} and \eqref{3-66}, 
we get
\begin{lem} For any $\alpha$, $\beta$, there exist constants $C_{\alpha\beta}>0$ such that
\begin{equation}
\begin{gathered}
|\partial_x^\alpha\partial_\xi^\beta({\mathcal S}_{\bar0\bar0}(t,s\,;x,\xi)-\langle x|\xi\rangle)|
\le C_{\alpha\beta}(1+|x|)^{(1-|\alpha|)_+}\delta_{0|\beta|}|t-s|\\
|\partial_x^\alpha\partial_\xi^\beta{\mathcal S}_{\bar1\bar0}(t,s\,;x,\xi)|
\le C_{\alpha\beta}|t-s|,\\ 
|\partial_x^\alpha\partial_\xi^\beta({\mathcal S}_{c_jd_j}(t,s\,;x,\xi)-1)|
\le C_{\alpha\beta}|t-s| \for j=1,2,\\
|\partial_x^\alpha\partial_\xi^\beta{\mathcal S}_{\bar0\bar1}(t,s\,;x,\xi)|
\le C_{\alpha\beta}|t-s|,\\ 
|\partial_x^\alpha\partial_\xi^\beta{\mathcal S}_{\bar1\bar1}(t,s\,;x,\xi)|
\le C_{\alpha\beta}|t-s|.
\end{gathered}
\label{rt3.84}\end{equation}
\end{lem}
\par{\it Proof. } 
As ${\mathcal S}_{\bar1\bar0}(t,s\,;x,\xi)={\mathcal S}_{\theta_2\theta_1}(t,s\,;x,\xi,0,0)
=\partial_{\theta_2}\rho_1(-t,s\,;x,\xi,0,0)$,
we have the desired one from Proposition 3.4 and \eqref{3-66}.
Other terms are calculated similarly.  $\qed$

%%%

%%\input s-Weyl3-3

\subsection{Continuity equation.} 

Defining ${\mathcal D}$ as in \eqref{rt2.15}, 
we get Theorem \ref{continuity} as in \cite{In98-1}.

Using the notation introduced in Theorem \ref{HJ-estimates}, we may decompose
\begin{equation}
{\mathcal D}(t,s\,;x,\xi,\theta,\pi)
=\sum_{|c|+|d|=even \ge 0}
{\mathcal D}_{cd}(t,s\,;x,\xi)\theta^{c} \pi^{d}
={\mathcal D}_{\mathrm B}(t,s\,;x,\xi)+{\mathcal D}_{\mathrm S}(t,s\,;x,\xi,\theta,\pi),
\end{equation}
where 
\begin{equation}
{\begin{aligned}
{\mathcal D}_{\mathrm B}&={\mathcal D}_{\bar 0 \bar 0}={\mathcal D}^{[0]},\\
{\mathcal D}_{\mathrm S}(t,s\,;x,\xi,\theta,\pi)
&={\mathcal D}_{\bar 1 \bar 0 }\theta_1\theta_2
+\sum_{|c|+|d|=2}
{\mathcal D}_{cd}(t,s\,;x,\xi)\theta^{c} \pi^{d}
+{\mathcal D}_{\bar 0 \bar 1}\pi_1\pi_2
+{\mathcal D}_{\bar 1 \bar 1}\theta_1\theta_2\pi_1\pi_2.
\end{aligned}}
\end{equation}

From the continuity equation \eqref{rt2.16}, we have
$$
\pdt{\mathcal D}_{\bar0\bar0}
+{\mathcal D}_{\bar0\bar0}\,
\widetilde{\partial_{\theta_k}{\mathcal H}_{\pi_k}}=0
%(t,s\,;x,{\mathcal S}_x,\theta,{\mathcal S}_\theta)
\with 
{\mathcal D}_{\bar0\bar0}(s,s\,;x,\xi)=1.
$$
As
$$
\begin{gathered}
\partial_{\theta_1}{\mathcal H}_{\pi_1}={\mathcal H}_{\theta_1\pi_1}
+\frac{\partial{\mathcal S}_{\theta_2}}{\partial \theta_1}{\mathcal H}_{\pi_2\pi_1}
+\frac{\partial{\mathcal S}_{x_j}}{\partial \theta_1}{\mathcal H}_{\xi_j\pi_1},\\
\partial_{\theta_2}{\mathcal H}_{\pi_2}={\mathcal H}_{\theta_2\pi_2}
+\frac{\partial{\mathcal S}_{\theta_1}}{\partial \theta_2}{\mathcal H}_{\pi_1\pi_2}
+\frac{\partial{\mathcal S}_{x_j}}{\partial \theta_2}{\mathcal H}_{\xi_j\pi_2},
\end{gathered}
$$
we have
$$
\widetilde{\partial_{\theta_k}{\mathcal H}_{\pi_k}}
=\partial_{\theta_k}{\mathcal H}_{\pi_k}\big|_{\theta=\pi=0}
=-\tilde{\mathcal H}_{\pi_1\theta_1}-\tilde{\mathcal H}_{\pi_2\theta_2}
-2{\mathcal S}_{\bar1\bar0}\tilde{\mathcal H}_{\pi_2\pi_1}=-2w_0(t,s\,;x,\xi),
$$
and we get
\begin{equation}
{\mathcal D}_{\bar0\bar0}(t,s\,;x,\xi)=e^{2\int_s^t dr\,w_0(r,s\,;x,\xi)}.
\end{equation}

Instead of ${\mathcal D}$, we should study the properties of a function
${\mathcal A}={\mathcal D}^{1/2}$:
Putting
\begin{equation}
\begin{aligned}
{\mathcal A}(t,s\,;x,\theta,\xi,\pi)
&=\sum_{|c|+|d|=even \ge 0}
{\mathcal A}_{cd}(t,s\,;x,\xi)\theta^{c} \pi^{d}\\
&={\mathcal A}_{\bar 0 \bar 0}
+{\mathcal A}_{\bar 1 \bar 0 }\theta_1\theta_2
+{\mathcal A}_{c_1d_1}\theta_1\pi_1
+{\mathcal A}_{c_2d_2}\theta_2\pi_2\\
&\qquad
+{\mathcal A}_{c_1d_2}\theta_1\pi_2
+{\mathcal A}_{c_2d_1}\theta_2\pi_1
+{\mathcal A}_{\bar 0 \bar 1}\pi_1\pi_2
+{\mathcal A}_{\bar 1 \bar 1}\theta_1\theta_2\pi_1\pi_2,
\end{aligned}
\end{equation}
we define each coefficient ${\mathcal A}_{cd}(t,s\,;x,\xi)$ 
from ${\mathcal A}^2={\mathcal D}$ as
\begin{equation}
\begin{aligned}
&{\mathcal D}_{\bar 0 \bar 0}={\mathcal A}_{\bar 0 \bar 0}^2
\with {\mathcal A}_{\bar 0 \bar 0}(s,s\,;x,\xi)=1,\\
&{\mathcal D}_{\bar 1 \bar 0 }=2{\mathcal A}_{\bar 0 \bar 0}{\mathcal A}_{\bar 1\bar 0},
\quad
{\mathcal D}_{\bar 0 \bar 1 }=2{\mathcal A}_{\bar 0 \bar 0}{\mathcal A}_{\bar 0 \bar 1 },\\
&{\mathcal D}_{c_1d_1}=2{\mathcal A}_{\bar 0 \bar 0}{\mathcal A}_{c_1d_1},\quad
{\mathcal D}_{c_2d_2}=2{\mathcal A}_{\bar 0 \bar 0}{\mathcal A}_{c_2d_2},\\
&{\mathcal D}_{c_1d_2}=2{\mathcal A}_{\bar 0 \bar 0}{\mathcal A}_{c_1d_2},\quad
{\mathcal D}_{c_2d_1}=2{\mathcal A}_{\bar 0 \bar 0}{\mathcal A}_{c_2d_1},\\
&{\mathcal D}_{\bar 1 \bar 1}=2{\mathcal A}_{\bar 0 \bar 0}{\mathcal A}_{\bar 1 \bar 1}
+2{\mathcal A}_{\bar 1 \bar 0 }{\mathcal A}_{\bar 0\bar 1 }
-2{\mathcal A}_{c_1d_1}{\mathcal A}_{c_2d_2}
+2{\mathcal A}_{c_1d_2}{\mathcal A}_{c_2d_1}.
\end{aligned}
\end{equation}
More precisely, we define 
${\mathcal A}_{\bar 0 \bar 0}(t,s\,;q,p)=\sqrt{{\mathcal D}_{\bar 0 \bar 0}(t,s\,;q,p)}$ 
such that ${\mathcal A}_{\bar 0 \bar 0}(s,s\,;q,p)=1$
and ${\mathcal A}_{cd}(t,s\,;q,p)$ are defined from above, 
and then they are Grassmann continued to ${\fR}^{3|0}$.

Using the continuity equation \eqref{rt2.16}, we have the \eqref{A00}.

Remarking also
$$
\partial_{x_j}{\mathcal H}_{\xi_j}
=\frac{\partial{\mathcal S}_{\theta_1}}{\partial x_j}{\mathcal H}_{\pi_1\xi_j}
+\frac{\partial{\mathcal S}_{\theta_2}}{\partial x_j}{\mathcal H}_{\pi_2\xi_j},
$$
we have $\tilde{\mathcal H}_{\xi_j}=0$ and
$\widetilde{\partial_{x_j}{\mathcal H}_{\xi_j}}=0$,
from which we have
\begin{equation}
{\mathcal A}_{\bar0\bar0,t}-w_0{\mathcal A}_{\bar0\bar0}=0\with
{\mathcal A}_{\bar0\bar0}(s,s\,;x,\xi)=1.
\end{equation}
That is, as is desired, we have
\begin{equation}
{\mathcal A}_{\bar0\bar0}(t,s\,;x,\xi)=e^{\int_s^t dr\,w_0(r,s\,;x,\xi)}.
\end{equation}

On the other hand, for $\{\cdots\}$ in \eqref{A00}, as
$$
\begin{aligned}
\partial_{\theta_1}\{\cdots\}
&={\mathcal A}_{\theta_1x_j}{\mathcal H}_{\xi_j}
+{\mathcal A}_{x_j}\partial_{\theta_1}{\mathcal H}_{\xi_j}
-{\mathcal A}_{\theta_1}\partial_{\theta_1}{\mathcal H}_{\pi_1}
+{\mathcal A}_{\theta_1\theta_2}{\mathcal H}_{\pi_2}
-{\mathcal A}_{\theta_2}\partial_{\theta_1}{\mathcal H}_{\pi_2}\\
&\qquad
+\frac12{\mathcal A}_{\theta_1}
[\partial_{x_j}{\mathcal H}_{\xi_j}+\partial_{\theta_k}{\mathcal H}_{\pi_k}]
+\frac12{\mathcal A}
\partial_{\theta_1}
[\partial_{x_j}{\mathcal H}_{\xi_j}+\partial_{\theta_k}{\mathcal H}_{\pi_k}],
\end{aligned}
$$
we have
$$
\begin{aligned}
\partial_{\theta_2}\partial_{\theta_1}\{\cdots\}
&={\mathcal A}_{\theta_2\theta_1x_j}{\mathcal H}_{\xi_j}
-{\mathcal A}_{\theta_1x_j}\partial_{\theta_2}{\mathcal H}_{\xi_j}
+{\mathcal A}_{\theta_2x_j}\partial_{\theta_1}{\mathcal H}_{\xi_j}
+{\mathcal A}_{x_j}\partial_{\theta_2}\partial_{\theta_1}{\mathcal H}_{\xi_j}\\
&\qquad
-{\mathcal A}_{\theta_2\theta_1}\partial_{\theta_1}{\mathcal H}_{\pi_1}
+{\mathcal A}_{\theta_1}\partial_{\theta_2}\partial_{\theta_1}{\mathcal H}_{\pi_1}
+{\mathcal A}_{\theta_1\theta_2}\partial_{\theta_2}{\mathcal H}_{\pi_2}
+{\mathcal A}_{\theta_2}\partial_{\theta_2}\partial_{\theta_1}{\mathcal H}_{\pi_2}\\
&\qquad\qquad
+\frac12{\mathcal A}_{\theta_2\theta_1}
[\partial_{x_j}{\mathcal H}_{\xi_j}+\partial_{\theta_k}{\mathcal H}_{\pi_k}]
-\frac12{\mathcal A}_{\theta_1}\partial_{\theta_2}
[\partial_{x_j}{\mathcal H}_{\xi_j}+\partial_{\theta_k}{\mathcal H}_{\pi_k}]\\
&\qquad\qquad\qquad
+\frac12{\mathcal A}_{\theta_2}\partial_{\theta_1}
[\partial_{x_j}{\mathcal H}_{\xi_j}+\partial_{\theta_k}{\mathcal H}_{\pi_k}]
+\frac12{\mathcal A}\partial_{\theta_2}\partial_{\theta_1}
[\partial_{x_j}{\mathcal H}_{\xi_j}+\partial_{\theta_k}{\mathcal H}_{\pi_k}].
\end{aligned}
$$
Using 
$\partial_{\theta_2}\partial_{\theta_1}[\partial_{\theta_k}{\mathcal H}_{\theta_k}]=0$
and
$$
\begin{gathered}
\partial_{\theta_2}\partial_{\theta_1}{\mathcal H}_{\xi_j}
={\mathcal H}_{\theta_2\theta_1\xi_j}
+\frac{\partial{\mathcal S}_{\theta_1}}{\partial\theta_2}{\mathcal H}_{\pi_1\theta_1\xi_j}
+\frac{\partial{\mathcal S}_{\theta_2}}{\partial\theta_1}
\bigg(\frac{\partial{\mathcal S}_{\theta_1}}{\partial\theta_2}{\mathcal H}_{\pi_1\pi_2\xi_j}
+{\mathcal H}_{\theta_2\pi_2\xi_j}\bigg),\\
\partial_{\theta_2}\partial_{\theta_1}[\partial_{x_j}{\mathcal H}_{\xi_j}]
=\frac{\partial^2{\mathcal S}_{\theta_1}}{\partial\theta_2\partial x_j}
\bigg({\mathcal H}_{\theta_1\pi_1\xi_j}
+\frac{\partial{\mathcal S}_{\theta_2}}{\partial\theta_1}{\mathcal H}_{\pi_2\pi_1\xi_j}\bigg)
+\frac{\partial^2{\mathcal S}_{\theta_2}}{\partial\theta_1\partial x_j}
\bigg({\mathcal H}_{\theta_2\pi_2\xi_j}
+\frac{\partial{\mathcal S}_{\theta_1}}{\partial\theta_2}{\mathcal H}_{\pi_1\pi_2\xi_j}\bigg),
\end{gathered}
$$
we have
$$
\begin{aligned}
\partial_{\theta_2}\partial_{\theta_1}\{\cdots\}\big|_{\theta=\pi=0}
&={\mathcal A}_{\bar0\bar0,x_j}[\tilde{\mathcal H}_{\theta_2\theta_1\xi_j}
+{\mathcal S}_{\bar1\bar0}
(\tilde{\mathcal H}_{\pi_1\theta_1\xi_j}+\tilde{\mathcal H}_{\pi_2\theta_2\xi_j})
+{\mathcal S}_{\bar1\bar0}^2\tilde{\mathcal H}_{\pi_2\pi_1\xi_j}]\\
&\qquad
+{\mathcal A}_{\bar1\bar0}w_0(t,s\,;x,\xi)
+\frac12{\mathcal A}_{\bar0\bar0}{\mathcal S}_{\bar1\bar0,x_j}
(\tilde{\mathcal H}_{\pi_1\theta_1\xi_j}+\tilde{\mathcal H}_{\pi_2\theta_2\xi_j}
+2{\mathcal S}_{\bar1\bar0}\tilde{\mathcal H}_{\pi_2\pi_1\xi_j}).
\end{aligned}
$$
Therefore, we get
\begin{equation}
{\mathcal A}_{\bar1\bar0,t}+w_0{\mathcal A}_{\bar1\bar0}+w_2=0 
\with {\mathcal A}_{\bar1\bar0}(s,s\,;x,\xi)=0,
\end{equation}
where
$$
\begin{aligned}
w_2(t,s\,;x,\xi)&={\mathcal A}_{\bar0\bar0,x_j}
[\tilde{\mathcal H}_{\theta_2\theta_1\xi_j}
+{\mathcal S}_{\bar1\bar0}(\tilde{\mathcal H}_{\pi_1\theta_1\xi_j}
+\tilde{\mathcal H}_{\pi_2\theta_2\xi_j})
+{\mathcal S}_{\bar1\bar0}^2\tilde{\mathcal H}_{\pi_2\pi_1\xi_j}]\\
&\qquad
+\frac12{\mathcal A}_{\bar0\bar0}{\mathcal S}_{\bar1\bar0,x_j}
(\tilde{\mathcal H}_{\pi_1\theta_1\xi_j}+\tilde{\mathcal H}_{\pi_2\theta_2\xi_j}
+2{\mathcal S}_{\bar1\bar0}\tilde{\mathcal H}_{\pi_2\pi_1\xi_j})\\
&=
c({\mathcal A}_{\bar0\bar0,x_1}+i{\mathcal A}_{\bar0\bar0,x_2})
-2ic\hbar^{-1}{\mathcal A}_{\bar0\bar0,x_3}{\mathcal S}_{\bar1\bar0}
+c\hbar^{-2}{\mathcal S}_{\bar1\bar0}({\mathcal A}_{\bar0\bar0,x_1}-i{\mathcal A}_{\bar0\bar0,x_2})\\
&\qquad
+{\mathcal A}_{\bar0\bar0}
[c\hbar^{-2}{\mathcal S}_{\bar1\bar0}
({\mathcal S}_{\bar1\bar0,x_1}-i{\mathcal S}_{\bar1\bar0,x_2})
-ic\hbar^{-1}{\mathcal S}_{\bar1\bar0,x_3}].
\end{aligned}
$$

Simple but lengthy calculation yields
\begin{prop}
\begin{gather}
{\mathcal A}_{c_1d_1,t}+{\mathcal S}_{c_1d_1}
[{\mathcal A}_{\bar0\bar0,x_j}(\tilde{\mathcal H}_{\pi_1\theta_1,\xi_j}
+{\mathcal S}_{\bar1\bar0}\tilde{\mathcal H}_{\pi_1\pi_2,\xi_j})
+{\mathcal A}_{\bar1\bar0}\tilde{\mathcal H}_{\pi_2\pi_1}
+{\mathcal A}_{\bar0\bar0}{\mathcal S}_{\bar1\bar0,x_j}
\tilde{\mathcal H}_{\pi_1\pi_2,\xi_j}]=0,\\
{\mathcal A}_{c_2d_2,t}+{\mathcal S}_{c_2d_2}
[{\mathcal A}_{\bar0\bar0,x_j}(\tilde{\mathcal H}_{\pi_2\theta_2,\xi_j}
+{\mathcal S}_{\bar1\bar0}\tilde{\mathcal H}_{\pi_1\pi_2,\xi_j})
+{\mathcal A}_{\bar1\bar0}\tilde{\mathcal H}_{\pi_2\pi_1}
+{\mathcal A}_{\bar0\bar0}{\mathcal S}_{\bar1\bar0,x_j}
\tilde{\mathcal H}_{\pi_1\pi_2,\xi_j}]=0,\\
{\mathcal A}_{c_1d_2,t}=0 \et {\mathcal A}_{c_2d_1,t}=0,\quad \mbox{i.e.}\quad 
{\mathcal A}_{c_1d_2}=0 \et {\mathcal A}_{c_2d_1}=0.
\end{gather}
\end{prop}
Analogously, we have
\begin{prop}
\begin{equation}
{\mathcal A}_{\bar0\bar1,t}-w_0{\mathcal A}_{\bar0\bar1}+w_3=0 \with 
{\mathcal A}_{\bar0\bar1}(s,s\,;x,\xi)=0,
\end{equation}
where
$$
\begin{aligned}
w_3(t,s\,;x,\xi)&=({\mathcal A}_{c_1d_1}{\mathcal S}_{c_2d_2}+{\mathcal A}_{c_2d_2}{\mathcal S}_{c_1d_1}
-{\mathcal A}_{\bar0\bar0}{\mathcal S}_{\bar1\bar1})\tilde{\mathcal H}_{\pi_2\pi_1}\\
&\qquad
+[{\mathcal S}_{c_1d_1}{\mathcal S}_{c_2d_2}{\mathcal A}_{\bar0\bar0,x_j}
+{\mathcal A}_{\bar0\bar0}({\mathcal S}_{c_1d_1}{\mathcal S}_{c_2d_2})_{x_j}
-{\mathcal A}_{\bar0\bar0}{\mathcal S}_{\bar1\bar0}{\mathcal S}_{\bar0\bar1,x_j}]
\tilde{\mathcal H}_{\xi_j\pi_2\pi_1}.
\end{aligned}
$$
\end{prop}

\begin{prop}
\begin{equation}
{\mathcal A}_{\bar1\bar1,t}-w_0{\mathcal A}_{\bar1\bar1}+w_4=0 \with 
{\mathcal A}_{\bar1\bar1}(s,s\,;x,\xi)=0,
\end{equation}
where
$$
\begin{aligned}
w_4(t,s\,;x,\xi)&={\mathcal A}_{\bar0\bar1,x_j}\tilde{\mathcal H}_{\xi_j\theta_2\theta_1}
+{\mathcal S}_{\bar1\bar0}{\mathcal A}_{\bar0\bar1,x_j}
(\tilde{\mathcal H}_{\xi_j\pi_1\theta_1}+\tilde{\mathcal H}_{\xi_j\pi_2\theta_2}
+{\mathcal S}_{\bar1\bar0}\tilde{\mathcal H}_{\xi_j\pi_2\pi_1})\\
&\quad
+{\mathcal S}_{\bar1\bar1}{\mathcal A}_{\bar0\bar0,x_j}
(\tilde{\mathcal H}_{\xi_j\pi_1\theta_1}+\tilde{\mathcal H}_{\xi_j\pi_2\theta_2}
+2{\mathcal S}_{\bar1\bar0}\tilde{\mathcal H}_{\xi_j\pi_2\pi_1})
+{\mathcal S}_{c_1d_1}{\mathcal S}_{c_2d_2}{\mathcal A}_{\bar1\bar0,x_j}\tilde{\mathcal H}_{\xi_j\pi_2\pi_1}\\
&\qquad
-{\mathcal S}_{c_1d_1}{\mathcal A}_{c_2d_2,x_j}
(\tilde{\mathcal H}_{\xi_j\pi_1\theta_1}+{\mathcal S}_{\bar1\bar0}\tilde{\mathcal H}_{\xi_j\pi_2\pi_1})
-{\mathcal S}_{c_2d_2}{\mathcal A}_{c_1d_1,x_j}
(\tilde{\mathcal H}_{\xi_j\pi_2\theta_2}+{\mathcal S}_{\bar1\bar0}\tilde{\mathcal H}_{\xi_j\pi_2\pi_1})\\
&\quad
+{\mathcal A}_{\bar1\bar0}{\mathcal S}_{\bar1\bar0,x_j}
(\tilde{\mathcal H}_{\pi_1\theta_1\xi_j}+\tilde{\mathcal H}_{\pi_2\theta_2\xi_j}
+2{\mathcal S}_{\bar1\bar0}\tilde{\mathcal H}_{\pi_2\pi_1\xi_j})
-{\mathcal A}_{\bar1\bar0}({\mathcal S}_{c_1d_1}{\mathcal S}_{c_2d_2})_{x_j}
\tilde{\mathcal H}_{\pi_2\pi_1\xi_j}\\
&\qquad
+({\mathcal A}_{c_1d_1}{\mathcal S}_{c_2d_2}+{\mathcal A}_{c_2d_2}{\mathcal S}_{c_1d_1})
{\mathcal S}_{\bar1\bar0,x_j}\tilde{\mathcal H}_{\pi_2\pi_1\xi_j}
-{\mathcal A}_{c_1d_1}{\mathcal S}_{c_2d_2,x_j}
(\tilde{\mathcal H}_{\xi_j\pi_1\theta_1}+{\mathcal S}_{\bar1\bar0}\tilde{\mathcal H}_{\xi_j\pi_2\pi_1})\\
&\qquad\quad
-{\mathcal A}_{c_2d_2}{\mathcal S}_{c_1d_1,x_j}
(\tilde{\mathcal H}_{\xi_j\pi_2\theta_2}+{\mathcal S}_{\bar1\bar0}\tilde{\mathcal H}_{\xi_j\pi_2\pi_1})\\
&\quad
+\frac12({\mathcal A}_{\bar0\bar0}{\mathcal S}_{\bar1\bar1,x_j}
+{\mathcal A}_{\bar0\bar1}{\mathcal S}_{\bar1\bar0,x_j}
-{\mathcal A}_{\bar1\bar0}{\mathcal S}_{\bar1\bar0,x_j})
(\tilde{\mathcal H}_{\pi_1\theta_1\xi_j}+\tilde{\mathcal H}_{\pi_2\theta_2\xi_j}
+2{\mathcal S}_{\bar1\bar0}\tilde{\mathcal H}_{\pi_2\pi_1\xi_j})\\
&\qquad
+[{\mathcal A}_{\bar1\bar0}({\mathcal S}_{c_1d_1}{\mathcal S}_{c_2d_2})_{x_j}
-{\mathcal A}_{c_1d_1}{\mathcal S}_{c_2d_2}{\mathcal S}_{\bar1\bar0,x_j}
-{\mathcal A}_{c_2d_2}{\mathcal S}_{c_1d_1}{\mathcal S}_{\bar1\bar0,x_j}]\tilde{\mathcal H}_{\pi_2\pi_1\xi_j}.
\end{aligned}
$$
\end{prop}

Therefore, we get the estimates in \eqref{rt3.99}.

%%%

%%\input s-Weyl4-1

\section{Quantum part: Composition formulas of FIOs and $\Psi$DOs}
\subsection{Super differential operators associated with
${\mathcal H}(X,\Xi)$}
Given ${\mathcal H}(X,\Xi)$, we
consider the super (pseudo)-differential
operator of Weyl type: 
\begin{equation}
\hat{\mathcal H}(X,D_X)u(X)=c_{3,2}^2
\iint d\Xi dY\,e^{i\hbar^{-1}\langle{X-Y}|\Xi\rangle}
{\mathcal H}\left(\frac{X+Y}2,\Xi\right) u(Y)
\label{rt4-1}
\end{equation}
for $u\in\ccsl_{{\mathrm{SS}},0}({\fR}^{3|2})$.
Here, we use the abbreviation
$$
c_{m,n}=(2\pi\hbar)^{-m/2}\hbar^{n/2}e^{i\pi n(n-2)/4},\;
X=(x,\theta),\;Y=(y,\omega),\;\Xi=(\xi,\pi),\;
D_X=(-i{\hbar}{\partial_x},{\partial_\theta}).
$$
\par
{\it Remark. }
We may give the definite
meaning to above expression \eqref{rt4-1} 
as oscillatory integrals:
As the right-hand side of \eqref{rt4-1} is not
necessarily `absolutely integrable' with respect to
$d\Xi dY=d\xi d\pi dy d\theta$, that is,
after integrating with respect to $d\pi d\theta$, the integrated function
is not necessarily absolutely integrable with respect to $d\xi dy$.
Therefore, it is necessary to give the definite meaning 
to the right hand side of \eqref{rt4-1}.
Let $\{\chi_\varepsilon(q,p)\}_{\varepsilon >0}$ be any bounded
sequence  of functions in ${\mathcal B}(\euc^{6})$ such that for
each  $\varepsilon >0$, 
$\chi_\varepsilon(q,p) \in {\mathcal S}(\euc^{6})$ 
and
$\lim_{\varepsilon\to 0}\chi_\varepsilon(q,p)=1$ 
in ${\mathcal E}(\euc^{6})$.  
Instead of \eqref{rt4-1}, we consider
\begin{equation}
(\hat{\mathcal H}_\varepsilon u)(X)
=c_{3,2}^2
\iint d\Xi dY\,
e^{i\hbar^{-1}\langle{X-Y}|\Xi\rangle} 
\chi_{\varepsilon}(y,\xi){\mathcal H}\Big(\frac{X+Y}2,\Xi\Big)
u(Y).
\label{rt4-2}
\end{equation}
As is easily seen that 
$\hat{\mathcal H}_\varepsilon u\in \ccsl_{{\mathrm{SS}}}({\fR}^{3|2})$
if $u\in \ccsl_{{\mathrm{SS}},0}({\fR}^{3|2})$,
we can write \eqref{rt4-2} as
\begin{equation}
(\hat{\mathcal H}_\varepsilon u)(x,\theta)
=\sum_a(\hat{\mathcal H}_\varepsilon u)_a(x)\theta^a, 
\label{4.3}
\end{equation}
where 
$(\hat{\mathcal H}_\varepsilon u)_a(x)
=\partial_\theta^a(\hat{\mathcal H}_\varepsilon u)(x,0)$. 
So, applying
the proof of bosonic case to $(\hat{\mathcal H}_{\varepsilon}u)_a$ as in
\cite{Kum76}, we get 
$\hat{\mathcal H}u(X)=\slim \hat{\mathcal H}_\varepsilon u(X)$
in a suitable sense. 
\par
Moreover, all integrals 
which appear hereafter should be considered in the
above sense (called, super oscillatory integral) if the integrand
is not absolutely integrable.  Moreover, if the calculas under
the integral sign is permitted using the above argument combined
with Lax' technique (of using integrations by parts repeatedly with
$\partial_{\xi_j} e^{i\phi(x,\xi)}
=i\partial_{\xi_j}\phi(x,\xi)e^{i\phi(x,\xi)}$), 
we do it without mentioning it. 

By simple calculation, we have
\begin{lem} For $j=1,2,3$ and $v(\theta)=v_0+v_1\theta_1\theta_2$,
putting
$$
\hat\sigma_j(\theta,\partial_\theta)v(\theta)=c_{3,2}^2
\int d\pi d\theta'\,
e^{i\hbar^{-1}\langle\theta-\theta'|\pi\rangle}
\sigma_j\Big(\frac{\theta+\theta'}2,\pi\Big) v(\theta'),
$$
we get
\begin{equation}
\begin{aligned}
&\flat \hat\sigma_j\sharp=\boldsymbol{\sigma}_j\;\;\text{or}\;\;
\hat\sigma_j=\sharp\boldsymbol{\sigma}_j\flat,\\
&\hat\sigma_j(\theta,\partial_\theta)v(\theta)
=c_{3,2}\int d\pi\,
e^{i\hbar^{-1}\langle\theta|\pi\rangle}\sigma_j(\theta,\pi)\hat v(\pi)
\;\;\text{for}\;\; j=1,2,\\
&\hat\sigma_3(\theta,\partial_\theta)v(\theta)
=c_{3,2}\int d\pi\,
e^{i\hbar^{-1}\langle\theta|\pi\rangle}(1-\sigma_3(\theta,\pi))\hat v(\pi).
\end{aligned}
\end{equation}
\end{lem}

\begin{lem} %\label{Proposition 2.5}
Let ${\mathcal H}(t,X,\Xi)$ be derived from 
${\Bbb H}(t,q,-i\hbar\partial_q)$ in \eqref{rt1.1}. Then, 
we have
\begin{equation}
{\Bbb H}\bigg(t,q,\frac{\hbar}i\pdq\bigg)=
\flat \hat{\mathcal H}\sharp
:C_0^\infty(\euc^3:{\mathbb C}^2)\to
C^\infty(\euc^3:{\mathbb C}^2)\quad\mbox{for each}\quad t. 
\label{rt4-4}
\end{equation}
We have also
\begin{equation}
\hat{\mathcal H}(t)u(X)=c_{3,2}\int d\Xi\, e^{i\hbar^{-1}\langle X|\Xi\rangle}
{\mathcal H}^0(t,X,\Xi)\hat u(\Xi),
\label{4.4+1}
\end{equation}
with
\begin{equation}
{\mathcal H}^0(t,X,\Xi)={\mathcal H}(t,X,\Xi)+c\xi_3-{\elec}A_3(t,x).
\end{equation}
\end{lem}

%%%

%%\input s-Weyl4-2

\subsection{Fourier Integral Operators associated with
${\mathcal H}(t,X,\Xi)$}

After reordering $({\mixdata})$ as $({\barx,\bartheta,\unbxi,\unbpi})$
and denoting them by $(x,\theta,\xi,\pi)$,
we consider an integral transformation ${\mathcal U}(t,s)$ on
$\ccsl_{{\mathrm{SS}},0}({\fR}^{3|2})$
where
${\mathcal S}(t,s\,;x,\theta,\xi,\pi)$ and
${\mathcal A}(t,s\,;x,\theta,\xi,\pi)$ 
are defined in \S2:
\begin{equation*}
{\mathcal U}(t,s)u(x,\theta)=({\mathcal U}(t,s)u)(x,\theta)=c_{3,2}
\int d\xi d\pi \,
{\mathcal A}(t,s\,;x,\theta,\xi,\pi)
e^{i\hbar^{-1}{\mathcal S}(t,s\,;x,\theta,\xi,\pi)}
{\mathcal F}u(\xi,\pi)
\end{equation*}
or simply, we write it as
\begin{equation}
{\mathcal U}(t,s)u(X)=({\mathcal U}(t,s)u)(X)=c_{3,2}\int d\Xi\,
{\mathcal A}(t,s\,;X,\Xi)
e^{i\hbar^{-1}{\mathcal S}(t,s\,;X,\Xi)}
{\mathcal F}u(\Xi).
\label{rt4.6}
\end{equation}
%\vskip 2mm
\begin{thm}\label{thm:bdd}
We have ${\mathcal U}(t,s)u \in \ccsl_{{\mathrm{SS}}}({\fR}^{3|2})$
for $u \in \ccsl_{{\mathrm{SS}},0}({\fR}^{3|2})$.
Moreover, there exists a constant C such that
\begin{equation}
\Vert {\mathcal U}(t,s)u\Vert\leq 2^2(1+C|t-s|)
\Vert u\Vert \leq 2^2 e^{C|t-s|} \Vert u \Vert .
\label{rt4.7}
\end{equation}
\end{thm}
\par
{\it Proof.}
Using
${\mathcal F}u(\xi,\pi)=
\hbar{\hat u_{\bar 1}}(\xi)+\hbar^{-1}{\hat u_{\bar 0}}(\xi)\pi_1\pi_2$, 
we rewrite \eqref{rt4.6} as 
\begin{equation}
\begin{aligned}
{\mathcal U}(t,s)u(x,\theta) =
c_{3,2}
&\int d\xi d\pi \,
({\mathcal A}_{\mathrm B}(t,x,\xi)+{\mathcal A}_{\mathrm S}(t,s\,;x,\theta,\xi,\pi))
e^{i\hbar^{-1}{\mathcal S}_{\mathrm B}(t,s\,;x,\xi)}
\\
&\qquad\times 
\Big[\sum_{\ell=0}^2
{(i\hbar^{-1})^\ell \over {\ell!}}
{\mathcal S}_{\mathrm S}(t,s\,;x,\theta,\xi,\pi)^\ell\Big]
\big(\hbar{\hat u_{\bar 1}}(\xi)
+\hbar^{-1}{\hat u_{\bar 0}}(\xi)\pi_1\pi_2\big).
\end{aligned}
\label{rt4.14}
\end{equation}
Here, we remark by \eqref{rt3.70} that
\begin{equation}
{\mathcal S}_{\mathrm S}(t,s\,;x,\theta,\xi,\pi)^\ell 
=
\begin{cases}
1 & \text{when $\ell=0$},\\
{\mathcal S}_{\bar 1 \bar 0 }\theta_1\theta_2
+\sum_{|c|=|d|=1}{\mathcal S}_{cd}\theta^c\pi^d
+{\mathcal S}_{\bar 0 \bar 1}\pi_1\pi_2
+{\mathcal S}_{\bar 1 \bar 1}\theta_1\theta_2\pi_1\pi_2
 & \text{when $\ell=1$},\\
2[{\mathcal S}_{\bar 1 \bar 0 }{\mathcal S}_{\bar 0 \bar 1}
-{\mathcal S}_{c_1d_1}{\mathcal S}_{c_2d_2}
]\theta_1\theta_2\pi_1\pi_2
 & \text{when $\ell=2$},\\ %+{\mathcal S}_{c_1d_2}{\mathcal S}_{c_2d_1}
0 & \text{when $\ell\ge 3$}.
\end{cases}
\end{equation}

After integrating \eqref{rt4.14} with respect to $d\pi$,
we have 
\begin{equation}
\begin{split}
{\mathcal U}(t,s)u(x,\theta)&=\sum_{|b|=0,2} v_b(t,s\,;x)\theta^b\\
&=\sum_b \[ \sum_a
(2\pi\hbar)^{-3/2}\int  d\xi\,
{\mathcal B}_{ba}(t,s\,;x,\xi)
e^{i\hbar^{-1}{\mathcal S}_{\mathrm B}(t,s\,;x,\xi)}
{\hat u}_a(\xi)\]\theta^b,
\end{split}\label{rt4.15}
\end{equation}
where
\begin{equation}
\left\{
\begin{split}
{\mathcal B}_{\bar 1 \bar 1}&=\hbar^2{\mathcal A}_{\bar 1 \bar 1}
+i\hbar({\mathcal A}_{\mathrm B}{\mathcal S}_{\bar 1 \bar 1}
+{\mathcal A}_{\bar 1 \bar 0 }{\mathcal S}_{\bar 0 \bar 1}
+{\mathcal A}_{\bar 0 \bar 1 }{\mathcal S}_{\bar 1 \bar 0}
+{\mathcal A}_{c_1d_1}{\mathcal S}_{c_2d_2}-{\mathcal A}_{c_2d_2}{\mathcal S}_{c_1d_1})\\
%-{\mathcal A}_{c_1d_2}{\mathcal S}_{c_2d_1}+{\mathcal A}_{c_1d_2}{\mathcal S}_{c_2d_1}
&\qquad\qquad\qquad\qquad\qquad\qquad
-{\mathcal A}_{\mathrm B}
[{\mathcal S}_{\bar 1 \bar 0 }{\mathcal S}_{\bar 0 \bar 1}
-{\mathcal S}_{c_1d_1}{\mathcal S}_{c_2d_2}
],\\ %+{\mathcal S}_{c_1d_2}{\mathcal S}_{c_2d_1}
{\mathcal B}_{\bar 0 \bar 1}
&={\mathcal A}_{\bar 0 \bar 1 }
+i\hbar^{-1} {\mathcal A}_{\mathrm B}{\mathcal S}_{\bar 0 \bar 1},\\
{\mathcal B}_{\bar 1 \bar 0} 
&={\mathcal A}_{\bar 1 \bar 0}+
i\hbar^{-1}{\mathcal A}_{\mathrm B}{\mathcal S}_{\bar 1 \bar 0},\\
{\mathcal B}_{\bar 0 \bar 0} &={\mathcal A}_{\mathrm B}.
\end{split}
\right.
\label{rt4.16}
\end{equation}
In the above, arguments of ${\mathcal B}_{**}$,
${\mathcal A}_{**}$ and ${\mathcal S}_{**}$ are $(t,s\,;x,\xi)$.

\par
By using \eqref{rt3.84} and \eqref{rt3.99}, we have that
$\partial^\alpha_{x} \partial^\beta_{\xi}
{\mathcal B}_{ba}(t,s\,;x_{\mathrm B},\xi_{\mathrm B})\in\Bbb C $
and
\begin{equation}
\begin{cases}
| \partial^\alpha_{x} \partial^\beta_{\xi}
({\mathcal B}_{ba}(t,s\,;x_{\mathrm B},\xi_{\mathrm B})-1)| 
\le  C|t-s| &\text{for $b=a=\bar1$ or $b=q=\bar0$},\\
| \partial^\alpha_{x} \partial^\beta_{\xi}
{\mathcal B}_{ba}(t,s\,;x_{\mathrm B},\xi_{\mathrm B})| 
\le C|t-s|&\text{for $b\neq a$.}
\end{cases}
\label{QP4-17}
\end{equation}
\par
Putting
\begin{equation}
({\mathcal E}_{ba}(t,s)u_a)(x_{\mathrm B})
=
(2\pi\hbar)^{-3/2}
\int d\xi_{\mathrm B}\,
{\mathcal B}_{ba}(t,s\,;x_{\mathrm B},\xi_{\mathrm B})
e^{i\hbar^{-1}{\mathcal S}_{\mathrm B}(t,s\,;x_{\mathrm B},\xi_{\mathrm B})}
\hat{u}_a(\xi_{\mathrm B}),
\label{QP4-18}
\end{equation}
we have
\begin{equation}
\begin{aligned}
v_{\bar0}(t,s\,;x)
&={\mathcal E}_{\bar0\bar0}{\hat u}_{\bar0}(\xi)
+{\mathcal E}_{\bar0\bar1}{\hat u}_{\bar1}(\xi)\\
&=(2\pi\hbar)^{-3/2}\int  d\xi\,
e^{i\hbar^{-1}{\mathcal S}_{\mathrm B}(t,s\,;x,\xi)}
[{\mathcal B}_{\bar0\bar0}(t,s\,;x,\xi){\hat u}_{\bar0}(\xi)
+{\mathcal B}_{\bar0\bar1}(t,s\,;x,\xi){\hat u}_{\bar1}(\xi)],\\
v_{\bar1}(t,s\,;x)
&={\mathcal E}_{\bar1\bar0}{\hat u}_{\bar0}(\xi)
+{\mathcal E}_{\bar1\bar1}{\hat u}_{\bar1}(\xi)\\
&=(2\pi\hbar)^{-3/2}\int  d\xi\,
e^{i\hbar^{-1}{\mathcal S}_{\mathrm B}(t,s\,;x,\xi)}
[{\mathcal B}_{\bar1\bar0}(t,s\,;x,\xi){\hat u}_{\bar0}(\xi)
+{\mathcal B}_{\bar1\bar1}(t,s\,;x,\xi){\hat u}_{\bar1}(\xi)],
\end{aligned}
\label{QP4-18'}
\end{equation}

By applying Theorem 2.1 of {\cite{AF78}} to \eqref{QP4-18}, 
we have
\begin{equation} 
\begin{aligned}
&\| {\mathcal E}_{\bar0\bar0}(t,s)u_{\bar0}\|
\le (1+C|t-s|)\|u_0\|,\quad
\| {\mathcal E}_{\bar0\bar1}(t,s)u_{\bar1} \|
\le C|t-s|\|u_1\|,\\
&\| {\mathcal E}_{\bar1\bar0}(t,s)u_{\bar0} \|
\le C|t-s|\|u_0\|,\quad
\| {\mathcal E}_{\bar1\bar1}(t,s)u_{\bar1} \|
\le (1+C|t-s|)\|u_1\|,\quad
\end{aligned}
\end{equation} 
which implies
\begin{equation}
\| {\mathcal U}(t,s)u \|^2
=\|v_{\bar0}\|^2+\|v_{\bar1}\|^2
\le
(1+\sqrt{2}C|t-s|)^2(\|u_{\bar0}\|^2+\|u_{\bar1}\|^2)
=(1+\sqrt{2}C|t-s|)^2\| u \|^2 .
\label{QP4-20}
\end{equation}
Moreover,
$ \pi_{\mathrm B}(\sum_{a}{\mathcal E}_{ba}(t,s)u_a(X)) 
\in C^{\infty} ({\euc}^3) $,
i.e. ${\mathcal U}(t,s)u\in\ccsl_{{\mathrm{SS}}}({\fR}^{3|2})$.
\qed 

\par
{\it Remark. }
For $u\in\clsl_{{\mathrm{SS}}}^2({\fR}^{3|2})$, 
${\mathcal U}(t,s)u$ is defined as the limit of
a Cauchy sequence $\{{\mathcal U}(t,s)u_k\}$ in 
$\clsl_{{\mathrm{SS}}}^2({\fR}^{3|2})$
where $u_k\in\ccsl_{{\mathrm{SS}},0}({\fR}^{3|2})$ 
is a sequence converging to $u$ in $\clsl_{{\mathrm{SS}}}^2({\fR}^{3|2})$.

\begin{prop}\label{prop:3.2}
%(0) ${\mathcal U}(0)u=u$ for each $u\in \clsl_{{\mathrm{SS}}}^2({\fR}^{3|2})$.
%\newline
(1) For each $u\in \clsl_{{\mathrm{SS}}}^2({\fR}^{3|2})$, we have
\begin{equation}
\slim_{|t-s|\to0}{\mathcal U}(t,s)u=u\qquad  in \quad
\clsl_{{\mathrm{SS}}}^2({\fR}^{3|2}). 
\label{0cont}
\end{equation}
(2) Define $U(s,s)=I$.
For fixed $s$, the correspondence $t\to {\mathcal U}(t,s)u$ gives 
a strongly continuous function
with values in $\clsl_{{\mathrm{SS}}}^2({\fR}^{3|2})$. 
\end{prop}
\par
{\it Proof. }
To prove \eqref{0cont}, we need to claim
$$
\slim_{|t-s|\to0} v_{\bar0}=u_{\bar0}\et
\slim_{|t-s|\to0} v_{\bar1}=u_{\bar1}.
$$
We get the desired results by the standard method
applying to \eqref{QP4-18'}.  \qed
 
\par{\it Remark. }
In the above,
Theorem \ref{thm:bdd} and Proposition \ref{prop:3.2} are proved 
after integrating w.r.t. $d\pi$ 
and applying the standard method for pseudo-differential and
Fourier integral operators on the Euclidian space $\euc^3$.
But, this suggests us to the neccesity of developping those operator theories
on the superspace ${\fR}^{m|n}$.

%%%

%%\input s-Weyl4-31
\subsection{Composition of FIOs with $\Psi$DOs}

As $(t,s)$ is inessential in this subsection, 
we abbreviate it and denote ${\mathcal A}(X,\Xi)={\mathcal A}(t,s\,;x,\xi,\theta,\pi)$, 
${\mathcal S}(X,\Xi)={\mathcal S}(t,s\,;x,\xi,\theta,\pi)$, 
${\mathcal H}(X,\Xi)={\mathcal H}(t,X,\Xi)$
and 
\begin{equation}
\begin{gathered}
{\mathcal U}({\mathcal A},{\mathcal S})u(X)=c_{3,2}\int d\Xi\, 
{\mathcal A}(X,\Xi)e^{i\hbar^{-1}{{\mathcal S}}(X,\Xi)}{\hat{u}}(\Xi),\\
\hat{\mathcal H} u(X)=c_{3,2}^2\iint d\Xi dY\, 
e^{i\hbar^{-1}\langle X-Y|\Xi\rangle}{\mathcal H}\Big(\frac{X+Y}2,\Xi\Big)u(Y).
\end{gathered}
\end{equation}

\begin{thm} %\label{thm 3.7}  
Let ${\mathcal U}({\mathcal A},{\mathcal S})$ and $\hat{\mathcal H}$ be given as above.
There exists ${\mathcal B}_L$=${\mathcal B}_L(X,\Eta)$ such that
\begin{equation}
\hat{\mathcal H}\, {\mathcal U}({\mathcal A},{\mathcal S}) = {\mathcal U}({\mathcal B}_L,{{\mathcal S}}).
\label{s3.45}
\end{equation}
Moreover, ${\mathcal B}_L$ has the following expansion
\begin{equation}
{\mathcal B}_L={\mathcal H}{\mathcal A}-i\hbar \Big\{\partial_{\Xi_j}{\mathcal H}\cdot {\partial}_{X_j}{\mathcal A}
+{1\over2}\Big(\partial^2_{X_j \Xi_j} {\mathcal H}
+\partial^2_{X_j X_k}{{\mathcal S}}\cdot
\partial^2_{\Xi_k \Xi_j}{\mathcal H}\Big){\mathcal A}\Big\}+{\mathcal R}_L.
\label{s3.46}
\end{equation}
Here, 
the argument of ${\mathcal H}$ is $(X,\partial_X{{\mathcal S}})$
and that of  ${{\mathcal S}}$ and ${\mathcal R}_L$ is $(X,\Eta)$, and 
${\mathcal R}_L\in{\ccsl}_{S\!S}(\fR^{3|2}\times\fR^{3|2})$
satisfies
\begin{equation}
\sup|D_X^{\alpha}D_\Eta^{\beta}\,
{\mathcal R}_L(X_{\mathrm B},\Eta_{\mathrm B})|\le 
C_{\alpha\beta}<\infty.
\label{s3.47}
\end{equation}
\end{thm}
\par
{\it Proof. }
By definition, we have
\begin{equation}
\begin{aligned}
{\hat{\mathcal H}}\,{\mathcal U}(A,{\mathcal S}) u(X)
&=
c_{3,2}^3\int d\Xi\,dY\,d\Eta\,
{\mathcal H}\Big(\frac{X+Y}2,\Xi\Big) {\mathcal A}(Y,\Eta)
e^{i\hbar^{-1}
(\langle X-Y|\Xi\rangle +{{\mathcal S}} (Y,\Eta))} 
{\hat{u}}(\Eta)\\  
&=
c_{3,2}\int d\Eta\,{\mathcal B}_L(X,\Eta)
e^{i\hbar^{-1}{{\mathcal S}}(X,\Eta)}{\hat{u}}(\Eta).
\end{aligned}\label{s3.48}
\end{equation}
Here, we put
\begin{equation}
{\mathcal B}_L(X,\Eta)
=
c_{3,2}^2 \int d\Xi\,dY\,
{\mathcal Q}(X,\Xi,Y,\Eta)
e^{i\hbar^{-1}\psi(X,\Xi,Y,\Eta)}
\label{s3.49}\end{equation}
with
\begin{equation}
\psi(X,\Xi,Y,\Eta)
=\langle X-Y|\Xi\rangle
+{{\mathcal S}} (Y,\Eta) -{{\mathcal S}} (X,\Eta),
\label{s3.50}
\end{equation}
\begin{equation}
{\mathcal Q}(X,\Xi,Y,\Eta)
={\mathcal H}\Big(\frac{X+Y}2,\Xi\Big) {\mathcal A}(Y,\Eta).
\label{s3.51}
\end{equation}
(I) Before giving the full proof, we calculate rather formally which yields
\eqref{s3.46}.
As
\begin{equation}
{{\mathcal S}} (Y,\Eta) -{{\mathcal S}} (X,\Eta)=
(Y_j-X_j){\widetilde{\partial_{X_j}{{\mathcal S}}}}(X,Y-X,\Eta)=
\langle Y-X|{\widetilde{\partial_X{{\mathcal S}}}}(X,Y-X,\Eta)\rangle
\label{s3.52}
\end{equation}
where
$$
{\widetilde{\partial_{X_j}}}{{\mathcal S}}(X,Y-X,\Eta)
=\int_0^1 d\tau \,
\partial_{X_j}{{\mathcal S}}(X+\tau (Y-X),\Eta),
$$
we introduce a change of variables by
\begin{equation}
{\left\{
{\begin{aligned}
&{\tilde Y}=Y-X,\\
&{\tilde \Xi}=\Xi-{\widetilde{\partial_X{{\mathcal S}}}}(X,Y-X,\Eta),
\end{aligned}}\right.}
\longleftrightarrow
{\left\{
{\begin{aligned}
& Y={\tilde Y}+X,\\
& \Xi={\tilde \Xi}+{\widetilde{\partial_X{{\mathcal S}}}}(X,{\tilde Y},\Eta).
\end{aligned}}
\right.}
\label{s3.53}
\end{equation}
As $\sdet\Big(\frac{\partial(Y,\Xi)}{\partial({\tilde Y},{\tilde\Xi})}\Big)=1$,
we may apply the formula of change of variables under integral sign,
see, Theorem 3.8 of \cite{IM91} or Theorem 1.14 of \cite{In92},
to \eqref{s3.49} getting
\begin{equation}
{\mathcal B}_L(X,\Eta)=c_{3,2}^2\int d{\tilde \Xi}d{\tilde Y}\,
e^{-i\hbar^{-1} \langle{\tilde Y}|{\tilde \Xi}\rangle }
{\mathcal H}\Big(X+\frac{\tilde Y}2,{\tilde \Xi}
+{\widetilde{\partial_X{{\mathcal S}}}}(X,{\tilde Y},\Eta)\Big)
{\mathcal A}(X+{\tilde Y},\Eta).\label{s3.54}
\end{equation}
By Taylor's formula w.r.t. $\tilde\Xi$, we have
\begin{equation}
\begin{aligned}
&{\mathcal H}\Big(X+\frac{\tilde Y}2,{\tilde \Xi}
+{\widetilde{\partial_X{{\mathcal S}}}}(X,{\tilde Y},\Eta)\Big)
={\mathcal H}\Big(X+\frac{\tilde Y}2,{\widetilde{\partial_X{{\mathcal S}}}}(X,{\tilde Y},\Eta)\Big)\\
&\qquad\qquad+{\tilde\Xi_j}
{\partial_{\Xi_j}}
{\mathcal H}\Big(X+\frac{\tilde Y}2,{\widetilde{\partial_X{{\mathcal S}}}}(X,{\tilde Y},\Eta)\Big)\\
&\qquad\qquad\qquad\qquad
+{\tilde\Xi_j}{\tilde\Xi_k}\int_0^1  d\tau_1 (1-\tau_1)
\partial^2_{\Xi_k \Xi_j}{\mathcal H}\Big(X+\frac{\tilde Y}2,
\tau_1{\tilde\Xi}+{\widetilde{\partial_X{{\mathcal S}}}}(X,{\tilde Y},\Eta)\Big).
\end{aligned}
\label{s3.55}\end{equation}
In the above, we abbreviate summation sign and $j\neq k$
because $\partial_{\xi_j}^2{\mathcal H}=\partial_{\pi_k}^2{\mathcal H}=0$.

Now, we remark
\begin{equation}
c_{3,2}^2\int_{\fR^{3|2}} d{\tilde \Xi}\,
e^{-i\hbar^{-1}\langle{\tilde Y}|{\tilde \Xi}\rangle }=\delta({\tilde Y}),\quad
c_{3,2}^2\int_{\fR^{3|2}} d{\tilde Y}\,
e^{-i\hbar^{-1}\langle{\tilde Y}|{\tilde \Xi}\rangle }=\delta({\tilde \Xi}),
\label{s-delta}
\end{equation}
and
\begin{equation}
{\tilde\Xi_j}\,e^{-i\hbar^{-1} \langle{\tilde Y}|{\tilde \Xi}\rangle }
=i\hbar\partial_{\tilde Y_j}
e^{-i\hbar^{-1} \langle{\tilde Y}|{\tilde \Xi}\rangle },\quad
{\tilde Y_j}\,e^{-i\hbar^{-1} \langle{\tilde Y}|{\tilde \Xi}\rangle }
=i\hbar(-1)^{p({\tilde \Xi}_j)}\partial_{\tilde \Xi_j}
e^{-i\hbar^{-1} \langle{\tilde Y}|{\tilde \Xi}\rangle }.
\label{bp}
\end{equation}

From the first equality of \eqref{s-delta},
we get easily that
\begin{equation}
\begin{aligned}
c_{3,2}^2\int d{\tilde Y}d{\tilde \Xi}\,
e^{-i\hbar^{-1} \langle{\tilde Y}|{\tilde \Xi}\rangle }
{\mathcal H}\Big(X+\frac{\tilde Y}2,{\widetilde{\partial_X{{\mathcal S}}}}(X,{\tilde Y},\Eta)\Big)
&{\mathcal A}(X+{\tilde Y},\Eta)\\
&={\mathcal H}(X,\partial_X{{\mathcal S}}(X,\Eta))
{\mathcal A}(X,\Eta).
\end{aligned}
\label{s3.61}
\end{equation}
Using the first equalty of \eqref{bp}
and applying the first equalty of \eqref{s-delta} after integration by parts,
%using \eqref{s3.57} and \eqref{s3.59}, 
we have
\begin{equation}
\begin{aligned}
&c_{3,2}^2\int d{\tilde Y}d{\tilde \Xi}\,
e^{-i\hbar^{-1} \langle{\tilde Y}|{\tilde \Xi}\rangle }
{\tilde\Xi}_j {\partial_{\Xi_j}}
{\mathcal H}\Big(X+\frac{\tilde Y}2,{\widetilde{\partial_X{{\mathcal S}}}}(X,{\tilde Y},\Eta)\Big)
{\mathcal A}(X+{\tilde Y},\Eta)\\
&\qquad
=-i\hbar c_{3,2}^2\int d{\tilde Y}d{\tilde \Xi}\,
e^{-i\hbar^{-1} \langle{\tilde Y}|{\tilde \Xi}\rangle }
{\partial_{{\tilde Y}_j}}
\Big[{\partial_{\Xi_j}}{\mathcal H}\Big(X+\frac{\tilde Y}2,{\widetilde{\partial_X{{\mathcal S}}}}(X,{\tilde Y},\Eta)\Big)
{\mathcal A}(X+{\tilde Y},\Eta)\Big]\\
&\qquad
=-i\hbar {\partial_{{\tilde Y}_j}}
\Big[{\partial_{\Xi_j}}{\mathcal H}
\big(X+\frac{\tilde Y}2,{\widetilde{\partial_X{{\mathcal S}}}}(X,{\tilde Y},\Eta)\Big)
{\mathcal A}(X+{\tilde Y},\Eta)\Big]\Big|_{\tilde Y=0}\\
&\qquad 
=-i\hbar
\Big\{{\partial}_{X_j}{\mathcal A}(*)\,\partial_{\Xi_j}{\mathcal H}(**)
+{1\over2}\Big(\partial^2_{X_j \Xi_j}{\mathcal H}(**)
+\partial^2_{X_j X_k}{{\mathcal S}}(*)\,
\partial^2_{\Xi_k \Xi_j}{\mathcal H}(**)\Big){\mathcal A}(*)\Big\}.
\end{aligned}
\label{s3.61-1}
\end{equation}
In the last line above, we put 
$$
(*)=(X,\Eta)\et (**)=(X,\partial_X{{\mathcal S}}(X,\Eta)),
$$
respectively.
Thus, we get the main terms of \eqref{s3.46} formally.
\par
The  remainder term is derived from
\begin{equation} 
\begin{aligned}
{\mathcal R}_L(X,\Eta)=&c_{3,2}^2\int d{\tilde Y}d{\tilde \Xi}\,
e^{-i\hbar^{-1} \langle{\tilde Y}|{\tilde \Xi}\rangle }
{\tilde\Xi_j}{\tilde\Xi_k}\\
&\qquad\times\Big[\int_0^1  d\tau_1 (1-\tau_1)
\partial^2_{\Xi_k \Xi_j}{\mathcal H}\Big(X+\frac{\tilde Y}2,
\tau_1{\tilde\Xi}+{\widetilde{\partial_X{{\mathcal S}}}}(X,{\tilde Y},\Eta)\Big)\Big]
{\mathcal A}(X+{\tilde Y},\Eta).
\end{aligned}
\end{equation}
As $\partial^4_{\Xi_m \Xi_\ell \Xi_k \Xi_j}{\mathcal H}=0$, we have,
for any $\tau_1\in(0,1)$, $\tilde \Xi\in\fR^{3|2}$,
\begin{equation}
\begin{aligned}
{\partial}_{\Xi_j \Xi_k}^2{\mathcal H}\Big(X+\frac{\tilde Y}2,
\tau_1{\tilde \Xi}+{\widetilde{\partial_X{{\mathcal S}}}}(X,{\tilde Y},\Eta)\Big)
=&\partial^2_{\Xi_k \Xi_j}{\mathcal H}\Big(X+\frac{\tilde Y}2,
{\widetilde{\partial_X{{\mathcal S}}}}(X,{\tilde Y},\Eta)\Big)\\
&\qquad
+\tau_1{\tilde\Xi}_\ell\partial^3_{\Xi_\ell \Xi_k \Xi_j}
{\mathcal H}\Big(X+\frac{\tilde Y}2,
{\widetilde{\partial_X{{\mathcal S}}}}(X,{\tilde Y},\Eta)\Big).
\end{aligned}
\label{s3.59}
\end{equation}
Therefore,
\begin{equation}
{\mathcal R}_L(X,\Eta)={\mathcal R}_{L1}(X,\Eta)+{\mathcal R}_{L2}(X,\Eta)
\end{equation}
with
$$\begin{aligned}
{\mathcal R}_{L1}(X,\Eta)&=c_{3,2}^2\int d{\tilde Y}d{\tilde \Xi}\,
e^{-i\hbar^{-1} \langle{\tilde Y}|{\tilde \Xi}\rangle }\,
{\tilde \Xi}_j{\tilde \Xi}_k\frac12
\partial^2_{\Xi_k \Xi_j}{\mathcal H}(\widetilde{**})\,{\mathcal A}(\tilde{*})\\
{\mathcal R}_{L2}(X,\Eta)&=c_{3,2}^2\int d{\tilde Y}d{\tilde \Xi}\,
e^{-i\hbar^{-1} \langle{\tilde Y}|{\tilde \Xi}\rangle }\,
{\tilde \Xi}_j{\tilde \Xi}_k
\frac16{\tilde \Xi}_\ell\partial^3_{\Xi_\ell \Xi_k \Xi_j}{\mathcal H}
(\widetilde{**})\,{\mathcal A}(\tilde{*})
\end{aligned}
$$
where we put $(\widetilde{**})=\big(X+\frac{\tilde Y}2,
{\widetilde{\partial_X{{\mathcal S}}}}(X,{\tilde Y},\Eta)\Big)$ and
$(\tilde{*})=(X+{\tilde Y},\Eta)$, respectively.
Using \eqref{bp} and integration by parts, we get
\begin{equation}
\begin{aligned}
{\mathcal R}_{L1}(*)=&\frac{-i\hbar}{2} c_{3,2}^2\int d{\tilde Y}d{\tilde \Xi}\,
e^{-i\hbar^{-1} \langle{\tilde Y}|{\tilde \Xi}\rangle }
{\tilde \Xi}_k\,(-1)^{p({\tilde Y}_j)p({\tilde\Xi}_k)}\\
&\qquad\qquad\times
\Big[\partial_{X_j}{\mathcal A}(\tilde{*})\,
\partial^2_{\Xi_k \Xi_j}{\mathcal H}(\widetilde{**})
+\partial_{{\tilde Y}_j}{\widetilde{\partial_{X_\ell}{{\mathcal S}}}(X,{\tilde Y},\Eta)}
\partial^3_{\Xi_\ell \Xi_k \Xi_j}{\mathcal H}(\widetilde{**})
{\mathcal A}(\tilde{*})\Big]\\
=&-\frac{\hbar^2}2(-1)^{p({\tilde Y}_j)p({\tilde\Xi}_k)}
\Big[
\Big(\frac13{\mathcal A}(*)\partial^3_{X_k X_j X_\ell}{{\mathcal S}}(*)
+\partial_{X_k}{\mathcal A}(*)\frac12\partial^2_{X_j X_\ell}{{\mathcal S}}(*)\Big)
\partial^3_{\Xi_\ell \Xi_k \Xi_j}{\mathcal H}(**)\\
&\qquad\qquad\qquad\qquad\qquad\qquad
+\partial^2_{X_k X_j}{\mathcal A}(*)
\partial^2_{\Xi_k \Xi_j}{\mathcal H}(**)
\Big].
\end{aligned}\label{4.45}
\end{equation}
Here, we used
$$
\begin{gathered}
\partial_{{\tilde Y}_j}{\widetilde{\partial_{X_\ell}{{\mathcal S}}}(X,{\tilde Y},\Eta)}
\Big|_{{\tilde Y}=0}
=\int_0^1 d\tau \,\tau\,
\partial^2_{X_j X_\ell}{{\mathcal S}}(X+\tau \tilde{Y},\Eta)\Big|_{{\tilde Y}=0}
=\frac12 \partial^2_{X_j X_\ell}{{\mathcal S}}(X,\Eta),\\
\partial_{{\tilde Y}_j}\partial_{{\tilde Y}_k}{\widetilde{\partial_{X_\ell}{{\mathcal S}}}(X,{\tilde Y},\Eta)}
\Big|_{{\tilde Y}=0}
=\int_0^1 d\tau \,\tau^2\,
\partial^3_{X_k X_j X_\ell}{{\mathcal S}}(X+\tau \tilde{Y},\Eta)\Big|_{{\tilde Y}=0}
=\frac13 \partial^2_{X_k X_j X_\ell}{{\mathcal S}}(X,\Eta),\\
\end{gathered}
$$
Calculating analogously, we get
\begin{equation}
{\mathcal R}_{L2}(*)=\frac{(-i\hbar)^3}6
(-1)^{p({\tilde Y}_j)p({\tilde\Xi}_k)+p({\tilde Y}_j)p({\tilde\Xi}_\ell)
+p({\tilde Y}_k)p({\tilde\Xi}_\ell)}
\partial^3_{X_\ell X_k X_j}{\mathcal A}(*)
\partial^3_{\Xi_\ell\Xi_k\Xi_j}{\mathcal H}(**).
\label{4.46}
\end{equation}

(II) To make the above procedure rigorous, we need to justify the usages
of the changing the order of integration and those of delta functions.
But, these are readily justified by using oscillatory integrals (see,
Kumano-go \cite{Kum81}).
Moreover, the estimate \eqref{s3.47} is obtained easily. 
For example, we consider the first term of \eqref{4.45}
$$
{\mathcal A}(*)\partial^3_{X_k X_j X_\ell}{\mathcal S}(*)
\partial^3_{\Xi_\ell \Xi_k \Xi_j}{\mathcal H}(**).
$$
By the structure of ${\mathcal H}$, we have terms as
$$
{\mathcal A}(X)\partial^3_{\theta_1\theta_2 x_j}{\mathcal S}(X)
\partial^3_{\pi_1 \pi_2 \xi_j}{\mathcal H}(X,\partial_X{\mathcal S}(X,\Eta)).
$$
The derivatives $\partial_X^\alpha\partial_\Eta^\beta$ of these terms 
have clearly bounded body terms by \eqref{rt3.84} and \eqref{rt3.99}.
 \qed
\par
{\it Remark. }
The main term is easily obtained from
\begin{equation} %\begin{aligned}
\sum_{|\alpha|=0}^1\frac{(-i\hbar)^{|\alpha|}}{\alpha!}\partial_{\tilde Y}^{\alpha}
\Big(\partial_{\tilde\Xi}^{\alpha}{\mathcal H}(X+\frac12{\tilde Y},
{\tilde\Xi}+{\widetilde{\partial_X{{\mathcal S}}}}(X,{\tilde Y},\Eta))\cdot
{\mathcal A}(X+\tilde Y,\Eta)\Big)
\Bigg|{\begin{Sb}{\tilde Y}=0,\\{\tilde \Xi}=0\end{Sb}}.
\end{equation}

%%%

%%\input s-Weyl4-32

The following theorem is given for the future use.
\begin{thm}\label{thm 3.5}
Let ${\mathcal U}({\mathcal A},{\mathcal S})$, ${\hat {\mathcal H}}$ be as above.  
Then,
there exists ${\mathcal B}_R={\mathcal B}_R(X,\Eta)$ such that
\begin{equation}
{\mathcal U}({\mathcal A},{\mathcal S}){\hat{\mathcal H}}= {\mathcal U}({\mathcal B}_R,{\mathcal S}).
\label{s3.83}
\end{equation}
Moreover, ${\mathcal B}_R$ has the following expansion:
\begin{equation}
{\mathcal B}_R={\mathcal A}{\mathcal H}-(-1)^{p(\Eta_j)}
i\hbar\Big\{\partial_{\Eta_j}{\mathcal A}\cdot\partial_{X_j}{\mathcal H}
+{1\over2}(-1)^{p(\Eta_k)}{\mathcal A}\Big(\partial^2_{\Eta_j X_j}{\mathcal H}
+\partial^2_{\Eta_j \Eta_k}{\mathcal S}\cdot\partial^2_{X_k X_j}{\mathcal H}\Big)\Big\}
+ {\mathcal R}_R
\label{s3.84}
\end{equation}
where arguments of ${\mathcal B}_R$, ${\mathcal A}$ and ${\mathcal S}$ are
$(X,\Eta)$ and that of ${\mathcal H}$ is
$((-1)^{p(\Eta)}\partial_{\Eta}{\mathcal S}(X,\Eta),\Eta)$. 
Furthermore, ${\mathcal R}_R(X,\Eta)$ has the following from:
\begin{equation}
{\mathcal R}_R(X,\Eta)={\mathcal R}_{R,i}^{(1)}(X,\Eta)\Eta_i+{\mathcal R}_R^{(0)}(X,\Eta).
\end{equation}
\end{thm}
\par
{\it Proof. } As before, it is enough to calculate formally
which yields \eqref{s3.84}.
\newline
(i) Remarking $\hat{\mathcal H}$ is represented as \eqref{rt4-4}, we have 
\begin{equation}
\begin{aligned}
{\mathcal U}({\mathcal A},{\mathcal S}){\hat{\mathcal H}}u(X)
&=
c_{3,2}^3 
\int d\Xi dY d\Eta\,{\mathcal A}(X,\Xi)
e^{i\hbar^{-1}({\mathcal S}(X,\Xi)+\langle Y|\Eta-\Xi\rangle)} 
{\mathcal H}^0(Y,\Eta)\hat{u}(\Eta)\\ 
&=
c_{3,2}\int d\Eta\,e^{i\hbar^{-1}{\mathcal S}(X,\Eta)} {\mathcal B}_R(X,\Eta)
{\hat u}(\Eta),
\end{aligned}\label{s3.86}
\end{equation}
with
\begin{equation}
{\mathcal B}_R(X,\Eta)
=c_{3,2}^2\int  d\Xi dY\,
{\mathcal A}(X,\Xi)
e^{i\hbar^{-1}({\mathcal S}(X,\Xi)-{\mathcal S}(X,\Eta)+\langle Y|\Eta-\Xi\rangle)}
{\mathcal H}^0(Y,\Eta).
\label{s3.87}\end{equation}
Using
$$
{\mathcal S}(X,\Xi)-{\mathcal S}(X,\Eta)
=(\Xi_j-\Eta_j){\widetilde{\partial_{\Eta_j}{\mathcal S}}}(X,\tilde\Xi,\Eta)
=\langle{\widetilde{{D}_{\Eta}{\mathcal S}}}(X,\tilde\Xi,\Eta)|\Xi-\Eta\rangle,
$$
where
$$
\begin{gathered}
{\widetilde{\partial_{\Eta_j}}}{\mathcal S}(X,\Xi -\Eta,\Eta)=
\int_0^1d\tau \,\partial_{\Eta_j}{\mathcal S}(X,\Eta+\tau(\Xi -\Eta)),\\
{\widetilde{{D}_{\Eta_j}}}{\mathcal S}(X,\Xi -\Eta,\Eta)=(-1)^{p(\Eta_j)}
\int_0^1d\tau \,\partial_{\Eta_j}{\mathcal S}(X,\Eta+\tau(\Xi -\Eta)),
\end{gathered}
$$
we define a change of variables as
\begin{equation}
{\left\{
{\begin{aligned}
&\tilde\Xi = \Xi -\Eta ,\\
&\tilde{Y}=Y- {\widetilde{{D}_{\Eta}{\mathcal S}}}(X,\Xi -\Eta,\Eta)
\end{aligned}}\right.}
\longleftrightarrow
{\left\{
{\begin{aligned}
&\Xi=\Eta+\tilde\Xi,\\
&Y=\tilde{Y}+{\widetilde{{D}_{\Eta}{\mathcal S}}}(X, \tilde\Xi,\Eta).
\end{aligned}}\right.}
\label{s3.90}\end{equation}
Rewriting \eqref{s3.86},
we get
\begin{equation}
{\mathcal B}_R(X,\Eta)=
c_{3,2}^2 \int d{\tilde\Xi} d{\tilde{Y}}\,
e^{-i\hbar^{-1}
\langle{\tilde{Y}}|{\tilde\Xi}\rangle}
{\mathcal A}(X,\tilde\Xi+\Eta)
{\mathcal H}^0\Big({\tilde{Y}}+
{\widetilde{{D}_{\Eta}{\mathcal S}}}(X,\tilde\Xi,\Eta),\Eta\Big).
\label{s3.91}
\end{equation}
(ii) Using Taylor's expansion for ${\mathcal H}^0(\cdots)$
w.r.t. $\tilde{Y}$, we decompose
\begin{equation*}
\begin{aligned}
{\mathcal H}^0(\tilde{Y}+
{\widetilde{{D}_{\Eta}{\mathcal S}}}(X,\tilde\Xi,\Eta),\Eta)
={\mathcal H}^0({\widetilde{{D}_{\Eta}{\mathcal S}}}(X,\tilde\Xi,\Eta),\Eta)
&+{\tilde{Y}}_j
{\partial}_{X_j}
{\mathcal H}^0({\widetilde{{D}_{\Eta}{\mathcal S}}}(X,\tilde\Xi,\Eta),\Eta)\\
&\qquad\qquad
+{\tilde{Y}}_j{\tilde{Y}}_k
{\widetilde{\partial^2_{X_kX_j}{\mathcal H}^0}}
(X,{\tilde\Xi},{\tilde{Y}},\Eta)
\end{aligned}
\end{equation*}
where
\begin{equation}
{\widetilde{\partial^2_{X_kX_j}{\mathcal H}^0}}
(X,{\tilde\Xi},{\tilde{Y}},\Eta)=
\int_0^1 d\tau_1(1-\tau_1)
\partial^2_{X_kX_j}{\mathcal H}(\tau_1{\tilde{Y}}+
{\widetilde{{D}_{\Eta}{\mathcal S}}}(X,\tilde\Xi,\Eta),\Eta).
\label{tild}\end{equation}
So, we put
$$
{\mathcal B}_R(X,\Eta)
=I_1+I_2+I_3,
$$
where
\begin{equation}
I_1=
c_{3,2}^2 \int  d{\tilde\Xi} d{\tilde{Y}}\,
e^{-i\hbar^{-1}\langle{\tilde{Y}}|{\tilde\Xi}\rangle}
{\mathcal A}(X,\tilde\Xi+\Eta)
{\mathcal H}^0\Big(
{\widetilde{{D}_{\Eta}{\mathcal S}}}(X,\tilde\Xi,\Eta),\Eta\Big),
\label{s3.92}
\end{equation}
\begin{equation}
I_2=c_{3,2}^2 \int  d{\tilde\Xi} d{\tilde{Y}} \,
e^{-i\hbar^{-1}\langle{\tilde{Y}}|{\tilde\Xi}\rangle}
{\mathcal A}(X,\tilde\Xi+\Eta)
\tilde{Y}_j{\partial}_{X_j}
{\mathcal H}^0\Big(
{\widetilde{{D}_{\Eta}{\mathcal S}}}(X,\tilde\Xi,\Eta),\Eta\Big),
\label{s3.93}
\end{equation}
and
\begin{equation}
I_3=c_{3,2}^2 \int  d{\tilde\Xi} d{\tilde{Y}}\,
e^{-i\hbar^{-1}\langle{\tilde{Y}}|{\tilde\Xi}\rangle}
{\mathcal A}(X,\tilde\Xi+\Eta)
{\tilde{Y}}_j{\tilde{Y}}_k
{\widetilde{\partial^2_{X_kX_j}{\mathcal H}^0}}
(X,{\tilde\Xi},{\tilde{Y}},\Eta).
\label{s3.94}
\end{equation}
(iii) Using \eqref{s-delta},
we get readily
\begin{equation}
I_1=
{\mathcal A}(X,\tilde\Xi+\Eta)
{\mathcal H}^0({\widetilde{{D}_{\Eta}{\mathcal S}}}(X,\tilde\Xi,\Eta),\Eta)
\big|_{\tilde\Xi=0}=
{\mathcal A}(X,\Eta){\mathcal H}^0({\partial_{\Eta}}{\mathcal S}(X,\Eta),\Eta).
\label{s3.95}
\end{equation}
Remarking the second equality of \eqref{bp},
integration by parts and applying \eqref{s-delta}, we get
\begin{equation}
\begin{aligned}
I_2&=(-1)^{1-p(\tilde\Xi_j)}{i\hbar}
\partial_{\tilde\Xi_j}\Big[{\mathcal A}(X,\tilde\Xi+\Eta)
\partial_{X_j}
{\mathcal H}^0({\widetilde{{D}_{\Eta}{\mathcal S}}}
(X,\tilde\Xi,\Eta),\Eta)
\Big]\Big|_{\tilde\Xi=0}\\
&=(-1)^{1-p(\tilde\Xi_j)}i\hbar
\Big[\partial_{\Eta_j}{\mathcal A}\cdot \partial_{X_j}{\mathcal H}^0
+\frac 12 (-1)^{p(\Eta_k)}{\mathcal A}\cdot\partial^2_{\Eta_j \Eta_k}{\mathcal S}
\cdot \partial^2_{X_kX_j}{\mathcal H}^0\Big]
\end{aligned}
\end{equation}
with arguments of ${\mathcal A}$, ${\mathcal S}$ are $(X,\Eta)$ and that of ${\mathcal H}^0$ is
$(\partial_{\Eta}{\mathcal S}(X,\Eta),\Eta)$.
Therefore, we have the main terms of \eqref{s3.84}
by adding $I_1+I_2$.
\newline
(iv) Using the second equality of \eqref{bp} twice, we get
\begin{equation}
\begin{aligned}
I_3&=c_{3,2}^2 \int  d{\tilde\Xi} d{\tilde{Y}}\,
e^{-i\hbar^{-1}\langle{\tilde{Y}}|{\tilde\Xi}\rangle}
{\mathcal A}(X,\tilde\Xi+\Eta)
{\tilde{Y}}_j{\tilde{Y}}_k
{\widetilde{\partial^2_{X_kX_j}{\mathcal H}^0}}
(X,{\tilde\Xi},{\tilde{Y}},\Eta)\\
&=-(-1)^{p(\tilde\Xi_j)(1+p(\Eta_k))}i\hbar
c_{3,2}^2 \int d{\tilde\Xi} d{\tilde{Y}} \,
e^{-i\hbar^{-1}\langle{\tilde{Y}}|{\tilde\Xi}\rangle}
{\tilde{Y}}_k 
\Big[\partial_{\tilde\Xi_j}{\mathcal A}(X,\tilde\Xi+\Eta)
{\widetilde{\partial^2_{X_kX_j}{\mathcal H}^0}}
(X,{\tilde\Xi},{\tilde{Y}},\Eta)\\
&\qquad\qquad\qquad\qquad\qquad\qquad\qquad\qquad
+{\mathcal A}(X,\tilde\Xi+\Eta)
\partial_{\tilde\Xi_j}[{\widetilde{\partial^2_{X_kX_j}{\mathcal H}^0}}
(X,{\tilde\Xi},{\tilde{Y}},\Eta)]\Big]\\
&=(-1)^{(1+p(\tilde\Xi_j)(1+p(\tilde\Xi_k))}\hbar^2
c_{3,2}^2 \int  d{\tilde\Xi} d{\tilde{Y}} \,
e^{-i\hbar^{-1}\langle{\tilde{Y}}|{\tilde\Xi}\rangle}
\Big\{
\partial^2_{\tilde\Xi_k \tilde\Xi_j}{\mathcal A}\cdot
{\widetilde{\partial^2_{X_kX_j}{\mathcal H}^0}}\\
&\qquad\qquad\qquad\qquad\qquad\qquad\qquad\qquad
+2\partial_{\tilde\Xi_j}{\mathcal A}\cdot\partial_{\tilde\Xi_k}
[{\widetilde{\partial^2_{X_kX_j}{\mathcal H}^0}}]
+{\mathcal A}\,\partial^2_{\tilde\Xi_j \tilde\Xi_k}
[{\widetilde{\partial^2_{X_kX_j}{\mathcal H}^0}}]
\Big\}
\end{aligned}%\label{s3.}
\end{equation}
where in the last equality, the argument of ${\mathcal A}$ is $(X,\tilde\Xi+\Eta)$ 
and that of ${\widetilde{\partial^2_{X_kX_j}{\mathcal H}^0}}$ is 
$(X,{\tilde\Xi},{\tilde{Y}},\Eta)$.

Remarking \eqref{s-delta} once more, we have
\begin{equation} 
\partial_{\tilde\Xi_j}[{\widetilde{\partial^2_{X_kX_j}{\mathcal H}^0}}
(X,{\tilde\Xi},{\tilde{Y}},\Eta)]
=\partial_{\tilde\Xi_j}[{\widetilde{{D}_{\Eta_\ell}{\mathcal S}}}]\cdot\!
\int_0^1d\tau_1(1-\tau_1)\partial^3_{X_\ell X_kX_j}
{\mathcal H}^0\Big(\tau_1{\tilde{Y}}+
{\widetilde{{D}_{\Eta}{\mathcal S}}}(X,\tilde\Xi,\Eta),\Eta\Big),
\end{equation}
and
\begin{equation}\begin{aligned}
&\partial^2_{\tilde\Xi_j \tilde\Xi_k}[{\widetilde{\partial^2_{X_kX_j}{\mathcal H}^0}}
(X,{\tilde\Xi},{\tilde{Y}},\Eta)]\\
&\qquad=\partial^2_{\tilde\Xi_j \tilde\Xi_k}[{\widetilde{{D}_{\Eta_\ell}{\mathcal S}}}]
\int_0^1d\tau_1(1-\tau_1)\partial^3_{X_\ell X_kX_j}
{\mathcal H}^0\Big(\tau_1{\tilde{Y}}+
{\widetilde{{D}_{\Eta}{\mathcal S}}}(X,\tilde\Xi,\Eta),\Eta\Big)\\
&\qquad\qquad
+(-1)^{p(\tilde\Xi_k)(p(\Eta_\ell)+p(\tilde\Xi_j))}\partial_{\tilde\Xi_j}[{\widetilde{{D}_{\Eta_\ell}{\mathcal S}}}]\cdot
\partial_{\tilde\Xi_k}[{\widetilde{{D}_{\Eta_n}{\mathcal S}}}]\\
&\qquad\qquad\qquad
\times\int_0^1d\tau_1(1-\tau_1)
\partial^4_{X_n X_\ell X_kX_j}
{\mathcal H}^0\Big(\tau_1{\tilde{Y}}+
{\widetilde{{D}_{\Eta}{\mathcal S}}}(X,\tilde\Xi,\Eta),\Eta\Big).
\end{aligned}
\end{equation}

Therefore, we get
\begin{equation}
{\mathcal B}_R={\mathcal A}{\mathcal H}^0-(-1)^{p(\tilde\Xi_j)}i\hbar\Big\{\partial^2_{\Eta_j X_j}{\mathcal H}^0
+(-1)^{p(\tilde\Eta_k)}{1\over2}{\mathcal A}\cdot\Big(\partial^2_{\Eta_jX_j}{\mathcal H}^0
+\partial^2_{\Eta_j \Eta_k}{\mathcal S}\cdot\partial^2_{X_kX_j}{\mathcal H}^0\Big)\Big\}
+ {\mathcal R}_R
\label{s3.102}
\end{equation}
with
$$
{\mathcal R}_R(X,\Eta)=I_{31}+I_{32}+I_{33}
$$
where
\begin{equation}
\begin{aligned}
I_{31}&=(-1)^{(1+p(\tilde\Xi_j))(1+p(\tilde\Xi_k))}\hbar^2
c_{3,2}^2 \int  d{\tilde\Xi} d{\tilde{Y}} \,
e^{-i\hbar^{-1}\langle{\tilde{Y}}|{\tilde\Xi}\rangle}
\partial^2_{\tilde\Xi_k \tilde\Xi_j}{\mathcal A}\cdot
{\widetilde{\partial^2_{X_kX_j}{\mathcal H}^0}},\\
I_{32}&=(-1)^{(1+p(\tilde\Xi_j))(1+p(\tilde\Xi_k))}\hbar^2
c_{3,2}^2 \int  d{\tilde\Xi} d{\tilde{Y}} \,
e^{-i\hbar^{-1}\langle{\tilde{Y}}|{\tilde\Xi}\rangle}
2\partial_{\tilde\Xi_j}{\mathcal A}\cdot\partial_{\tilde\Xi_k}
[{\widetilde{\partial^2_{X_kX_j}{\mathcal H}^0}}],\\
I_{33}&=(-1)^{(1+p(\tilde\Xi_j))(1+p(\tilde\Xi_k))}\hbar^2
c_{3,2}^2 \int  d{\tilde\Xi} d{\tilde{Y}} \,
e^{-i\hbar^{-1}\langle{\tilde{Y}}|{\tilde\Xi}\rangle}
{\mathcal A}\,\partial^2_{\tilde\Xi_j \tilde\Xi_k}
[{\widetilde{\partial^2_{X_kX_j}{\mathcal H}^0}}].
\end{aligned}
\end{equation}

Finally, we estimate ${\mathcal R}_R(X,\Eta)$ using the structure of ${\mathcal H}^0$.
As
$$
\partial^2_{X_4X_5}{\mathcal H}^0(X,\Xi)
=\partial^2_{\theta_1\theta_2}{\mathcal H}^0(X,\Xi)
=-c(\xi_1+i\xi_2),
$$
using \eqref{tild}, we get
$$
{\widetilde{\partial^2_{X_4X_5}{\mathcal H}^0}}
(X,{\tilde\Xi},{\tilde{Y}},\Eta)
=-\frac{c}2(\eta_1+i\eta_2).
$$

Therefore,
$$
I_{31}=I_{311}+I_{312}
$$
with
$$
\begin{aligned}
I_{311}&=\hbar^2
c_{3,2}^4 \int  d{\tilde\Xi} d{\tilde{Y}} \,
e^{-i\hbar^{-1}\langle{\tilde{Y}}|{\tilde\Xi}\rangle}
\partial^2_{\tilde\Xi_k \tilde\Xi_j}{\mathcal A}\cdot
{\mathcal S}_{ijk}(X,{\tilde\Xi},{\tilde{Y}},{\tilde{Z}},\Eta)
({\tilde\Eta}_i+{\tilde\Xi}_i+\Eta_i),\\
I_{312}&=\hbar^2
c_{3,2}^4 \int  d{\tilde\Xi} d{\tilde{Y}} \,
e^{-i\hbar^{-1}\langle{\tilde{Y}}|{\tilde\Xi}\rangle}
\partial^2_{\tilde\Xi_k \tilde\Xi_j}{\mathcal A}\cdot
\Psi_{jk}(X,{\tilde\Xi},{\tilde{Y}},{\tilde{Z}},\Eta).
\end{aligned}
$$
By integration by parts, we get
$$
\begin{aligned}
I_{311}&=-\frac{i\hbar^3}2
c_{3,2}^2 \int  d{\tilde\Xi} d{\tilde{Z}}\,
e^{-i\hbar^{-1}\langle{\tilde{Z}}|{\tilde\Xi}\rangle}
\partial^2_{\tilde\Xi_k \tilde\Xi_j}{\mathcal A}\cdot
\tilde{\mathcal S}_{jk}(X,{\tilde\Xi},0,{\tilde{Z}},\Eta)\\
&\qquad+\hbar^2
c_{3,2}^2 \int  d{\tilde\Xi} d{\tilde{Z}}\,
e^{-i\hbar^{-1}\langle{\tilde{Z}}|{\tilde\Xi}\rangle}
\partial^2_{\tilde\Xi_k \tilde\Xi_j}{\mathcal A}\cdot
{\mathcal S}_{ijk}(X,{\tilde\Xi},0,{\tilde{Z}},\Eta)
\,\Eta_i
\end{aligned}
$$
where
$$
\tilde{\mathcal S}_{jk}(X,{\tilde\Xi},0,{\tilde{Z}},\Eta)
=\int_0^1 d\tau_1\tau_1(1-\tau_1)
\partial^3_{X_iX_kX_j}
A_i\Big(\tau_1{\tilde{Z}}+
{\widetilde{{D}_{\Eta}{\mathcal S}}}(X,\tilde\Xi,\Eta)\Big),
$$
and also
$$
I_{312}=\hbar^2
c_{3,2}^2 \int  d{\tilde\Xi} d{\tilde{Z}}\,
e^{-i\hbar^{-1}\langle{\tilde{Z}}|{\tilde\Xi}\rangle}
\partial^2_{\tilde\Xi_k \tilde\Xi_j}{\mathcal A}\cdot
\Psi_{jk}(X,{\tilde\Xi},0,{\tilde{Z}},\Eta).
$$
Therefore, applying the same procedures to $I_{32}$ and $I_{33}$, we get
$$
{\mathcal R}_{R,i}^{(1)}(X,\Eta)=\hbar^2
c_{3,2}^2 \int  d{\tilde\Xi} d{\tilde{Z}}\,
e^{-i\hbar^{-1}\langle{\tilde{Z}}|{\tilde\Xi}\rangle}
\partial^2_{\tilde\Xi_k \tilde\Xi_j}{\mathcal A}\cdot
{\mathcal S}_{ijk}(X,{\tilde\Xi},0,{\tilde{Z}},\Eta)
+\cdots,
$$
$$
{\mathcal R}_R^{(0)}(X,\Eta)=-\frac{i\hbar^3}2
c_{3,2}^2 \int  d{\tilde\Xi} d{\tilde{Z}}\,
e^{-i\hbar^{-1}\langle{\tilde{Z}}|{\tilde\Xi}\rangle}
\partial^2_{\tilde\Xi_k \tilde\Xi_j}{\mathcal A}\cdot
\tilde{\mathcal S}_{jk}(X,{\tilde\Xi},0,{\tilde{Z}},\Eta)
+I_{312}+\cdots.  \qed
$$

%%%

%%\input s-Weyl5-1
\section{Proofs of Theorem 2.6 and Theorem 2.7.}
\subsection{The infinitesimal generator}

Let $u\in\ccsl_{{\mathrm{SS}},0}(\fR^{3|2})$. 
As
$$
i\hbar \pdt \big({\mathcal A}e^{i\hbar^{-1}{\mathcal S}}\big)
=(i\hbar{\mathcal A}_t-{\mathcal S}_t{\mathcal A})e^{i\hbar^{-1}{\mathcal S}},
$$
using the Hamilton-Jacobi equation, the continuity equation and 
the composition formula \eqref{s3.46},
we have
\begin{equation}
\begin{aligned}
i\hbar  \pdt {\mathcal U}(t,s)u(X)
&=c_{3,2}\int_{{\fR}^{3|2}}d\Xi\,
\big\{{\mathcal H}{\mathcal A}-i\hbar\big[{\mathcal A}_{X_j}{\mathcal H}_{\Xi_j}
+\frac12{\mathcal A}({\mathcal H}_{X_j\Xi_j}
+{\mathcal S}_{X_j\Xi_k}{\mathcal H}_{\Xi_k\Xi_j})\big]\big\}
e^{i\hbar^{-1}{\mathcal S}}\hat u(\Xi)\\
&={\hat{\mathcal H}}(X,\partial_X){\mathcal U}(t,s)u(X)-{\mathcal R}_L(t,s)u(X)
\end{aligned}\label{5.1}
\end{equation}
with
$$
{\mathcal R}_L(t,s)u=c_{3,2}\int d\Xi\, 
{\mathcal R}_L(t,s\,;X,\Xi)e^{i\hbar^{-1}{\mathcal S}(t,s\,;X,\Xi)}
\hat u(\Xi).
$$

Moreover, we have
\begin{prop}
\begin{equation}
\Vert {\mathcal R}_L(t,s)u\Vert \le C|t-s|\Vert u\Vert .
\label{5.2}
\end{equation}
\end{prop}
\par{\it Proof. } %as the estimate \eqref{s3.47}
Using the estimates \eqref{rt3.84} and \eqref{rt3.99},
we get 
$$
|\pi_{\mathrm B}\partial_x^{\fa}\partial_\Xi^{\fb}
{\mathcal R}_L(t,s\,;X,\Xi)|\le C_{\fa,\fb}|t-s|.
$$
Therefore, we may proceed as we did
in proving Theorem 3?.3 to have \eqref{5.2}.  \qed

%%%

%%\input s-Weyl5-21
\subsection{Evolutional property}
The following theorem gives one of the main estimate
necessary to apply Theorem A.1 of \cite{In99-1} to our case.
\begin{thm}\label{evol}
Let $u\in\ccsl_{{\mathrm {SS}},0}(\fR^{3|2})$. 
If $|t-s|+|s-r|$ is sufficiently small, we have
\begin{equation}
\Vert {\mathcal U}(t,s){\mathcal U}(s,r)u
-{\mathcal U}(t,r)u\Vert \le C(|t-s|^2+|s-r|^2)\Vert u\Vert. 
\label{5.3}
\end{equation}
\end{thm}

By definition, we have
$$
{\mathcal U}(t,r)\,{\underline{u}}(X)
=c_{3,2}\int_{\fR^{3|2}}d\Xi\, %(2\pi\hbar)^{-3/2}\hbar
{\mathcal D}^{1/2}({t},r\,;{X,\Xi})
e^{i\hbar^{-1}{\mathcal S}({t},r\,;{X,\Xi})}
{\mathcal F}{\underline{u}}(\Xi),
$$
and
\begin{equation}
\begin{aligned}
{\mathcal U}(t,s){\mathcal U}(s,r)\,{\underline{u}}(X)
&=c_{3,2}\int_{\fR^{3|2}}d\Eta\,
{\mathcal D}^{1/2}(t,s\,;{X,\Eta})
e^{i\hbar^{-1}{\mathcal S}(t,s\,;{X,\Eta})}
{\mathcal F}({\mathcal U}(s,r)\,{\underline{u}})(\Eta)\\
&=c_{3,2}^3\int_{\fR^{3|2}\times\fR^{3|2}\times\fR^{3|2}}
d\Eta\,dY\,d\Xi\,
{\mathcal D}^{1/2}(t,s\,;{X,\Eta}){\mathcal D}^{1/2}(s,r\,;{Y,\Xi})\\
&\qquad\qquad\qquad\times
e^{i\hbar^{-1}
({\mathcal S}(t,s\,;{X,\Eta})-\langle Y|\Eta\rangle+{\mathcal S}(s,r\,;{Y,\Xi}))}
{\mathcal F}{\underline{u}}(\Xi)].
\end{aligned}
\label{5.4}
\end{equation}

To prove Theorem \ref{evol}, we follow the procedure of Taniguchi \cite{Tan99}.
For the sake of notational simplicity, 
we use the following abbreviation:
\begin{equation}
\begin{gathered}
\phi_1({X,\Xi})={\mathcal S}(t,s\,;{X,\Xi}),\;
\phi_2({X,\Xi})={\mathcal S}(s,r\,;{X,\Xi}),\;
\phi({X,\Xi})={\mathcal S}(t,r\,;{X,\Xi}),\\
\mu_1({X,\Xi})={\mathcal D}^{1/2}(t,s\,;{X,\Xi}),\;
\mu_2({X,\Xi})={\mathcal D}^{1/2}(s,r\,;{X,\Xi}),\;
\mu({X,\Xi})={\mathcal D}^{1/2}(t,r\,;{X,\Xi}),\\
\lambda_1=|t-s|,\;\lambda_2=|s-r|,\; \lambda=\lambda_1+\lambda_2.
\end{gathered}\label{5.5}
\end{equation}

First of all, we prepare
\begin{lem}
There exists a unique solution  $(\tilde X, \tilde \Xi)$, 
$\tilde X=(\tilde x,\tilde\theta)$, $\tilde\Xi=(\tilde\xi,\tilde\pi)$ of
\begin{equation}
\left\{
\begin{aligned}
\tilde X_A&=(-1)^{p(\Xi_A)}\partial_{\Xi_A}\phi_1(X,\tilde\Xi),\\
\tilde\Xi_A&=\partial_{X_A}\phi_2(\tilde X,\Xi), 
%\for A=1,2,\cdots,3+2, %(-1)^{p(X)}
\end{aligned}\right.
\mbox{i.e.}
\left\{
\begin{aligned}
&\tilde x_j=\partial_{\xi_j}\phi_1(X,\tilde\Xi),\;
\tilde\theta_k=-\partial_{\pi_k}\phi_1(X,\tilde\Xi),\\
&\tilde\xi_j=\partial_{x_j}\phi_2(\tilde X,\Xi),\;
\tilde\pi_k=\partial_{\theta_k}\phi_2(\tilde X,\Xi). 
\end{aligned}\right.
\label{5.6}\end{equation}
Moreover, we have, for ${\frak a}=(\alpha,a), {\frak b}=(\beta,b)$,
\begin{equation}
\begin{aligned}
&|\pi_{\mathrm B}\partial_X^{\frak a}\partial_\Xi^{\frak b}(\tilde X-X)|
\le C_{\fa,\fb}\lambda(1+|x_{\mathrm B}|+|\xi_{\mathrm B}|)^{(1-|\alpha+\beta|)_+},\\
&|\pi_{\mathrm B}\partial_X^{\frak a}\partial_\Xi^{\frak b}(\tilde \Xi-\Xi)|
\le C_{\fa,\fb}\lambda(1+|x_{\mathrm B}|+|\xi_{\mathrm B}|)^{(1-|\alpha+\beta|)_+}.
\end{aligned}
\label{5.7}
\end{equation}
\end{lem}
\par{\it Proof. }
Putting 
$$
\begin{gathered}
J_\alpha(X,\Xi)=\phi_\alpha(X,\Xi)-\langle X|\Xi\rangle \et
\tilde X=X+Y,\; \tilde\Xi=\Xi+\Eta,\\
\with X=(x,\theta),\; Y=(y,\omega),\; \Xi=(\xi,\pi),\; \Eta=(\eta,\rho),
\end{gathered}
$$ 
we rewrite \eqref{5.6} as
\begin{equation}
\left\{
\begin{aligned}
&y_j=\partial_{\xi_j}J_1(X,\Xi+\Eta),\quad
\omega_k=-\partial_{\pi_k}J_1(X,\Xi+\Eta),\\
%&Y_A=(-1)^{p(\Xi_A)}\partial_{\Xi_A} J_1(X,\Xi+\Eta),\\
&\eta_j=\partial_{x_j}J_2(X+Y,\Xi),\quad
\rho_k=\partial_{\theta_k}J_2(X+Y,\Xi),\\
%&\Eta_A=\partial_{X_A} J_2(X+Y,\Xi), %(-1)^{p(X_A)}
\end{aligned}\right.
\label{5.8}
\end{equation}
where
$$
y=\sum_{\ell=0}^\infty y^{[2\ell]},\; y^{(2\ell)}=\sum_{j=0}^\ell y^{[2j]},\;
\omega=\sum_{\ell=0}^\infty \omega^{[2\ell+1]},\; 
\omega^{(2\ell+1)}=\sum_{j=0}^\ell \omega^{[2j+1]},\;
etc.
$$
Defining the map 
${\mathcal T}:(Y,\Eta)\to
((-1)^{p(\Xi)}\partial_\Xi J_1(X,\Xi+\Eta),\partial_X J_2(X+Y,\Xi))$,
we claim that there exists $\delta_0$, 
such that if $\lambda\le\delta_0$, 
then there exists a fixed point of the map ${\mathcal T}$.

To prove this claim, we decompose, for $\alpha=1,2$,
$$
J_\alpha(X,\Xi)=J_{\alpha,\bar0\bar0}
+J_{\alpha,\bar1\bar0}\theta_1\theta_2
+\sum_{j,k=1}^2J_{\alpha,c_j d_k}\theta^{c_j}\pi^{d_k}
+J_{\alpha,\bar0\bar1}\pi_1\pi_2
+J_{\alpha,\bar1\bar1}\theta_1\theta_2\pi_1\pi_2
$$
where $J_{\alpha,**}=J_{\alpha,**}(x,\xi)$ and
$\bar 0=(0,0)$, $\bar 1=(1,1)$, $c_1=(1,0)=d_1$,
$c_2=(0,1)=d_2\in\{0,1\}^2$.

{\bf Existence}:
We consider the body part of \eqref{5.8}.
$$
\left\{
\begin{aligned}
&y_j^{[0]}=\partial_{\xi_j} J_1(x^{[0]},0,\xi^{[0]}+\eta^{[0]},0)
=\partial_{\xi_j} J_{1,\bar0\bar0}(x^{[0]},\xi^{[0]}+\eta^{[0]}),
\; j=1,2,3,\\
&\eta_j^{[0]}=\partial_{x_j} J_2(x^{[0]}+y^{[0]},0,\xi^{[0]},0)
=\partial_{x_j}J_{2,\bar0\bar0}(x^{[0]}+y^{[0]},\xi^{[0]}),
\; j=1,2,3,
\end{aligned}\right.
$$
with 
$J_{\alpha,\bar0\bar0}(x^{[0]},\xi^{[0]})
=\phi_{\alpha\mathrm B}(x^{[0]},\xi^{[0]})
-x^{[0]}\cdot\xi^{[0]}$ for $(\alpha=1,2)$.
%In the above, we abbuse symbols in the sense that
%for notational simplicity, we put $y$ as representing $y_{\mathrm B}$, etc.
By Theorem 1.7 of \cite{KTT}, there exists $\delta_0$, 
such that if $|t-s|+|s-r|\le\delta_0$, 
the unique existence of solutions $y^{[0]}$, $\eta^{[0]}$ of this equation
is garanteed.
Moreover, the estimate is also established in Theorem 1.7${}'$ of \cite{KTT}.

Substituting these $y^{[0]}$, $\eta^{[0]}$ into \eqref{5.8}, we get
$$
\left\{
\begin{aligned}
\omega_1^{[1]}&=-\partial_{\pi_1} J_1(x^{[0]},\theta^{[1]},
\xi^{[0]}+\eta^{[0]},\pi^{[1]}+\rho^{[1]})\\
&=\sum_{j=1}^2J_{1,c_jd_1}(x^{[0]},\xi^{[0]}+\eta^{[0]})
(\theta^{c_j})^{[1]}
-J_{1,\bar0\bar1}(x^{[0]},\xi^{[0]}+\eta^{[0]})
(\pi_2^{[1]}+\rho_2^{[1]}),\\
\omega_2^{[1]}&=-\partial_{\pi_2} J_1(x^{[0]},\theta^{[1]},
\xi^{[0]}+\eta^{[0]},\pi^{[1]}+\rho^{[1]})\\
&=\sum_{j=1}^2J_{1,c_jd_2}(x^{[0]},\xi^{[0]}+\eta^{[0]})
(\theta^{c_j})^{[1]}
+J_{1,\bar0\bar1}(x^{[0]},\xi^{[0]}+\eta^{[0]})
(\pi_1^{[1]}+\rho_1^{[1]}),\\
\rho_1^{[1]}&=\partial_{\theta_1} J_2(x^{[0]}+y^{[0]},
\theta^{[1]}+\omega^{[1]},\xi^{[0]},\pi^{[1]})\\
&=J_{2,\bar1\bar0}(x^{[0]}+y^{[0]},\xi^{[0]})(\theta_2^{[1]}+\omega_2^{[1]})
+\sum_{k=1}^2J_{2,c_1d_k}(x^{[0]}+y^{[0]},\xi^{[0]})(\pi^{d_k})^{[1]},\\
\rho_2^{[1]}&=\partial_{\theta_2} J_2(x^{[0]}+y^{[0]},
\theta^{[1]}+\omega^{[1]},\xi^{[0]},\pi^{[1]})\\
&=-J_{2,\bar1\bar0}(x^{[0]}+y^{[0]},\xi^{[0]})(\theta_1^{[1]}+\omega_1^{[1]})
+\sum_{k=1}^2J_{2,c_2d_k}(x^{[0]}+y^{[0]},\xi^{[0]})(\pi^{d_k})^{[1]}.
\end{aligned}\right.
$$
Clearly, the components of the right-hand side above are the given data.

For the part of degree 2, we have
$$
\left\{
\begin{aligned}
y_i^{[2]}&=\partial_{\xi_i} J_1(x^{(2)},\theta^{[1]},
\xi^{(2)}+\eta^{(2)},\pi^{[1]}+\rho^{[1]})\\
&=\partial_{\xi_i}J_{1,\bar0\bar0}(x,\xi+\eta)^{[2]}+
\partial_{\xi_i}J_{1,\bar1\bar0}(x^{[0]},\xi^{[0]}+\eta^{[0]})
\theta_1^{[1]}\theta_2^{[1]}\\
&\qquad\qquad
+\sum_{j,k=1}^2\partial_{\xi_i}J_{1,c_jd_k}(x^{[0]},\xi^{[0]}+\eta^{[0]})
(\theta^{c_j}(\pi+\rho)^{d_k})^{[2]}
+\partial_{\xi_i}J_{1,\bar0\bar1}(x^{[0]},\xi^{[0]}+\eta^{[0]})
(\pi_1^{[1]}+\rho_1^{[1]})(\pi_2^{[1]}+\rho_2^{[1]}),\\
\eta_i^{[2]}&=\partial_{x_i}J_2(x^{(2)}+y^{(2)},\theta^{[1]}+\omega^{[1]},
\xi^{(2)},\pi^{[1]})\\
&=\partial_{x_i}J_{2,\bar0\bar0}(x+y,\xi)^{[2]}+
\partial_{x_i}J_{2,\bar1\bar0}(x^{[0]}+y^{[0]},\xi^{[0]})
(\theta_1^{[1]}+\omega_1^{[1]})(\theta_2^{[1]}+\omega_2^{[1]})\\
&\qquad\qquad
+\sum_{j,k=1}^2\partial_{x_i}J_{2,c_jd_k}(x^{[0]}+y^{[0]},\xi^{[0]})
((\theta+\omega)^{c_j}\pi^{d_k})^{[2]}
+\partial_{x_i}J_{1,\bar0\bar1}(x^{[0]}+y^{[0]},\xi^{[0]})
\pi_1^{[1]}\pi_2^{[1]}.
\end{aligned}\right.
$$
Here, $\partial_{\xi_j}J_{1,\bar0\bar0}(x,\xi)^{[2]}=\sum_{i=1}^m
\big[
\partial_{\xi_j}\partial_{x_i}J_{1,\bar0\bar0}(x^{[0]},\xi^{[0]})x_i^{[2]}
+\partial_{\xi_j}\partial_{\xi_i}J_{1,\bar0\bar0}(x^{[0]},\xi^{[0]})\xi_i^{[2]}
\big]$, etc,
and the right-hand sides are represented by the given data.

Then, analogously we have, for $\ell\ge1$,
$$
\left\{
\begin{aligned}
&\omega^{[2\ell+1]}=\partial_\pi J_1(x^{(2\ell)},\theta^{(2\ell+1)},
\xi^{(2\ell)}+\eta^{(2\ell)},\pi^{(2\ell+1)}+\rho^{(2\ell+1)}),\\
&\rho^{[2\ell+1]}=\partial_\theta J_2(x^{(2\ell)}+y^{(2\ell)},
\theta^{(2\ell+1)}+\omega^{(2\ell+1)},\xi^{(2\ell)},\pi^{(2\ell+1)}),
\end{aligned}
\right.
$$
and
$$
\left\{
\begin{aligned}
&y^{[2\ell+2]}=\partial_\xi J_1(x^{(2\ell+2)},\theta^{(2\ell+1)},
\xi^{(2\ell+2)}+\eta^{(2\ell+2)},\pi^{(2\ell+1)}+\rho^{(2\ell+1)}),\\
&\eta^{[2\ell+2]}=\partial_x J_2(x^{(2\ell+2)}
+y^{(2\ell+2)},\theta^{(2\ell+1)}+\omega^{(2\ell+1)},
\xi^{(2\ell+2)},\pi^{(2\ell+1)}).
\end{aligned}
\right.
$$
\noindent
{\bf Estimate}:
We proceed as we did in proving Proposition 3?.3.  
\qed

Putting the $\#$-product $\phi_1\#\phi_2$ of $\phi_1$ and $\phi_2$ by
$$
\phi_1\#\phi_2(X,\Xi)=\phi_1(X,\tilde\Xi)-\langle\tilde X|\tilde\Xi\rangle
+\phi_2(\tilde X,\Xi), %\Phi(X,\Xi)=
$$
we have
\begin{lem}
\begin{equation}
\left\{
\begin{aligned}
\partial_{X_A}\phi(X,\Xi)&=\partial_{\Xi_A}\phi_1(X,\tilde\Xi),\\
\partial_{\Xi_A}\phi(X,\Xi)&=\partial_{X_A}\phi_2(\tilde X,\Xi)
\for A=1,\cdots,5.
\end{aligned}\right.
\label{5.9}
\end{equation}
\end{lem}
\par{\it Proof. }
By the above definition of $\phi_1\#\phi_2$, we get
$$
\begin{aligned}
\partial_{X_A}\phi_1\#\phi_2(X,\Xi)&=\partial_{X_A}\phi_1(X,\tilde\Xi)
+\partial_{X_A}\tilde\Xi_C\,\partial_{\Xi_C}\phi_1(X,\tilde\Xi)
-\partial_{X_A}\tilde X_C\,\tilde{\Xi_C}
-\partial_{X_A}\tilde\Xi_C\, (-1)^{p(\Xi_C)}\tilde X_C\\
&\qquad\qquad\qquad\qquad\qquad\qquad
+\partial_{X_A}\tilde X_C\,\partial_{X_C}\phi_2(\tilde X,\Xi)\\
&=\partial_{X_A}\phi_1(X,\tilde\Xi)
+\partial_{X_A}\tilde\Xi_C\,(-1)^{p(\Xi_C)}\tilde X_C
-\partial_{X_A}\tilde X_C\,\tilde\Xi_C
-\partial_{X_A}\tilde\Xi_C\,(-1)^{p(\Xi_C)}\tilde X_C\\
&\qquad\qquad\qquad\qquad\qquad\qquad
+\partial_{X_A}\tilde X_C\,\tilde\Xi_C %(-1)^{p(X)}
=\partial_{X_A}\phi_1(X,\tilde\Xi).
\end{aligned}
$$
Same holds for $\partial_\Xi\phi(X,\Xi)$.   \qed

\begin{lem}
\begin{equation}
\phi_1\#\phi_2(X,\Xi)={\mathcal S}(t,r\,;{X,\Xi})=\phi({X,\Xi})
\label{5.10}
\end{equation}
\end{lem}
\par{\it Proof. }
We differentiate $\phi_1\#\phi_2(X,\Xi)$ w.r.t. $s$
to have
$$
\pds \phi_1\#\phi_2(X,\Xi)=0.
$$
As $\phi_1\#\phi_2(X,\Xi)|_{s=r}=\phi_1\#\phi_2(X,\Xi)|_{s=t}
={\mathcal S}(t,r,X,\Xi)$, we get the result. 
More precisely, repeat the arguments in proving Lemma 5.2
of \cite{In99-1} with necessary modifications. \qed

Now, we have the following:
\begin{prop}
Under the same assumption as above, we have
\begin{equation}
|\pi_{\mathrm B}\partial_X^{\fa}\partial_\Xi^{\fb}
(\mu_1(X,\tilde\Xi)\mu_2(\tilde X,\Xi)-\mu(X,\Xi))|
\le C_{\fa, \fb}\lambda^2.
\label{5.11}
\end{equation}
\end{prop}

{\it Notation}:
In the following, for a small parameter $\delta>0$,
we denote by ${\mathcal S}^0_0[\delta]$, the class of functions
$p(X,\Xi)\in\ccsl_{{\mathrm S}\!{\mathrm S},0}(\fR^{3|2})$ satisfying
$$
|\pi_{\mathrm B}\partial_X^{\fa}\partial_\Xi^{\fb}p(X,\Xi)|
\le C_{\fa,\fb}\delta
$$
with a constant $C_{\fa,\fb}$ independent of $\delta$.

\par{\it Proof. }
Differentiate the first equation of \eqref{5.9} w.r.t. $\Xi$ to have
$$
\partial_{\Xi}\partial_{X}\phi(X,\Xi)=\partial_{\Xi}\tilde\Xi
\partial_{\Xi}\partial_{X}\phi_1(X,\tilde\Xi).
$$
On the other hand, from \eqref{5.6}, we have
$$
\left\{
\begin{aligned}
\partial_{\Xi}\tilde{X}&=(-1)^{p(\Xi)}\partial_{\Xi}\tilde\Xi
\partial_{\Xi}\partial_{\Xi}\phi_1(X,\tilde\Xi),\\
\partial_{\Xi}\tilde{\Xi}&=\partial_{\Xi}\tilde{X}
\partial_{X}\partial_{X}\phi_2(\tilde X,\Xi)+
\partial_{\Xi}\partial_{X}\phi_2(\tilde X,\Xi).
\end{aligned}\right.
$$
Substituting the first equation above into the second one, we get
$$
\partial_{\Xi}\tilde{\Xi}[I-
(-1)^{p(\Xi)}\partial_{\Xi}\partial_{\Xi}\phi_1(X,\tilde\Xi)
\partial_{X}\partial_{X}\phi_2(\tilde X,\Xi)]
=\partial_{\Xi}\partial_{X}\phi_2(\tilde X,\Xi).
$$
Because of
$$
|\pi_{\mathrm B}(-1)^{p(\Xi)}\partial_{\Xi}\partial_{\Xi}\phi_1(X,\tilde\Xi)
\partial_{X}\partial_{X}\phi_2(\tilde X,\Xi)|
\le \delta_0<1,
$$
we have
$$
\begin{aligned}
\partial_{\Xi}\partial_{X}\phi(X,\Xi)&=
\partial_{\Xi}\partial_{X}\phi_2(\tilde X,\Xi)\\
&\qquad\times[I-
(-1)^{p(\Xi)}\partial_{\Xi}\partial_{\Xi}\phi_1(X,\tilde\Xi)
\partial_{X}\partial_{X}\phi_2(\tilde X,\Xi)]^{-1}
\partial_{\Xi}\partial_{X}\phi_1(X,\tilde\Xi).
\end{aligned}
$$

We prove that there exists $q_1(X,\Xi)$ such that
$$
\sdet(I-(-1)^{p(\Xi)}\partial_{\Xi}\partial_{\Xi}\phi_1(X,\tilde\Xi)
\partial_{X}\partial_{X}\phi_2(\tilde X,\Xi))=1+q_1(X,\Xi).
$$
Moreover, by the same argument of Proposition 1.5 of \cite{KTT}, we have
$$
\pi_{\mathrm B}(1+q_1(X,\Xi))\ge (1-\delta_0)^m>0,
$$
which yields
$$
\sdet\partial_{\Xi}\partial_{X}\phi(X,\Xi)
=\sdet(\partial_{\Xi}\partial_{X}\phi_2(\tilde X,\Xi))\cdot
(1+q_1(X,\Xi))^{-1}\cdot
\sdet(\partial_{\Xi}\partial_{X}\phi_1(X,\tilde\Xi)).
$$
Taking the square root of both sides, and 
remarking the elements of the right-hand side are even, we have
$$
\mu(X,\Xi)=\mu_1(X,\tilde\Xi)\mu_2(\tilde X,\Xi)+q_2(X,\Xi)
$$
with
$$
\begin{aligned}
q_2(X,\Xi)&=\mu_1(X,\tilde\Xi)\mu_2(\tilde X,\Xi)
[(1+q_1(X,\Xi))^{-1/2}-1]\\
&=-\mu_1(X,\tilde\Xi)\mu_2(\tilde X,\Xi)
\frac{q_1(X,\Xi)}{{\sqrt{1+q_1(X,\Xi)}}(1+\sqrt{1+q_1(X,\Xi)})}. 
\end{aligned}
$$
Then, we have readily that $q_2(X,\Xi)\in{\mathcal S}^0_0[\lambda^2]$. \qed

Rewriting \eqref{5.6} by
$$
Y=\tilde X+\bar Y, \; \Eta=\tilde\Xi+\bar\Eta,
$$
we have
$$
\phi_1({X,\Eta})-\langle Y|\Eta\rangle+\phi_2({Y,\Xi})-\phi(X,\Xi)
=-\langle \bar Y|\tilde\Eta\rangle+{\mathcal R}(X,\Xi,\bar Y,\bar \Eta)
$$
where
$$
{\mathcal R}(X,\Xi,\bar Y,\bar \Eta)=\Psi_1(X,\Xi,\bar\Eta)+\Psi_2(X,\Xi,\bar Y)
$$
with
$$
\begin{aligned}
\Psi_1(X,\Xi,\bar\Eta)&=\Psi_1(\bar\Eta)=\phi_1(X,\tilde\Xi+\bar \Eta)-\phi_1(X,\tilde\Xi)
-\langle \tilde X|\bar \Eta\rangle,\\
\Psi_2(X,\Xi,\bar Y)&=\Psi_2(\bar Y)=\phi_2(\tilde X+\bar Y,\Xi)-\phi_2(\tilde X,\Xi)
-\langle \bar Y|\tilde\Xi\rangle.
\end{aligned}
$$

Therefore, we may rewrite \eqref{5.6} as
$$
{\mathcal U}(t,s){\mathcal U}(s,r)u(X) %-{\mathcal U}(t,r)u(X)
=c_{3,2}\int_{\fR^{3|2}}d\Xi\,{\mathcal B}(X,\Xi)e^{i\hbar^{-1}\phi(X,\Xi)}
{\mathcal F}u(\Xi),
$$
with
$$
{\mathcal B}(X,\Xi)=c_{3,2}^2\int_{\fR^{3|2}\times\fR^{3|2}}
d\bar \Eta\,d\bar Y\,
\mu_1(X,\tilde\Xi+\bar \Eta)\mu_2(\tilde X+\bar Y,\Xi)
e^{i\hbar^{-1}({\mathcal R}(X,\Xi,\bar Y,\bar \Eta)
-\langle \bar Y|\bar \Eta\rangle)}.
%-\mu(X,\Xi).
$$
Now, we want to prove
\begin{prop}
\begin{equation}
|\pi_{\mathrm B}\partial_X^{\fa}\partial_\Xi^{\fb}
({\mathcal B}(X,\Xi)-\mu_1(X,\tilde\Xi)\mu_2(\tilde X,\Xi))|
\le C_{\fa, \fb}\lambda^2.
\label{5.13}
\end{equation}
\end{prop}

%%%

%%\input s-Weyl5-22

To prove this Proposition, we remark
\begin{lem}
$\Psi_1(X,\Xi,\bar\Eta)$ and $\Psi_2(X,\Xi,\bar Y)$ satisfy
\begin{equation}
\begin{aligned}
&|\pi_{\mathrm B}\partial_X^{\fa}\partial_\Xi^{\fb}\Psi_1(X,\Xi,\bar\Eta)|
\le C_{\fa,\fb}|\pi_{\mathrm B}\bar\Eta|^2,\\
&|\pi_{\mathrm B}\partial_X^{\fa}\partial_\Xi^{\fb}\Psi_2(X,\Xi,\bar{Y})|
\le C_{\fa,\fb}|\pi_{\mathrm B}\bar{Y}|^2,
\end{aligned}
\label{5.14}
\end{equation}
\begin{equation}
\begin{aligned}
&|\pi_{\mathrm B}\partial_X^{\fa}\partial_\Xi^{\fb}
\partial_{\bar\Eta}^{\fc}\partial_{\bar\Eta}\Psi_1(X,\Xi,\bar\Eta)|
\le C_{\fa,\fb,\fc}\langle\pi_{\mathrm B}\bar\Eta\rangle,\\
&|\pi_{\mathrm B}\partial_X^{\fa}\partial_\Xi^{\fb}
\partial_{\bar\Eta}^{\fc}\partial_{\bar{Y}}\Psi_2(X,\Xi,\bar{Y})|
\le C_{\fa,\fb,\fc}\langle\pi_{\mathrm B}\bar{Y}\rangle.
\end{aligned}
\label{5.15}
\end{equation}
and for $\fc\ne1$,
\begin{equation}
\begin{aligned}
&|\pi_{\mathrm B}\partial_{\bar\Eta}^{\fc}\partial_{\bar\Eta}
\Psi_1(X,\Xi,\bar\Eta)|
\le C_{\fc},\\
&|\pi_{\mathrm B}\partial_{\bar\Eta}^{\fc}\partial_{\bar{Y}}
\Psi_2(X,\Xi,\bar{Y})|
\le C_{\fc}.
\end{aligned}
\label{5.16}
\end{equation}
\end{lem}
\par{\it Proof. } 
As $\Psi_1(\bar\Eta)$ is represented by
$$
\Psi_1(\bar\Eta)=\bar \Eta\bar \Eta
\int_0^1d\tau(1-\tau)\phi_{1,\Xi\Xi}(X,\tilde\Xi+\tau\bar \Eta),
$$
we get readily the first inequalities in \eqref{5.14}. Rewriting 
\begin{equation}
\begin{aligned}
\partial_{\bar\Eta}\Psi_1(\bar\Eta)
&=\partial_\Xi\phi_1-(-1)^{p(\bar\Eta)}\tilde X
=\partial_{\bar\Eta}J_1-(\tilde X-X)\\
&=\bar\Eta\int_0^1d\tau
\partial_\Xi\partial_\Xi\phi_1(X,\tilde\Xi+\tau\bar\Eta).
\end{aligned}
\label{5.16?}
\end{equation}
Then, we get the first inequality of \eqref{5.15}
from the last expression of \eqref{5.16?}
and the first inequality of \eqref{5.16}
from the third expression of \eqref{5.16?}.
Similar arguments work for other inequalities.  \qed

Remarking
$$
e^{i\hbar^{-1}\Psi_2}
=1+i\hbar^{-1}\Psi_2\int_0^1 d\tau e^{i\tau\hbar^{-1}\Psi_2},
$$
we rewrite
\begin{equation}
\begin{aligned}
{\mathcal B}(X,\Xi)=c_{3,2}^2\int_{\fR^{3|2}\times\fR^{3|2}}
d\bar \Eta\,d\bar Y\, &
\mu_1(X,\tilde\Xi+\bar \Eta)\mu_2(\tilde X+\bar Y,\Xi)
e^{i\hbar^{-1}(\Psi_1(X,\Xi,\bar \Eta)
-\langle \bar Y|\bar \Eta\rangle)}\\
&\qquad\qquad
+i\hbar^{-1}\int_0^1d\tau {\mathcal B}_{1,\tau}(X,\Xi)
\end{aligned}\label{5.17}
\end{equation}
where
$$
\begin{aligned}
{\mathcal B}_{1,\tau}(X,\Xi)=c_{3,2}^2\int_{\fR^{3|2}\times\fR^{3|2}}
d\bar \Eta\,d\bar Y\, &
\Psi_2(X,\Xi,\bar{Y})\mu_1(X,\tilde\Xi+\bar \Eta)\mu_2(\tilde X+\bar Y,\Xi)\\
&\qquad\times
e^{i\hbar^{-1}(\Psi_1(X,\Xi,\bar \Eta)+\tau\Psi_2(X,\Xi,\bar{Y})
-\langle \bar Y|\bar \Eta\rangle)}.
\end{aligned}
$$
\begin{lem}
The symbol ${\mathcal B}_{1,\tau}(X,\Xi)$ belongs to 
${\mathcal S}_0^0[\lambda^2]$
uniformly with respect to $\tau$.
\end{lem}
\par{\it Proof. }
Remarking
$$
\Psi_2(\bar Y)=\sum \bar Y_A\tilde\Psi_{2,A}(\bar Y)
\with
\tilde\Psi_{2,A}(\bar Y)=\int_0^1 d\tau\partial_{\bar Y_A}\Psi_2(\tau\bar Y),
$$
we have
$$
\begin{aligned}
{\mathcal B}_{1,\tau}(X,\Xi)&=\sum_{a=1}^5{\mathcal B}_{1,\tau,A}(X,\Xi),\\
{\mathcal B}_{1,\tau,A}(X,\Xi)&=c_{3,2}^2\int_{\fR^{3|2}\times\fR^{3|2}}
d\bar \Eta\,d\bar Y\, 
\bar Y_A\tilde\Psi_{2,A}(X,\Xi,\bar{Y})
\mu_1(X,\tilde\Xi+\bar \Eta)\mu_2(\tilde X+\bar Y,\Xi)
e^{i\hbar^{-1}\Psi^\tau}\\
&=c_{3,2}^2\int_{\fR^{3|2}\times\fR^{3|2}}
d\bar \Eta\,d\bar Y\, \tilde{\mathcal B}_{1,\tau,A}(X,\Xi,\bar Y,\bar\Eta)
e^{i\hbar^{-1}\Psi^\tau}.
\end{aligned}
$$
Here, we put
$$
\begin{gathered}
\Psi^\tau=\Psi_1(X,\Xi,\bar \Eta)+\tau\Psi_2(X,\Xi,\bar{Y})
-\langle \bar Y|\bar \Eta\rangle,\\
\tilde{\mathcal B}_{1,\tau,A}(\bar Y,\bar\Eta)
=\tilde\Psi_{2,A}(\bar Y)
\{\partial_{\bar\Eta}\Psi_1(\Bar \Eta)\cdot\mu_1+i\partial_{\bar\Eta}\mu_1\}
\mu_2,
\end{gathered}
$$
and we used
$$
\partial_{\bar\Eta}(\mu_1e^{i\hbar^{-1}\Psi^\tau})=
\{\partial_{\bar\Eta}\mu_1
+i\hbar^{-1}
(\partial_{\bar\Eta}\Psi_1-(-1)^{p(\bar\Eta)}\bar Y) \mu_1\}e^{i\hbar^{-1}\Psi^\tau}.
$$

Setting
$$
{\mathcal L}_\tau=\frac{1-i\partial_{\bar\Eta}\Psi^\tau\cdot\partial_{\bar\Eta}
-i\partial_{\bar Y}\Psi^\tau\cdot\partial_{\bar Y}}
{1+\hbar^{-1}|\partial_{\bar\Eta}\Psi^\tau|^2+\hbar^{-1}|\partial_{\bar Y}\Psi^\tau|^2},
$$
we have
$$
{\mathcal L}_\tau\Psi^\tau=\Psi^\tau.
$$
We apply the Lax's technique.
That is,
we have
$$
{\mathcal B}_{1,\tau}(X,\Xi)=c_{3,2}^2\int_{\fR^{3|2}\times\fR^{3|2}}
d\bar \Eta\,d\bar Y\, e^{i\hbar^{-1}\Psi^\tau}
\big({\mathcal L}_\tau^*\big)^{2m+3}\tilde{\mathcal B}_{1,\tau,A}(\bar Y,\bar \Eta)\\
$$
We have
$$
|\pi_{\mathrm B}\partial_{\bar Y}^{\fc}
\partial_{\Xi}^{\fb}\partial_{X}^{\fa}\Psi_2(X,\Xi,\bar{Y})|
\le C_{\fa,\fb,\fc}\lambda_2\langle\pi_{\mathrm B}\bar Y\rangle
$$
On the other hand, from
$$
\begin{aligned}
1+|\pi_{\mathrm B}\partial_{\bar\Eta}\Psi^\tau|+|\pi_{\mathrm B}\partial_{\bar Y}\Psi^\tau|
&\ge C\{1+(|\pi_{\mathrm B}\bar Y|-\lambda_1|\pi_{\mathrm B}\bar \Eta|)
+(|\pi_{\mathrm B}\bar\Eta|-\lambda_2|\pi_{\mathrm B}\bar Y|)\}\\
&\ge C(1+|\pi_{\mathrm B}\bar Y|+|\pi_{\mathrm B}\bar\Eta|).
\end{aligned}
$$
and
$$
|\pi_{\mathrm B}\partial_{\bar\Eta}^{\fc}
\{\partial_{\bar\Eta_A}\Psi_1(\bar\Eta)\mu_1+i\partial_{\bar\Eta}\mu_1\}|
\le C_{\fc}\lambda_1\langle\pi_{\mathrm B}\bar\Eta\rangle,
$$
we get
$$
|\pi_{\mathrm B}
\big({\mathcal L}_\tau^*\big)^{2m+3}{\mathcal B}_{1,\tau}(\bar Y,\bar \Eta)|
\le C\lambda^2(1+|\pi_{\mathrm B}\bar Y|+|\pi_{\mathrm B}\bar \Eta|)^{-(2m+1)}
$$
This proves
$$
|\pi_{\mathrm B}\partial_X^{\fa}\partial_\Xi^{\fb}{\mathcal B}_{1,\tau,A}(X,\Xi)|
\le C_{\fa,\fb}\lambda^2.   \qed
$$

Using the Taylor expansion, we may rewrite the first term of \eqref{5.17} as
$$
\begin{aligned}
&c_{3,2}^2\int_{\fR^{3|2}\times\fR^{3|2}}
d\bar \Eta\,d\bar Y\,
\mu_1(X,\tilde\Xi+\bar \Eta)\mu_2(\tilde X+\bar Y,\Xi)
e^{i\hbar^{-1}(\Psi_1(X,\Xi,\bar \Eta)
-\langle \bar Y|\bar \Eta\rangle)}\\
&\qquad
=
c_{3,2}^2\int_{\fR^{3|2}\times\fR^{3|2}}
d\bar \Eta\,d\bar Y\,
\mu_1(X,\tilde\Xi+\bar \Eta)\mu_2(\tilde X,\Xi)
e^{i\hbar^{-1}(\Psi_1(X,\Xi,\bar \Eta)
-\langle \bar Y|\bar \Eta\rangle)}
+i\hbar^{-1}\int_0^1d\tau \sum_{A=1}^5 {\mathcal B}_{2,\tau,A}(X,\Xi)
\end{aligned}
$$
with
$$
\begin{aligned}
{\mathcal B}_{2,\tau,A}(X,\Xi)&=
c_{3,2}^2\int_{\fR^{3|2}\times\fR^{3|2}}
d\bar \Eta\,d\bar Y\,
\mu_1(X,\tilde\Xi+\bar \Eta)
\bar Y_A\partial_{X_A}\mu_2(\tilde X+\tau\bar Y,\Xi)
e^{i\hbar^{-1}(\Psi_1(X,\Xi,\bar \Eta)
-\langle \bar Y|\bar \Eta\rangle)}\\
&=c_{3,2}^2\int_{\fR^{3|2}\times\fR^{3|2}}
d\bar \Eta\,d\bar Y\,
\tilde{\mathcal B}_{2,\tau,A}(\bar Y,\bar\Eta)e^{i\hbar^{-1}\Psi^0},
\end{aligned}.
$$
and
because of
$$
\partial_{\bar\Eta}(\mu_1e^{i\hbar^{-1}\Psi^0})=
\{\partial_{\bar\Eta}\mu_1
+i\hbar^{-1}
(\partial_{\bar\Eta}\Psi_1-(-1)^{p(\bar\Eta)}\bar Y) \mu_1\}
e^{i\hbar^{-1}\Psi^0},
$$
$$
\tilde{\mathcal B}_{2,\tau,A}(\bar Y,\bar\Eta)
=\{\partial_{\bar\Eta_A}\Psi_1\cdot\mu_1
+i\hbar^{-1}\partial_{\bar\Eta_A}\mu_1\}\partial_{X_A}\mu_2.
$$
\begin{lem}
The symbol ${\mathcal B}_{2,\tau,A}(X,\Xi)$ belongs to 
${\mathcal S}_0^0[\lambda^2]$
uniformly with respect to $\tau$.
\end{lem}
\par{\it Proof. } Set
$$
{\mathcal L}_0=\frac{1-i\partial_{\bar\Eta}\Psi^0\cdot\partial_{\bar\Eta}
-i\partial_{\bar Y}\Psi^0\cdot\partial_{\bar Y}}
{1+\hbar^{-1}|\partial_{\bar\Eta}\Psi^0|^2+\hbar^{-1}|\partial_{\bar Y}\Psi^0|^2},
\quad
{\mathcal L}_0e^{i\hbar^{-1}\Psi^0}=e^{i\hbar^{-1}\Psi^0}
$$
and apply the Lax's technique.
We have
$$
{\mathcal B}_{2,\tau}(X,\Xi)=c_{3,2}^2\int_{\fR^{3|2}\times\fR^{3|2}}
d\bar \Eta\,d\bar Y\, e^{i\hbar^{-1}\Psi^0}
\big({\mathcal L}_0^*\big)^{2m+2}\tilde{\mathcal B}_{2,\tau,A}(\bar Y,\bar \Eta).
$$
As
$$
|\pi_{\mathrm B}\partial_{\bar Y}^{\fc}
\{\partial_{\bar\Xi_A}\mu_2(\tilde X+\tau\bar Y,\Xi)\}|
\le C_{\fc}\lambda_2
$$
we get
$$
|\pi_{\mathrm B}
\big({\mathcal L}_0^*\big)^{2m+2}\tilde{\mathcal B}_{2,\tau,A}(\bar Y,\bar \Eta)|
\le C\lambda^2(1+|\pi_{\mathrm B}\bar Y|+|\pi_{\mathrm B}\bar \Eta|)^{-(2m+1)}
$$
This proves
$$
|\pi_{\mathrm B}\partial_X^{\fa}\partial_\Xi^{\fb}{\mathcal B}_{2,\tau,A}(X,\Xi)|
\le C_{\fa,\fb}\lambda^2.   \qed
$$

\begin{cor}
\begin{equation}
|\pi_{\mathrm B}\partial_X^{\fa}\partial_\Xi^{\fb}
({\mathcal B}(X,\Xi)-\mu(X,\Xi))|
\le C_{\fa, \fb}\lambda^2.
\label{5.19}
\end{equation}
\end{cor}
\par{\it Proof. } Combining \eqref{5.11} amd \eqref{5.13},
 we get the desired result.

\par{\it Proof of Proposition \ref{evol}. }
By \eqref{5.19}, we get the desired result \eqref{5.3}.  \qed

%%%

%%\input s-Weyl5-3

\subsection{Proof of Theorem 2.6}

Using Theorem A.1 in \cite{In98-1},
we have our Theorem 2.6 readily.   \qed

\subsection{Proof of Theorem 2.7}

From Theorem 2.6, using identifications $\sharp$ and $\flat$, 
we get the desired results.   \qed

%%%

%%\input s-Weyl6

\section{Concluding remarks}

Though in this paper, we answer one of several problems in Inoue \cite{In98-1},
we give other problems which should be solved:

(i) Propagation of singularity: How one can extend Egorov's theorem
to the system of PDE (see \cite{Cor82})?

(ii) How does the Aharonov-Bohm effect and Berry's phase
appear when Schr\"odinger equation is replaced 
by the Weyl equation (see \cite{Par95})? 
This should be studied by constructing the
fundamental solution of \eqref{rt1.1} using \#-product introduced in
Kumano-go \cite{Kum76}.

%%%

%%\input s-Weylref

%%%

\end{document}